\documentclass[11pt,a4paper]{article}
\pdfoutput=1

\usepackage{jheppub}
\usepackage{graphicx}
\usepackage{xspace}
\usepackage{booktabs}

\makeatletter
\g@addto@macro\bfseries{\boldmath}
\makeatother

\newcommand{\GeV}{\,\mathrm{GeV}}
\newcommand{\TeV}{\,\mathrm{TeV}}

\newcommand{\LLR}{\text{LL$_R$}\xspace}
\newcommand{\NLLR}{\text{NLL$_R$}\xspace}
\newcommand{\LO}{\text{LO}\xspace}
\newcommand{\NLO}{\text{NLO}\xspace}
\newcommand{\NNLO}{\text{NNLO}\xspace}
\newcommand{\NNLOR}{\text{NNLO$_R$}\xspace}
\newcommand{\NNLORK}{\text{NNLO$_{R,K}$}\xspace}
\newcommand{\NNLOLLR}{\text{NNLO+LL$_R$}\xspace}
\newcommand{\NNLORLLR}{\text{NNLO$_R$+LL$_R$}\xspace}
\newcommand{\NNLORKLLR}{\text{NNLO$_{R,K}$+LL$_R$}\xspace}
\newcommand{\NLOLLR}{\text{NLO+LL$_R$}\xspace}

\newcommand{\powheg}{\texttt{POWHEG}\xspace}
\newcommand{\powhegbox}{\texttt{POWHEG BOX V2}\xspace}
\newcommand{\pythia}{\texttt{Pythia}\xspace}
\newcommand{\herwig}{\texttt{Herwig}\xspace}
\newcommand{\herwigpp}{\texttt{Herwig++}\xspace}
\newcommand{\nlojet}{\texttt{NLOJet++}\xspace}

\newcommand{\as}{\alpha_s}
\newcommand{\order}[1]{{\cal O}\left(#1\right)}

\definecolor{darkgreen}{rgb}{0,0.5,0}
\definecolor{darkblue}{rgb}{0,0,0.7}
\definecolor{darkred}{rgb}{0.5,0,0.0}
\definecolor{darkorange}{rgb}{0.8,0.4,0.0}

\newcommand{\la}{\langle}
\newcommand{\ra}{\rangle}
\newcommand{\mean}[1]{\left\la\smash{#1}\right\ra}

\title{Inclusive jet spectrum for small-radius jets}

\author[a]{Mrinal Dasgupta,}
\author[b,c,d]{Fr\'ed\'eric A. Dreyer,}
\author[d,*]{Gavin P.~Salam,\note[*]{On leave from CNRS, UMR 7589, LPTHE, F-75005, Paris, France}}
\author[e]{and Gregory Soyez}
\affiliation[a]{Consortium for
  Fundamental Physics, School of Physics \& Astronomy, University
  of Manchester, Manchester M13 9PL, United Kingdom}
\affiliation[b]{Sorbonne Universit\'es, UPMC
  Univ Paris 06, UMR 7589, LPTHE, F-75005, Paris, France}
\affiliation[c]{CNRS, UMR 7589, LPTHE, F-75005, Paris, France}
\affiliation[d]{CERN, Theoretical Physics Department, CH-1211 Geneva 23, Switzerland}
\affiliation[e]{IPhT, CEA Saclay, CNRS UMR 3681, F-91191 Gif-sur-Yvette, France}

\preprint{CERN-TH/2016-020}


\abstract{ 
  Following on our earlier work on leading-logarithmic
  (LL$_R$) resummations for the properties of jets with a small
  radius, $R$, we here examine the phenomenological considerations for the
  inclusive jet spectrum.
  We discuss how to match the NLO predictions with small-$R$
  resummation. 
  As part of the study we propose a new, physically-inspired
  prescription for fixed-order predictions and their uncertainties.
  We investigate the $R$-dependent part of the next-to-next-to-leading
  order (NNLO) corrections, which is found to be substantial, and
  comment on the implications for scale choices in inclusive jet
  calculations.
  We also examine hadronisation corrections, identifying potential
  limitations of earlier analytical work with regards to their
  $p_t$-dependence.
  Finally we assemble these different elements in order to compare
  matched (N)NLO+\LLR predictions to data from ALICE and ATLAS,
  finding improved consistency for the $R$-dependence of the results
  relative to NLO predictions.
}

\begin{document}
\maketitle

\section{Introduction}

Jets are used in a broad range of physics analyses at colliders and
notably at CERN's large hadron collider (LHC).
The study of jets requires the use of a jet definition and for
many of the algorithms in use today that jet definition involves a
``radius'' parameter, $R$, which determines how far in angle a jet
clusters radiation.
A limit of particular interest is that where $R$ is taken small.
One reason is that contamination from multiple simultaneous $pp$
interactions (``pileup'') is minimised, as is the contribution from
the large underlying event when studying heavy-ion collisions.
Furthermore small-$R$ subjets are often a component of jet
substructure analyses used to reconstruct hadronic decays of highly
boosted weak bosons and top quarks.
Finally a powerful systematic cross-check in precision jet studies
comes from the variation of $R$, e.g.\ to verify that conclusions
about data--theory agreement remain stable.
Such a variation will often bring one into the small-$R$ limit.

In the small-$R$ limit the perturbative series involves terms
$\as^n \ln^n 1/R^2$, where the smallness of $\as^n$ is partially
compensated by large logarithms $\ln^n 1/R^2$.
The first order $\as \ln 1/R^2$ terms (together with the constant)
were first calculated long
ago~\cite{Aversa:1990ww,Aversa:1990uv,Guillet:1990ez} and have been
examined also more recently
\cite{Jager:2004jh,Mukherjee:2012uz,Kaufmann:2014nda}.
About a year ago, the whole tower of leading-logarithmic (\LLR) terms
was determined~\cite{Dasgupta:2014yra}, i.e. $\as^n \ln^n 1/R^2$ for
all $n$, for a range of observables (for related work, see also
Refs.~\cite{Catani:2013oma,Alioli:2013hba}).
Work is also ongoing towards next-to-leading logarithmic accuracy,
NLL$_R$~\cite{Becher:2015hka,Chien:2015cka}, however the concrete
results do not yet apply to hadron-collider jet algorithms.

From the point of view of phenomenological studies, there has so far
been one investigation of the impact of small-$R$ resummation in the
context of jet vetoes in Higgs production~\cite{Banfi:2015pju}. 
Though small-$R$ contributions have a significant impact on the
results, most of their effect (at the phenomenologically relevant $R$
value of $0.4$) is already accounted for in fixed-order and
$p_t$-resummed calculations.
Accordingly the resummation brings only small additional changes. 

In this article, we examine the phenomenological impact of small-$R$
terms for the archetypal hadron-collider jet observable, namely the
inclusive jet spectrum.
This observable probes the highest scales that are accessible at
colliders and is used for constraining parton distribution functions
(PDFs), determining the strong coupling and also in studies of
hard probes heavy-ion collisions.
Two factors can contribute to enhance small-$R$ effects in the
inclusive jet spectrum: firstly, it falls steeply as a
function of $p_t$, which magnifies the impact of any physical effect
that modifies the jet's energy.
Secondly, a wide range of $R$ values has been explored for this
observable, going down as far as $R=0.2$ in measurements in
proton--proton collisions by the ALICE
collaboration~\cite{Abelev:2013fn}.
That wide range of $R$ has been exploited also in studies of ratios
of cross sections at different $R$
values~\cite{Abelev:2013fn,Chatrchyan:2014gia,Ploskon:2009zd,Eckweiler:2011mna,Soyez:2011np}.
For $R=0.2$, \LLR small-$R$ effects can be responsible for up to $40\%$
modifications of the jet spectrum.

A first part of our study (section~\ref{sec:resum}) will be to
establish the region of $R$ where the small-$R$ approximation is valid
and to examine the potential impact of effects beyond the \LLR
approximation. 
This will point to the need to include at the least the subleading
$R$-enhanced terms that arise at next-to-next-to-leading order (NNLO)
and 
motivate us to devise schemes to match \LLR resummation with both NLO
and NNLO calculations (sections~\ref{sec:matching} and
\ref{sec:matching-NNLO} respectively). 
At NLO we will see indications of spuriously small scale dependence and
discuss how to resolve the issue.
Concerning NNLO, since the full calculation is work in
progress~\cite{Currie:2013dwa}, to move forwards we will introduce
an approximation that we refer to as ``NNLO$_R$''.
It contains the full $R$-dependence of the NNLO prediction but misses
an $R$-independent, though $p_t$-dependent, constant term.
By default we will take it to be zero at some reference radius
$R_\text{m}$, but we will also examine the impact of other choices.

In addition to the perturbative contributions at small-$R$, one must
take into account non-perturbative effects, which are enhanced roughly
as $1/R$ at small $R$ and grow large especially at smaller values of
$p_t$.
Two approaches exist for incorporating them, one based on analytic
calculations~\cite{Dasgupta:2007wa}, the other based on the
differences between parton and hadron-level results in Monte Carlo
event generators such as \pythia~\cite{Sjostrand:2006za,Sjostrand:2007gs} and
\herwig~\cite{Corcella:2000bw,Corcella:2002jc,Bahr:2008pv,Bellm:2013hwb}.
Based on our studies in section~\ref{sec:hadronisation}, we
adopt the Monte Carlo approach for our comparisons to data.
These are the subject of section~\ref{sec:data-comparisons}, where we
examine data from the ALICE collaboration~\cite{Abelev:2013fn} at
$R=0.2$ and $0.4$, and from the ATLAS~\cite{Aad:2014vwa} collaboration
at $R=0.4$ and $0.6$.

A broad range of dynamically-generated plots comparing different
theory predictions across a range of rapidities, transverse momenta
and $R$ values can be viewed online~\cite{OnlineTool}.
Furthermore some of the plots included in the arXiv source for this
paper contain additional information in the form of extra pages not
displayed in the manuscript.

\section{Small-$R$ resummation for the inclusive jet spectrum}
\label{sec:resum}

\subsection{Recall of the small-$R$ resummation formalism at \LLR accuracy}
\label{sec:resummation-formalism}

As for the inclusive hadron spectrum \cite{Ellis:1991qj}, the
small-$R$ inclusive ``microjet'' spectrum can be
obtained~\cite{Dasgupta:2014yra} from the convolution of the
leading-order inclusive spectrum of partons of flavour $k$ and
transverse momentum $p_t'$, $\frac{d\sigma^{(k)}}{dp'_t}$,
with the inclusive microjet fragmentation function,
$f^\text{incl}_{\text{jet}/k}(p_t/p'_t,t)$, for producing microjets
carrying a fraction $p_t/p'_t$ of the parton's
momentum, 
\begin{equation}
  \label{eq:LLR-master}
  \sigma^{\LLR}(p_t,R) \equiv
  \frac{d\sigma_\text{jet}^\LLR}{dp_t} =
  \sum_k\int_{p_t}\frac{dp'_t}{p'_t}\,
  f^\text{incl}_{\text{jet}/k}\left( \frac{p_t}{p'_t},t(R,R_0,\mu_R) \right) \,
  \frac{d\sigma^{(k)}}{dp'_t}\,.
\end{equation}
To keep the notation compact, we use $\sigma^{\LLR}(p_t,R)$ to denote
either a differential cross section, or the cross section in a given
$p_t$ bin, depending on the context.
At \LLR accuracy, the small-$R$ effects are entirely contained in the
fragmentation function, which depends on $R$ through the evolution
variable $t$, defined as
\begin{equation}
  \label{eq:t}
  t(R, R_0, \mu_R) = \int_{R^2}^{R_0^2} \frac{d\theta^2}{\theta^2}  
  \frac{\as(\mu_R\,
    \theta/R_0)}{2\pi}
  = \frac{1}{b_0} \ln \frac{1}{1 - \frac{\as(\mu_R)}{2\pi}\, b_0  \ln
    \frac{R_0^2}{R^2} } \,,
\end{equation}
with $b_0 = \frac{11 C_A - 4 T_R n_f}{6}$.\footnote{The choice of
  whether to use $\as(\mu_R\, \theta/R_0)$ or $\as(\mu_R\, \theta)$ in
  the integral is arbitrary. We choose the former because it ensures
  that $\as(\mu_R)$ factorises from the logarithm of $R$ in the
  right-hand side.}
Here, $R_0$ is the angular scale, of order $1$, at which 
the small-radius approximation starts to become valid. 
For $R=R_0$, or equivalently $t=0$, the fragmentation function has the
initial condition $f^\text{incl}_{\text{jet}/k}(z,0) = \delta(1-z)$.
It can be determined for other $t$ values by solving a DGLAP-like
evolution equation in $t$~\cite{Dasgupta:2014yra}.
These results are identical for any standard hadron-collider jet
algorithm of the generalised-$k_t$
\cite{KtHH,Ellis:1993tq,Dokshitzer:1997in,Wobisch:1998wt,Cacciari:2008gp}
and SISCone~\cite{Salam:2007xv} families, with differences between
them appearing only at subleading logarithmic order.
In this work we will restrict our attention to the anti-$k_t$
algorithm (as implemented in \texttt{FastJet}~v3.1.3~\cite{Cacciari:2011ma}).
The \LLR resummation is implemented with the aid of
\texttt{HOPPET}~\cite{Salam:2008qg}.

\begin{figure}
  \centering
  \includegraphics[width=0.48\textwidth]{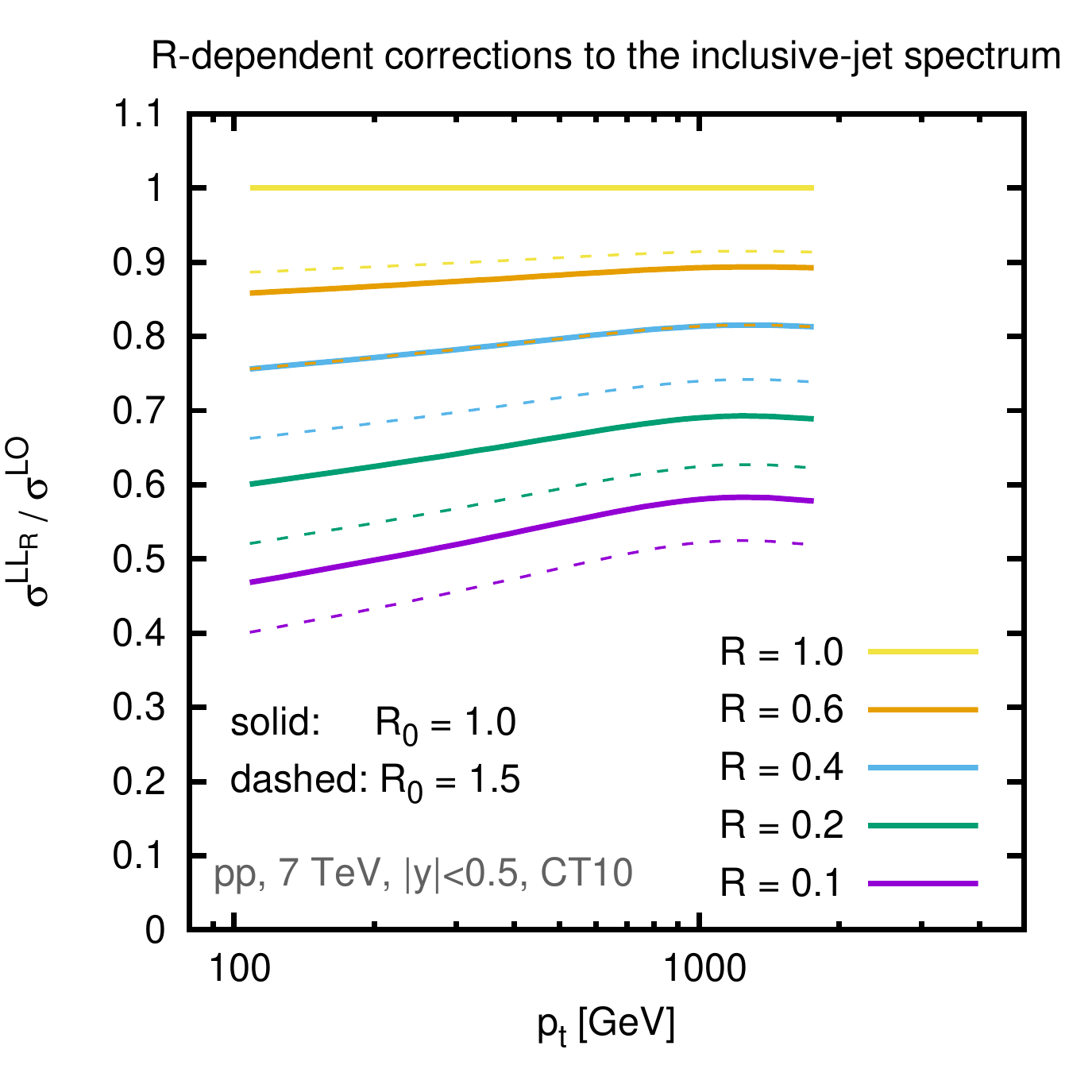} 
  \caption{
    Impact of $R$-dependent terms in the inclusive-jet spectrum,
    illustrated using the small-$R$ resummation factor obtained from
    the ratio of 
    $\sigma^\LLR$ in Eq.~(\ref{eq:LLR-master}) to the leading order 
    inclusive jet spectrum $\sigma^\LO$. It is shown as a function of the
    jet $p_t$ for different jet radius values.
    For each $R$ value, the plot illustrates the impact of two choices
    of $R_0$: $R_0= 1$ (our default) as solid lines and $R_0 = 1.5$ as
    dashed lines.
  }
  \label{fig:smallR-resummation-impact}
\end{figure}

The phenomenological relevance of the small-$R$ terms is illustrated
in Fig.~\ref{fig:smallR-resummation-impact}, which shows the ratio of
the jet spectrum with small-$R$ resummation effects to the LO jet
spectrum.\footnote{Although for this purpose we have used the
  small-$R$ resummed calculations, we could also have used NLO
  calculations which would also indicate similarly visible
  $R$-dependent effects. The differences between the small-$R$
  resummation and NLO predictions will be discussed later.}
For this and a number of later plots, the $p_t$ and rapidity ranges
and the collider energy choice correspond to those of ATLAS
measurements~\cite{Aad:2014vwa}, to which we will later compare our
results.
We show the impact of resummation for a range of $R$ values from $0.1$
to $1.0$ and two $R_0$ choices.
The smallest $R$ values typically in use experimentally are in the
range $0.2$--$0.4$ and one sees that the fragmentation of partons into
jets brings up to $40\%$ reduction in the cross section for the
smaller of these radii.
The fact that the small-$R$ effects are substantial is one of the
motivations for our work here.

From the point of view of phenomenological applications, the question
that perhaps matters more is the impact of corrections beyond NLO (or
forthcoming NNLO), since fixed order results are what are most
commonly used in comparisons to data.
%
This will be most easily quantifiable when we discuss matched results
in sections~\ref{sec:matching} and \ref{sec:matching-NNLO}.
Note that there was already some level of discussion of effects beyond
fixed order in Ref.~\cite{Dasgupta:2014yra}, in terms of an expansion
in powers of $t$.
However comparisons to standard fixed order refer to an expansion in
$\alpha_s$, which is what we will be using throughout this article.
A brief discussion of the different features of $t$ and $\alpha_s$
expansions is given in Appendix~\ref{sec:convergence-t-v-as}.
%
%

\subsection{Range of validity of the small-$R$ approximation and
  effects beyond \LLR}
\label{sec:smallR-validation}

In order to carry out a reliable phenomenological study of small-$R$
effects it is useful to ask two questions about the validity of our
\LLR small-$R$ approach.
Firstly we wish to know from what radii the underlying small-angle
approximation starts to work.
Secondly, we want to determine the potential size of small-$R$ terms
beyond \LLR accuracy.

\begin{figure}
  \centering
  \includegraphics[width=0.48\textwidth]{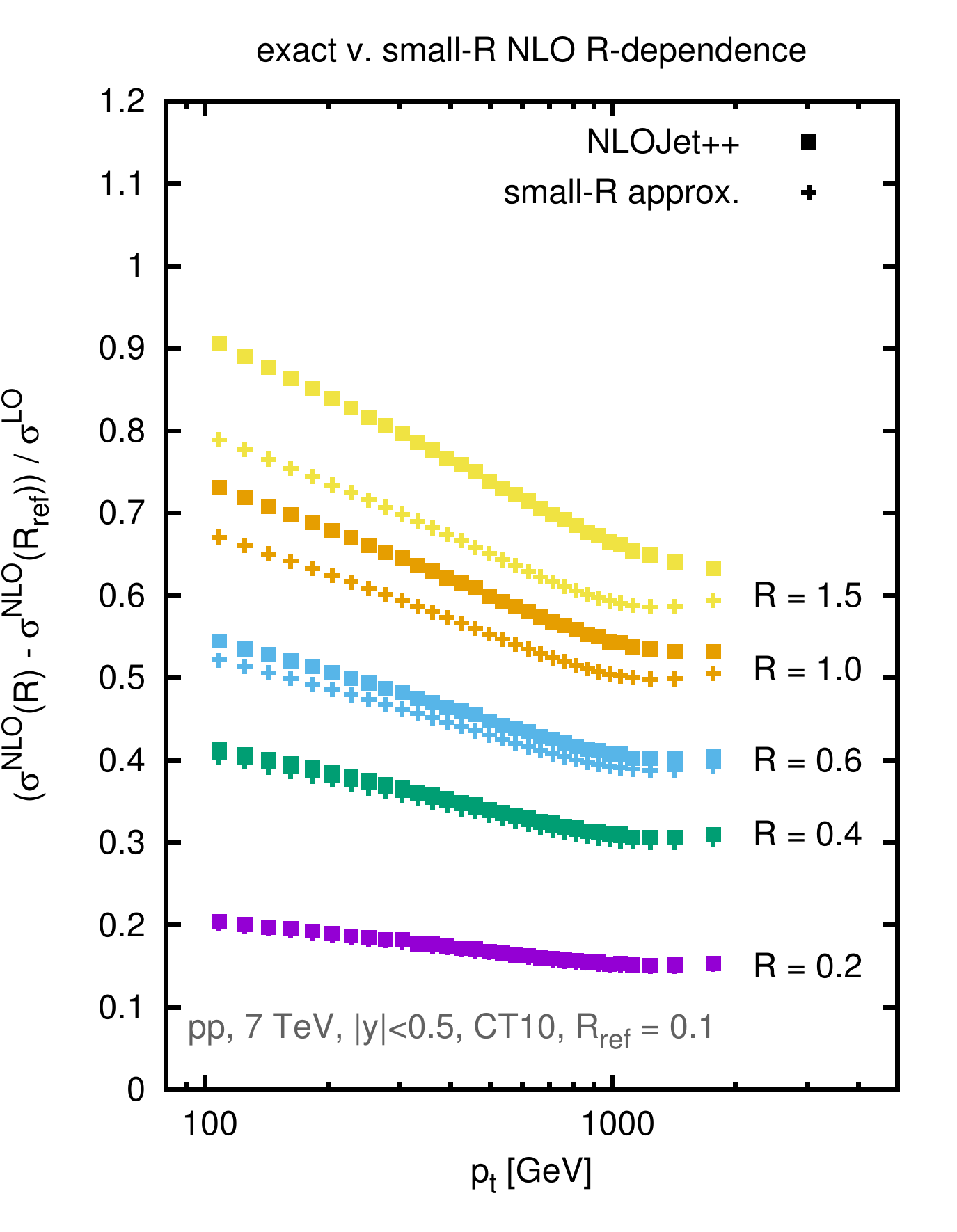}
  \includegraphics[width=0.48\textwidth]{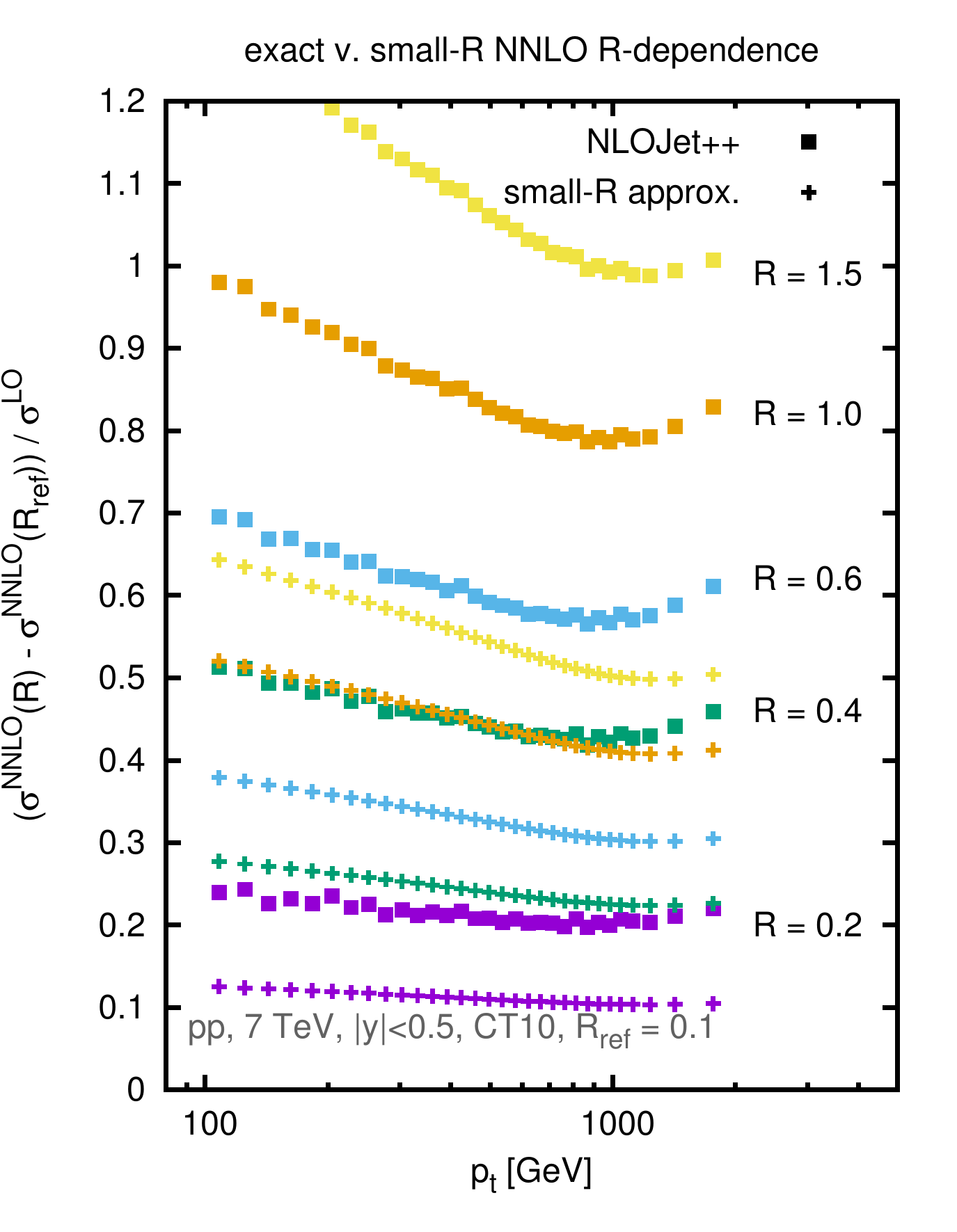}
  \caption{%
    Left: Comparison of the $R$ dependence in the exact and
    small-$R$ approximated NLO expansion, using
    Eq.~(\ref{eq:LLR-validation-ratio}), shown as a 
    function of jet transverse momentum $p_t$, for $\sqrt{s} = 7\TeV$ in the
    rapidity region $|y| < 0.5$.
    Right: comparison of $\Delta_{1+2}(p_t, R, R_\text{ref})$ and
    $\Delta_{1+2}^\LLR(p_t, R, R_\text{ref})$
    (cf.\ Eq.~(\ref{eq:LLR-validation-ratio-Delta12})). 
    In both plots CT10 NLO PDFs~\cite{Lai:2010vv} are used, while  
    the renormalisation and factorisation scales are set equal to the
    $p_t$ of the highest-$p_t$ $R=1$ jet in the event (this same scale
    is used for all  $R$ choices in  the final jet finding).
  }
  \label{fig:smallR-validation}
\end{figure}

To investigate the first question we take the full
next-to-leading-order (NLO) calculation for the inclusive jet spectrum
from the \nlojet program~\cite{nlojet}, and look at the 
quantity $\Delta_1(p_t,R,R_\text{ref})$, where
\begin{equation}
  \label{eq:LLR-validation-ratio}
  \Delta_i(p_t,R,R_\text{ref}) \equiv \frac{\sigma_i(p_t,R) -
    \sigma_i(p_t,R_\text{ref})}{\sigma_0(p_t)}\,. 
\end{equation}
Here $\sigma_i(p_t)$ corresponds to the order $\alpha_s^{2+i}$
contribution to the inclusive jet cross section in a given bin of
$p_t$.
This can be compared to a similar ratio,
$\Delta_1^{\LLR}(p_t,R,R_\text{ref})$, obtained from the NLO expansion
of Eq.~(\ref{eq:LLR-master}) instead of the exact NLO
result.\footnote{$\Delta_1^{\LLR}(p_t,R,R_\text{ref})$ is independent
  of $R_0$ because the $R_0$ cancels between the two terms in the
  numerator.}
The quantity $R_\text{ref}$ here is some small
reference radius at which one expects the small-$R$
approximation to be valid; we choose $R_\text{ref} = 0.1$.
Fig.~\ref{fig:smallR-validation} (left) shows the comparison of
$\Delta_1$ (filled squares) and $\Delta_1^{\LLR}$ (crosses) as a
function of $p_t$ for several different $R$ values.
One sees very good agreement between $\Delta_1$ and $\Delta_1^{\LLR}$
for the smaller $R$ values, while the agreement starts to break down
for $R$ in the vicinity of $1$--$1.5$.
This provides grounds for using the small-$R$ approximation for $R$
values $\lesssim 0.6$ and motivates a choice of $R_0$ in range
$1$--$1.5$. 
We will take $R_0 = 1$ as our default, and use $R_0=1.5$ as a probe of
resummation uncertainties.

Next let us examine effects of subleading small-$R$ logarithms, terms
that come with a factor $\as^n \ln^{n-1} R$ relative to the Born cross
section.
While there has been some work investigating such classes of terms in
Refs.~\cite{Becher:2015hka,Chien:2015cka}, those results do not apply
to hadron-collider jet algorithms.
Instead, here we examine the $R$ dependence in the NNLO part of the
inclusive jet cross section to evaluate the size of these terms.
Because the $R$ dependence starts only at order $\as^3$, we can use
the NLO 3-jet component of the \nlojet program to determine these
terms.
More explicitly, we use the fact that
\begin{equation}
  \label{eq:nlo3j-nnlo}
  \sigma^{\NNLO}(R)-\sigma^{\NNLO}(R_\text{ref})
  = \sigma^{\NLO_{3j}}(R) - \sigma^{\NLO_{3j}}(R_\text{ref})\,.
\end{equation}
To determine this difference in practice, for each event in the
\nlojet 3-jet run we apply the following procedure: we cluster the
event with radius $R$ and for each resulting jet add the event weight
to the jet's corresponding $p_t$ bin; we then recluster the particles
with radius $R_\text{ref}$ and for each jet subtract the event weight
from the corresponding $p_t$ bin.
For this procedure to give a correct answer, it is crucial not to have
any 3-jet phasespace cut in the NLO 3-jet calculation (i.e.\ there is
no explicit requirement of a 3rd jet). \footnote{%
  %
  Note that we have encountered issues with the convergence of the
  \nlojet calculation, with some bins showing extremely large
  excursions in individual runs.
  To obtain stable results, we perform a combination of a large number
  of runs (order $2000{-}4000$) in which each bin's weight from a
  given run is inversely proportional to the square of its statistical
  error.
  Given that such weighted combinations are known to give biased
  results, we then apply a global correction factor to
  $\Delta_{1+2}(p_t,R,R_{\text{ref}})$ across all bins.
  That factor is equal to the ratio of $\int_{p_{t,\min}}^{p_{t,\max}}
  dp_t\, \Delta_{1+2}(p_t,R,R_{\text{ref}})$,  as
  obtained from a bin-wise unweighted combination (with removal of a
  few percent of outlying runs in each bin) and a bin-wise weighted
  combination (an alternative approach to a similar issue was recently
  discussed in Ref.~\cite{Ridder:2016rzm}).
  We believe that the systematics associated with this procedure are
  at the level of a couple of percent.
}

Hence, we can then examine
\begin{equation}
  \label{eq:LLR-validation-ratio-Delta12}
  \Delta_{1+2}(p_t,R,R_\text{ref}) \equiv
  \Delta_{1}(p_t,R,R_\text{ref}) + 
  \Delta_{2}(p_t,R,R_\text{ref})
\end{equation}
and its corresponding \LLR approximation,
$\Delta_{1+2}^\LLR(p_t,R,R_\text{ref})$.
The reason for including both NLO and NNLO terms is to facilitate
comparison of the size of the results with that of the pure NLO piece.
The results for $\Delta_{1+2}$ (filled squares) and
$\Delta_{1+2}^\LLR$ (crosses) are shown in
Fig.~\ref{fig:smallR-validation} (right).
The difference between the crosses in the left-hand and right-hand
plots is indicative of the size of the NNLO \LLR contribution. 
At small $R$, the difference between the crosses and solid squares in
the right-hand plot gives the size of the NLL$_R$ contribution at
NNLO.
It is clear that this is a substantial contribution, of the same order
of magnitude as the \LLR contribution itself, but with the opposite sign.
Ideally one would therefore carry out a full NLL$_R$ calculation.
That, however, is beyond the scope of this article. 

Instead we will include a subset of the subleading $\ln R$ terms by
combining the \LLR resummation with the exact $R$ dependence up to
NNLO fixed order, i.e.\ the terms explicitly included in the solid
squares in Fig.~\ref{fig:smallR-validation}.

\section{Matching NLO and \LLR}
\label{sec:matching}

For phenomenological predictions, it is necessary to combine the
\LLR resummation with results from fixed-order calculations.
In this section we will first examine how to combine \LLR and NLO
results, and then proceed with a discussion of NNLO corrections.

\subsection{Matching prescriptions}
\label{sec:matching-NLO}

One potential approach for combining \LLR and NLO results would be to
use an additive type matching,
\begin{equation}
  \label{eq:additive-matching}
  \sigma^{\NLO+\LLR, \text{add.}}(R) = \sigma^\LLR(R) + \sigma_1(R) -  \sigma^{\LLR}_1(R)\,,
\end{equation}
where $\sigma_1(R)$ denotes the pure NLO contribution to the inclusive
jet spectrum (without the LO part, as in
section~\ref{sec:smallR-validation}) and $\sigma^{\LLR}_1(R)$ refers to
the pure NLO contribution within the \LLR resummation.
For compactness, the $p_t$ argument in the cross sections has been
left implicit.

A simple, physical condition that the matching must satisfy is that in
the limit $R\to 0$, the ratio of the matched result to the LO result
should tend to zero perturbatively, \footnote{ 
  Once non-perturbative
  effects are accounted for, $\sigma^\LLR(R=0)$ must coincide with the
  inclusive hadron spectrum.}
\begin{equation}
  \label{eq:matching-ratio-condition}
  \frac{\sigma^{\NLO+\LLR}}{\sigma_0}\, \to 0\qquad\text{for $R\to0$}\,.
\end{equation}
Eq.~(\ref{eq:additive-matching}) does not satisfy this property: while
$\sigma^\LLR/\sigma_0$ does tend to zero, the quantity
$(\sigma_1 - \sigma^{\LLR}_1)/\sigma_0$ instead tends to a constant
for small $R$.
We will therefore not use additive matching.

Another class of matching procedure is multiplicative matching.
One simple version of multiplicative matching is given by
\begin{equation}
  \label{eq:multiplicative-matching-schemeA}
  \sigma^{\NLO+\LLR, \text{mult.simple}} = \frac{\sigma^\LLR(R)}{\sigma_0}
  \times \left(\sigma_0 + \sigma_1(R) -  \sigma^{\LLR}_1(R)\right)\,.
\end{equation}
Because $\sigma^{\LLR}_1(R)$ contains the same logs as those in
$\sigma_1(R)$, the right hand bracket tends to a constant for small
$R$ and all the $R$ dependence comes from the $\sigma^\LLR(R)$ factor.
Since $\sigma^\LLR(R)$ tends to zero for $R\to 0$,
Eq.~(\ref{eq:multiplicative-matching-schemeA}) satisfies the condition
in Eq.~(\ref{eq:matching-ratio-condition}).
The matching formula that we actually use is
\begin{equation}
  \label{eq:multiplicative-matching-schemeB}
  \sigma^{\NLO+\LLR} = 
  \left(\sigma_0 + \sigma_1(R_0) \right)
  \times 
  \left[\frac{\sigma^\LLR(R)}{\sigma_0}
  \times \left(1 + \frac{\sigma_1(R)- \sigma_1(R_0) - \sigma^{\LLR}_1(R)}{
      \sigma_0}\right)\right],
\end{equation}
where $R_0$ is taken to be the same arbitrary radius of order $1$ that
appears in $\sigma^\LLR(R)$ as defined in Eq.~(\ref{eq:LLR-master}).
Compared to Eq.~(\ref{eq:multiplicative-matching-schemeA}), we have
explicitly separated out a factor $(\sigma_0 + \sigma_1(R_0))$.
As with Eq.~(\ref{eq:multiplicative-matching-schemeA}), at small-$R$
the entire $R$ dependence comes from the $\sigma^{\LLR}(R)$ factor, thus
ensuring that Eq.~(\ref{eq:matching-ratio-condition}) is satisfied.
%
%
%
%
Eq.~(\ref{eq:multiplicative-matching-schemeB}) has the
advantage over Eq.~(\ref{eq:multiplicative-matching-schemeA}) that
matching will be simpler to extend to NNLO+\LLR, which is why we make
it our default choice.

Eq.~(\ref{eq:multiplicative-matching-schemeB}) has a simple physical
interpretation: the left-hand factor is the cross section for
producing a jet of radius $R_0$ and is effectively a stand-in for the
normalisation of the (ill-defined) partonic scattering cross section,
i.e.\ we equate partons with jets with radius $R_0\sim 1$. %
The right hand factor (in square brackets) then accounts for the
effect of fragmentation on the cross section, including both the \LLR
contribution and an exact NLO remainder for the difference between the
cross sections at radii $R_0$ and $R$.
%

Even without a small-$R$ resummation, one can argue that the physical
separation that is embodied in Eq.~(\ref{eq:multiplicative-matching-schemeB})
is one that should be applied to normal NLO calculations.
This gives us the following alternative expression for the NLO cross
section
\begin{equation}
  \label{eq:multiplicative-NLO}
  \sigma^{\NLO\text{-mult.}} = 
  \left(\sigma_0 + \sigma_1(R_0)\right) \times
  \left(1 + \frac{\sigma_1(R) - \sigma_1(R_0)}{\sigma_0}\right)\,,
\end{equation}
i.e.\ the cross section for producing a small-radius jet should be
thought of as the cross section for the initial partonic scattering,
followed by the fragmentation of the parton to a jet.
As in Eq.~(\ref{eq:multiplicative-matching-schemeB}), we introduce a
radius $R_0 \sim 1$ to give meaning to the concept of a ``parton''
beyond leading order.
It is straightforward to see that Eq.~(\ref{eq:multiplicative-NLO})
differs from standard NLO only in terms of corrections at
order $\as^2$ relative to LO. 

\subsection{Unphysical cancellations in scale dependence}
\label{sec:scale-dep-cancellation}

\begin{figure}
  \centering
  \includegraphics[width=0.48\textwidth]{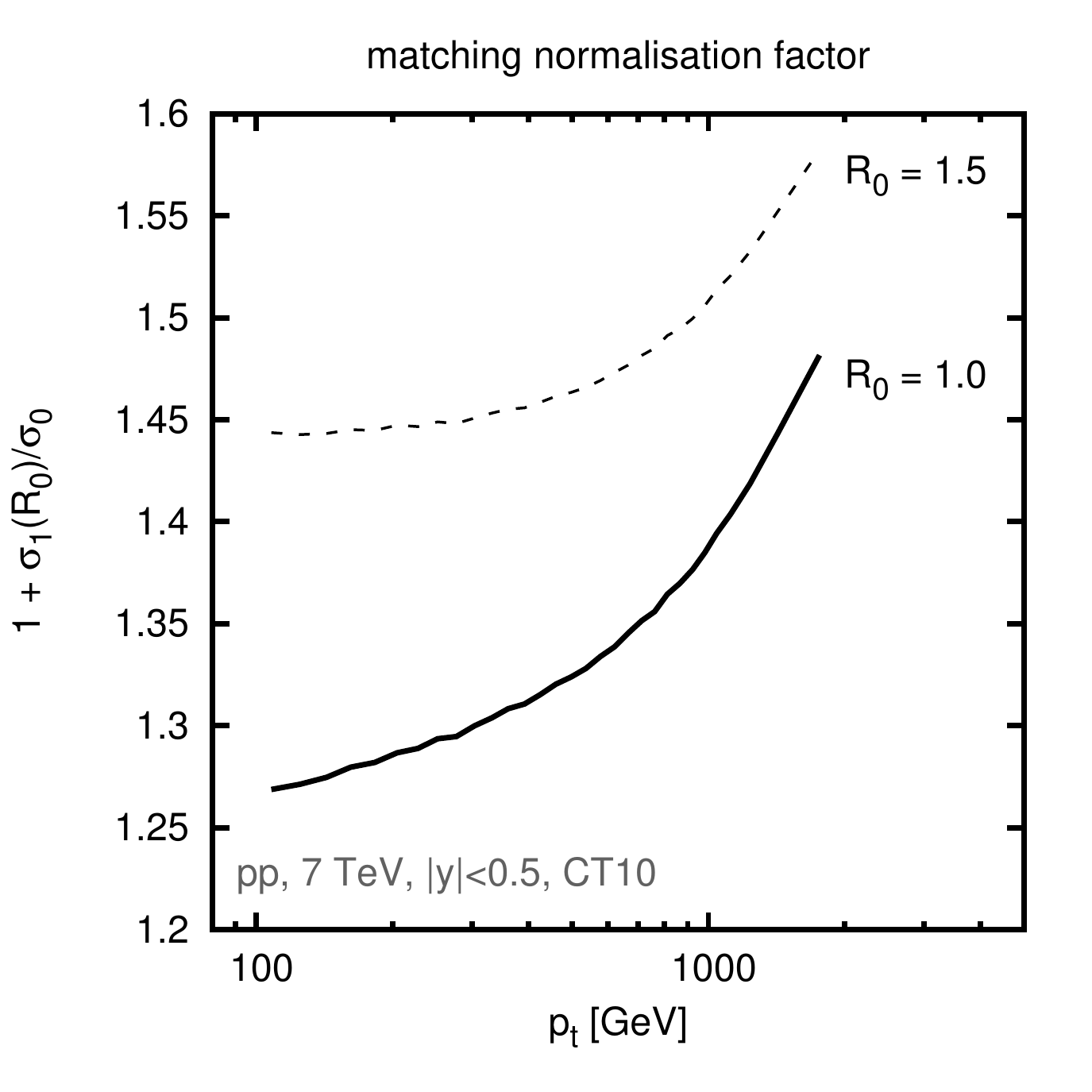}
  \hfill
  \includegraphics[width=0.48\textwidth]{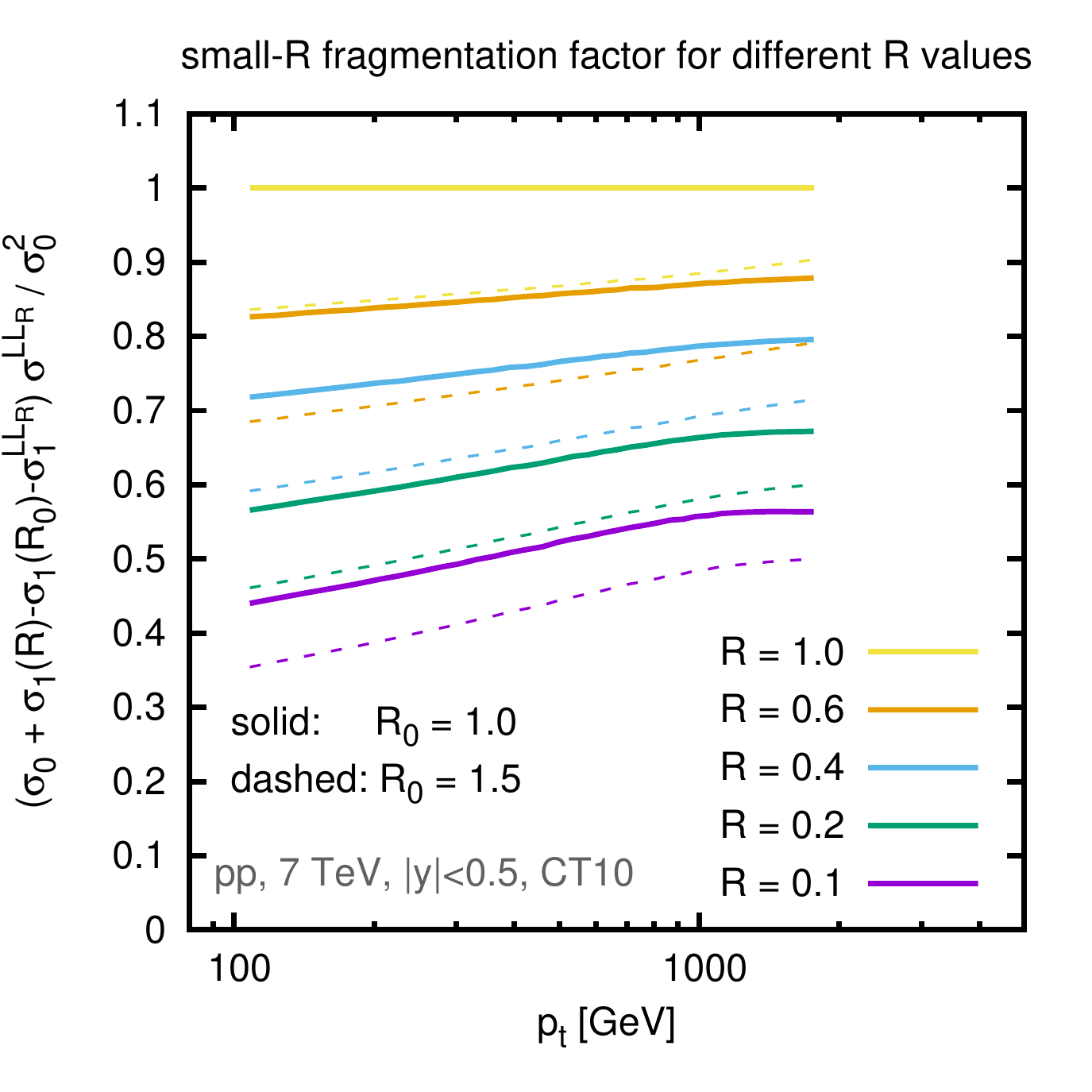} 
  \caption{
    Left: size of the matching normalisation factor (left-hand factor of 
    Eq.~(\ref{eq:multiplicative-matching-schemeB}), normalised to LO), 
    shown v. $p_t$ for various $R$ values and two $R_0$ choices. 
    Right: size of the matched small-$R$ fragmentation factor
    (right-hand factor of
    Eq.~(\ref{eq:multiplicative-matching-schemeB}); similar results are
    observed for the right-hand factor of Eq.~(\ref{eq:multiplicative-NLO})).
    The results are shown for the scale choice $\mu_R = \mu_F =
    p_{t,\max}$, where $p_{t,\max}$ is the transverse momentum of the
    hardest jet in the event.
  }
  \label{fig:smallR-matching-factors}
\end{figure}

Let us now return to the resummed matched prediction,
Eq.~(\ref{eq:multiplicative-matching-schemeB}). 
The left and right-hand factors in that 
formula are shown separately in
Fig.~\ref{fig:smallR-matching-factors}.
The left-hand factor, corresponding to an overall normalisation for
hard partonic scattering, is shown in the left-hand plot (divided by
the LO to ease visualisation), while the small-$R$ fragmentation
(i.e.\ right-hand) factor, which includes the resummation and
matching contributions, is shown on the right.
One sees that the two terms bring $K$-factors going in opposite
directions.
The overall normalisation has a $K$-factor that is larger than one and
grows with $p_t$.
Meanwhile the fragmentation effects generate a $K$-factor
that is substantially below one for smaller $R$ values,
with a relatively weak $p_t$ dependence.

The $p_t$ dependence of the two factors involves an interplay between
two effects: on one hand, the fraction of gluons decreases at large
$p_t$, as does $\as$; on the other hand the PDFs fall off more steeply
at higher $p_t$, which enhances (positive) threshold logarithms in the
normalisation factor and also increases the effect of small-$R$
logarithms in the fragmentation factor (i.e.\ reduces the
fragmentation factor).
We believe that the gentle increase of the fragmentation factor is due
to the decrease in gluon fraction, partially counteracted by the
increasing steepness of the PDFs.
A similar cancellation is probably responsible for the flatness of the
normalisation factor at low and moderate $p_t$'s, with threshold
logarithms induced by the PDFs' steepness driving the increase at the
highest $p_t$'s.

We note also that both factors in
Eq.~(\ref{eq:multiplicative-matching-schemeB}) depend significantly on
the choice of $R_0$, with two values being shown in
Fig.~\ref{fig:smallR-matching-factors}, $R_0=1$ (solid) and $R_0= 1.5$
(dashed).
However in the full results,
Eqs.~(\ref{eq:multiplicative-matching-schemeB}) and
(\ref{eq:multiplicative-NLO}), the $R_0$ dependence in cancels up to
NLO, leaving a residual $R_0$ dependence that corresponds only to
uncontrolled higher-order terms.

The partial cancellation between higher-order effects that takes place
between the small-$R$ effects and the residual matching correction is
somewhat reminiscent of the situation for jet vetoes in Higgs-boson
production.
There it has been argued that such a cancellation can be
dangerous when it comes to estimating scale uncertainties.
As a result, different schemes have been proposed to obtain a more
reliable and conservative estimate, notably the
Stewart-Tackmann~\cite{Stewart:2011cf} and
jet-veto-efficiency~\cite{Banfi:2012yh} methods.
Here we will take an approach that is similar in spirit to those
suggestions (though somewhat closer to the jet-veto-efficiency method)
and argue that for a reliable estimate of uncertainties,
scale-dependence should be evaluated independently for the left and
right-hand factors in Eqs.~(\ref{eq:multiplicative-matching-schemeB})
and (\ref{eq:multiplicative-NLO}) (and also in
Eq.~(\ref{eq:multiplicative-matching-schemeA})), and the resulting
relative uncertainties on those two factors should be added in
quadrature.
We will always verify that the $R_0$ dependence (for just the central
scale choice) is within our scale uncertainty band.

\subsection{NLO+\LLR matched results}

\begin{figure}
  \centering
  \includegraphics[width=0.49\textwidth,page=1]{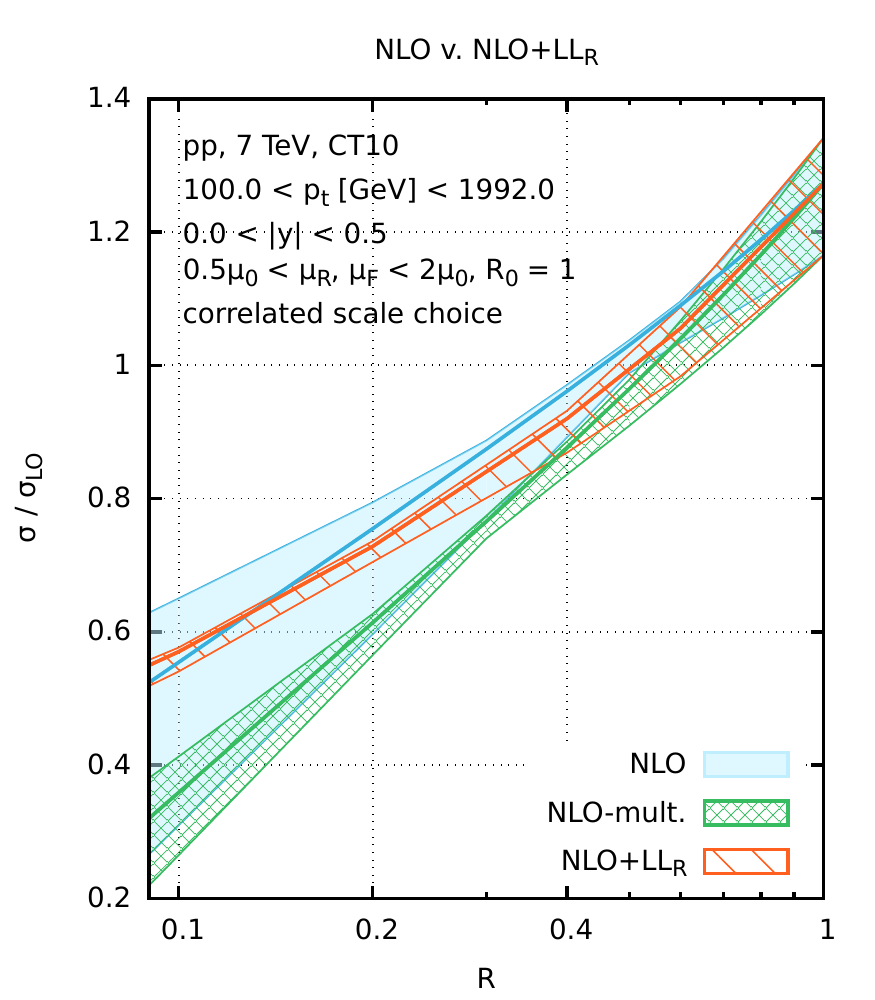}
  \hfill
  \includegraphics[width=0.49\textwidth,page=1]{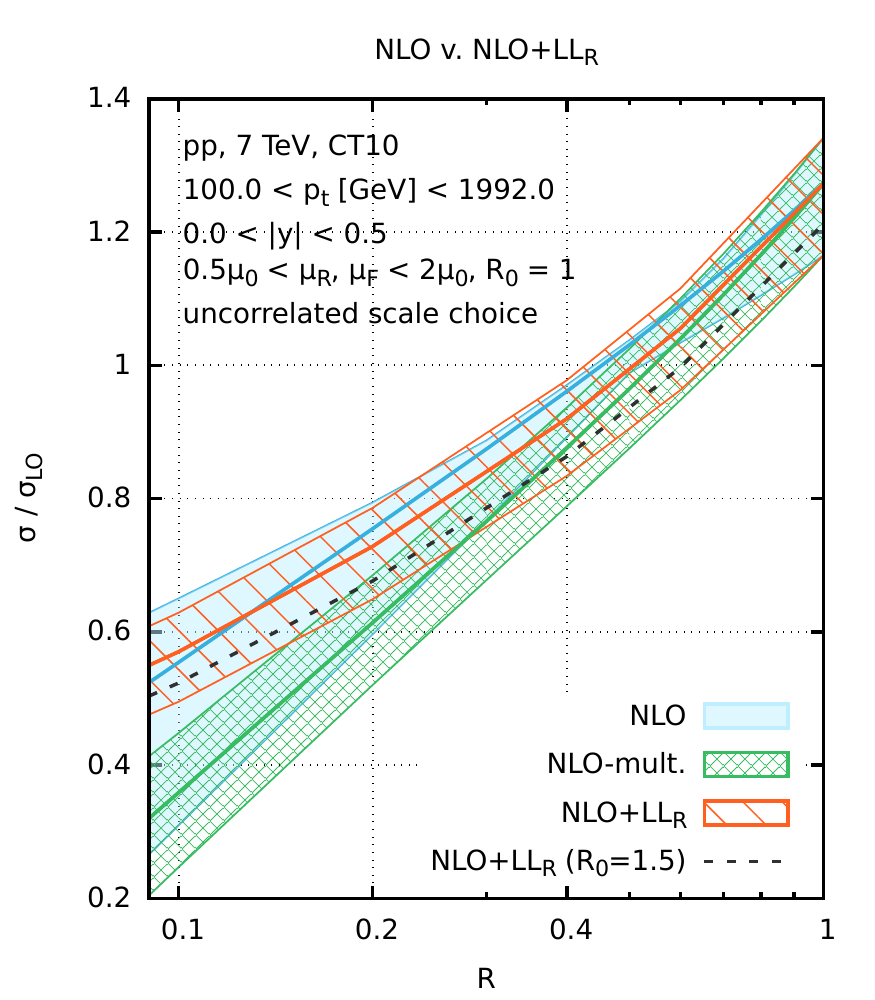}
  \caption{ %
    Inclusive jet cross section for $p_t > 100\GeV$, as a function of
    $R$, normalised to the ($R$-independent) leading-order result.  %
    Left: the standard NLO result, compared to the ``NLO-mult.''\
    result of Eq.~(\ref{eq:multiplicative-NLO}) and the NLO+\LLR
    matched result of
    Eq.~(\ref{eq:multiplicative-matching-schemeB}).  %
    The scale uncertainty here has been obtained within a prescription
    in which the scale is varied simultaneously in the left and
    right-hand factors of
    Eqs.~(\ref{eq:multiplicative-matching-schemeB}) and
    (\ref{eq:multiplicative-NLO}) (``\emph{correlated scale choice}'').  %
    Right: the same plot, but with the scale uncertainties determined
    separately the left and right-hand factors of
    Eqs.~(\ref{eq:multiplicative-matching-schemeB}) and
    (\ref{eq:multiplicative-NLO}), and then added in quadrature
    (``\emph{uncorrelated scale choice}'').
    The plot also shows the NLO+\LLR result for $R_0=1.5$ at our
    central scale choice.
  }
  \label{fig:correl-v-uncorrel-scale-choice}
\end{figure}

Fig.~\ref{fig:correl-v-uncorrel-scale-choice} shows the
inclusive jet cross section integrated from $100\GeV$ to $1992\GeV$
(the full range covered by the ATLAS data~\cite{Aad:2014vwa}), as a
function of $R$, normalised to the leading order result. 
The left-hand plot shows the standard NLO result (light blue band),
the ``NLO-mult.''\ result of Eq.~(\ref{eq:multiplicative-NLO}) (green
band) and the NLO+\LLR matched result of
Eq.~(\ref{eq:multiplicative-matching-schemeB}) (orange band).
To illustrate the issue of cancellation of scale dependence discussed in
section~\ref{sec:scale-dep-cancellation}, the scale uncertainty here
has been obtained within a prescription in which the scale is varied
in a correlated way in the left and right-hand factors of
Eqs.~(\ref{eq:multiplicative-matching-schemeB}) and
(\ref{eq:multiplicative-NLO}).
  We adopt the standard
  convention of independent $\mu_R = \{\frac12,1,2\}\mu_0$ and $\mu_F
  = \{\frac12,1,2\}\mu_0$ variations around a central scale $\mu_0$,
  with the additional condition $\frac12 \le \mu_R/\mu_F\le 2$.
  Our choice for $\mu_0$ is discussed below.

One sees that in each of the 3 bands, there is an $R$ value for which the
scale uncertainty comes close to vanishing, roughly $R=0.5$ for NLO,
$R=0.3$ for ``NLO-mult.''\ and $R=0.1{-}0.2$ for NLO+\LLR.
We believe that this near-vanishing is unphysical, an artefact of a
cancellation in the scale dependence between small-$R$ and overall
normalisation factors, as discussed in the previous paragraph.
One clear sign that the scale dependence is unreasonably small is
that the NLO-mult.\ and NLO+\LLR bands differ by substantially more
than their widths.

The right-hand plot of Fig.~\ref{fig:correl-v-uncorrel-scale-choice}
instead shows uncertainty bands when one separately determines the
scale variation uncertainty in the left-hand (normalisation) and the
right-hand (small-$R$ matching) factors and then adds those two
uncertainties in quadrature (``uncorrelated scale choice''; note that
the NLO band is unchanged).
Now the uncertainties remain fairly uniform over the whole range of
$R$ and if anything increase towards small $R$, as one might expect.
This uncorrelated scale variation is the prescription that we will
adopt for the rest of this article.

Intriguingly, the NLO+\LLR result is rather close to the plain NLO
prediction. 
Given the large difference between the NLO and NLO-mult.\ results,
this is, we believe, largely a coincidence:
if one examines yet smaller $R$, one finds that the NLO and NLO+\LLR
results separate from each other, with the NLO and NLO-mult.\ results
going negative for sufficiently small $R$, while the NLO+\LLR result
maintains a physical behaviour.

Fig.~\ref{fig:correl-v-uncorrel-scale-choice} (right) also shows the
impact of increasing $R_0$ to $1.5$. 
One sees a $5{-}10\%$ reduction in the cross section, however this
reduction is within the uncertainty band that comes from the
uncorrelated scale variation.

A comment is due concerning our choice of central scale, $\mu_0$.
At NLO, for each event, we take $\mu_0$ to be the $p_t$ of the hardest
jet in the event, $p_{t,\max}$.
In NLO-like configurations, with at most 3 final-state partons, this
hardest jet $p_t$ is independent of the jet radius and so we have a
unique scale choice that applies to all jet radii.
An alternative, widely used prescription is to use a separate scale
for each jet, equal to that jet's $p_t$.
We disfavour this alternative because it leads to a spurious
additional $R$ dependence, induced by inconsistent scale choices
in real and virtual terms.
Further details are given in Appendix~\ref{sec:central-scale-choice}.

\section{Matching to NNLO}
\label{sec:matching-NNLO}

\subsection{Matching prescription}
\label{sec:NNLO-prescription}

Given that full NNLO results for the inclusive cross section are
likely to be available soon~\cite{Currie:2013dwa}, here we propose
matching schemes for combining our small-$R$ resummed results with a
full NNLO result.
The direct analogue of Eq.~(\ref{eq:multiplicative-matching-schemeB})
is 
\begin{multline}
  \label{eq:multiplicative-matching-nnloB}
  \sigma^{\NNLO+\LLR} = 
  \left(\sigma_0 + \sigma_1(R_0) + \sigma_2(R_0)\right)
  \times
  \\
  \times
  \Bigg[\frac{\sigma^\LLR(R)}{\sigma_0}
  \times
  \Bigg(1 + \Delta_{1+2}(R,R_0) -
    \frac{\sigma^{\LLR}_1(R)
      +\sigma^{\LLR}_2(R)}{\sigma_0}
    \hspace{50mm}
    \\
      -  \frac{\sigma_1^\LLR(R) \left(\sigma_1(R) -
          \sigma^{\LLR}_1(R)\right)}{\sigma_0^2} 
      - \frac{\sigma_1(R_0)}{\sigma_0} \bigg(
        \Delta_1(R,R_0) - \frac{\sigma^{\LLR}_1(R)}{\sigma_0}\bigg)
  \Bigg)\Bigg]\,,
\end{multline}
where the functions $\Delta_1$ and $\Delta_{1+2}$ were defined in
Eq.~(\ref{eq:LLR-validation-ratio}) and
(\ref{eq:LLR-validation-ratio-Delta12}) and we recall that
$\sigma^\LLR$ and its expansion are functions both of $R$ and
$R_0$. 
As with our NLO+\LLR formula,
Eq.~(\ref{eq:multiplicative-matching-schemeB}), we have written
Eq.~(\ref{eq:multiplicative-matching-nnloB}) in terms of two factors:
an overall normalisation for producing $R_0$-jets, together with a
matched fixed-order and resummed result for the correction coming from
fragmentation of the $R_0$ jet into small-$R$ jets.
One comment here is that in 
Eq.~(\ref{eq:multiplicative-matching-schemeB}) the matching part
(big round brackets inside the square brackets) gave a finite result
for $R \to 0$.
The situation is different at NNLO because the \LLR resummation does
not capture the $\as^2 \ln 1/R^2$ (N\LLR) term that is present at
fixed order and so the matching term has a residual $\as^2 \ln 1/R^2$
dependence. 
This means that for sufficiently small-$R$,
Eq.~(\ref{eq:multiplicative-matching-nnloB}) can become unphysical.
We have not seen any evidence of this occurring in practice, but one
should keep in mind that for future work one might aim to resolve this
in-principle problem either by incorporating N\LLR
resummation%
%
or by choosing a different form of matching, for example one where the
$\order{\as^2}$ parts of the matching correction are exponentiated,
ensuring a positive-definite result.
Note that with N\LLR resummation one could also use a formula
analogous to Eq.~(\ref{eq:multiplicative-matching-schemeA}),
\begin{multline}
  \label{eq:multiplicative-matching-nnloA}
  \sigma^{\NNLO+\NLLR, \text{mult.simple}} = 
  \\
  =
  \Bigg(\sigma_0 + \sigma_1 + \sigma_2 -
    \sigma^{\NLLR}_1 -  \sigma^{\NLLR}_2
    - \frac{\sigma^{\NLLR}_1}{\sigma_0}\left(\sigma_1 - \sigma^{\NLLR}_1\right)
  \Bigg)
  \times
  \frac{\sigma^\NLLR}{\sigma_0}\,,
\end{multline}
where each of the terms is evaluated at radius $R$.

As well as a matched result, it can be instructive to study a
modification of the plain NNLO result, ``NNLO-mult.'', in analogy with
Eq.~(\ref{eq:multiplicative-NLO}). This remains a fixed order result,
but it factorises the production of large-$R_0$ jets from the
fragmentation to small-$R$ jets,
\begin{equation}
  \label{eq:multiplicative-NNLO}
  \sigma^{\NNLO\text{-mult.}} = 
  \left(\sigma_0 + \sigma_1(R_0) + \sigma_2(R_0)\right)
  \times
  \left(1 + \Delta_{1+2}(R,R_0) -
    \frac{\sigma_1(R_0)}{\sigma_0} \Delta_1(R,R_0)\right) 
   \,.
\end{equation}
It differs from $\sigma^\NNLO$ only by terms beyond NNLO.

As in section~\ref{sec:matching-NLO}, in
Eqs.~(\ref{eq:multiplicative-matching-nnloB})--(\ref{eq:multiplicative-NNLO})
we advocate varying scales separately in the normalisation and
fragmentation factors, and also studying the $R_0$ dependence of
the final result.

\subsection{A stand-in for NNLO: \NNLOR}
\label{sec:NNLOR}

We have seen in section~\ref{sec:smallR-validation} that NNLO terms of
the form $\as^2 \ln 1/R^2$ that are not accounted for in our \LLR
calculation can be large.
Insofar as they are known, they should however be included in 
phenomenological studies.
This specific class of terms can be taken into account in the context
of a stand-in for the full NNLO calculation which contains the exact
NNLO $R$ dependence and that we refer to as \NNLOR.
It is constructed as follows:
\begin{equation}
  \label{eq:sigma-nnlo}
  \sigma^\NNLOR(R,R_m) \equiv \sigma_0 + \sigma_1(R) + [\sigma_2(R) -
  \sigma_2(R_m)], 
\end{equation}
which depends on an arbitrary angular scale $R_m$. 
Though neither $\sigma_2(R)$ nor $\sigma_2(R_m)$ can be fully
determined currently, their difference can be obtained from the same
NLO 3-jet calculation that was used to examine
$\Delta_{1+2}(p_t,R,R_\text{ref})$ in Fig.~\ref{fig:smallR-validation}
(right).

Since the full NNLO result has the property 
\begin{equation}
  \label{eq:full-NNLO-from-NNLOR}
  \sigma^\NNLO(R) = \sigma^\NNLOR(R,R_m) + \sigma_2(R_m)\,,
\end{equation}
the use of $\sigma^\NNLOR(R,R_m)$ instead of $\sigma^\NNLO(R)$ is
equivalent to the assumption that $\sigma_2(R_m)$ vanishes.
In practice we will take $R_m = 1$, independently of $p_t$.

One point to be aware of is that $\sigma^\NNLOR(R,R_m)$ and
$\sigma^\NNLO(R)$ have parametrically different scale dependence.
On one hand, the $\sigma_2(R)$ term in $\sigma^\NNLO(R)$ fully cancels the
(relative) $\order{\as^2}$ scale variation that is left over from
$\sigma_0$ and $\sigma_1$, leaving just $\order{\as^3}$ dependence.
On the other, in $\sigma^\NNLOR(R,R_m)$ the use of the
$\sigma_2(R)-\sigma_2(R_m)$ means that some residual $\order{\as^2}$
dependence is left over.
In particular, for $R = R_m$ the scale dependence is identical to that
at NLO.
Accordingly, when estimating higher-order uncertainties in studies
that use \NNLOR results, we do not explicitly need to vary $R_m$, since
the $\order{\as^2}$ uncertainty that it brings should already be
accounted for in the scale variation.\footnote{Despite this statement,
  one may wish to examine the robustness of conclusions with
  respect to different possibles values of $\sigma_2(R_m)$. 
  This is the subject of section~\ref{sec:double-virtual-impact}.
}

Our central scale choice for any given event will be
$\mu _0 = p_{t,\max}^{R=1}$, the transverse momentum of the hardest
jet in the event as clustered with $R=1$.
This is analogous to the choice of $p_{t,\max}$ used at NLO, except
that at NNLO one needs to explicitly specify $R$ since $p_{t,\max}$
can depend on the jet clustering.
The logic for taking $p_{t,\max}$ at a fixed jet radius of $1$,
independently of the $R$ used in the clustering for the final jet
spectrum, is that one obtains a unique scale for the event as a whole
and avoids mixing scale-variation effects with
$R$ dependence.
%
Another potential choice that we did not investigate is to take the
averaged $p_t$ of the two hardest jets.
As long as the jets are obtained with a clustering radius $\sim 1$
such a choice is expected to be good at minimising the impact
both of initial-state and final-state radiation, whereas our
$p_{t,\max}$ choice has some sensitivity to initial-state radiation.
%

\subsection{Results at \NNLOR and NNLO$_R$+\LLR}
\label{sec:results-NNLOR}

\begin{figure}
  \centering
  \includegraphics[width=0.49\textwidth,page=1]{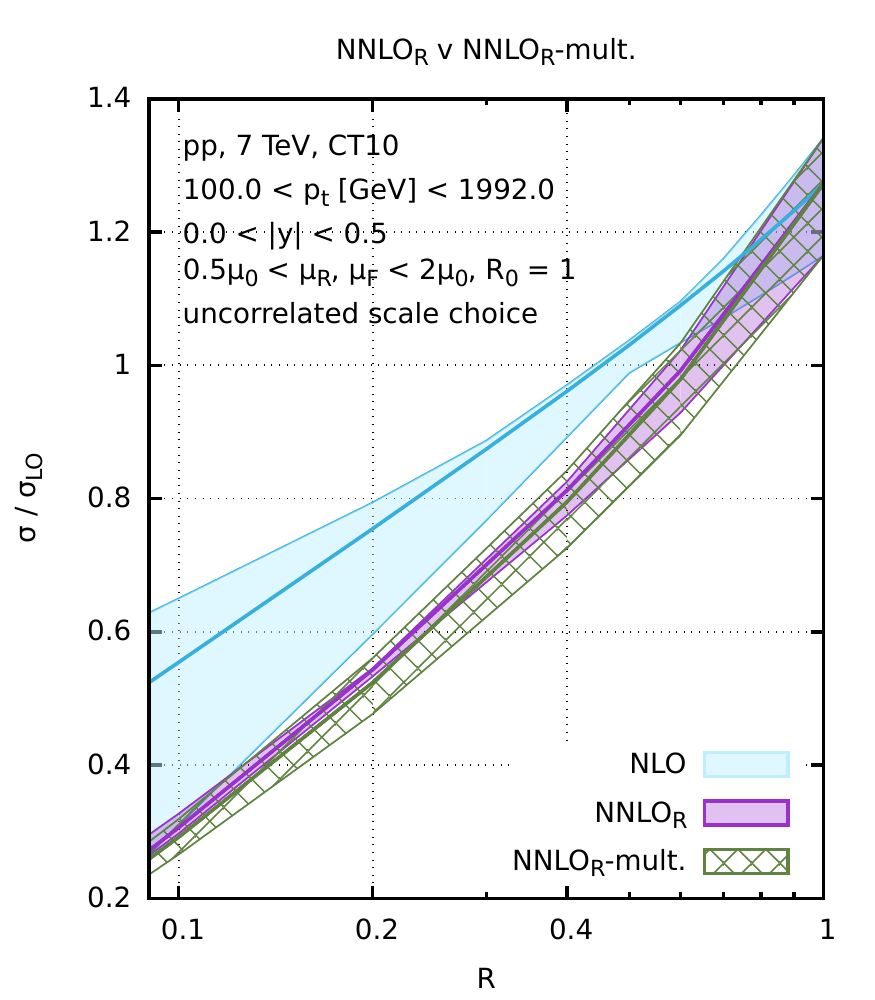}
  \hfill
  \includegraphics[width=0.49\textwidth,page=1]{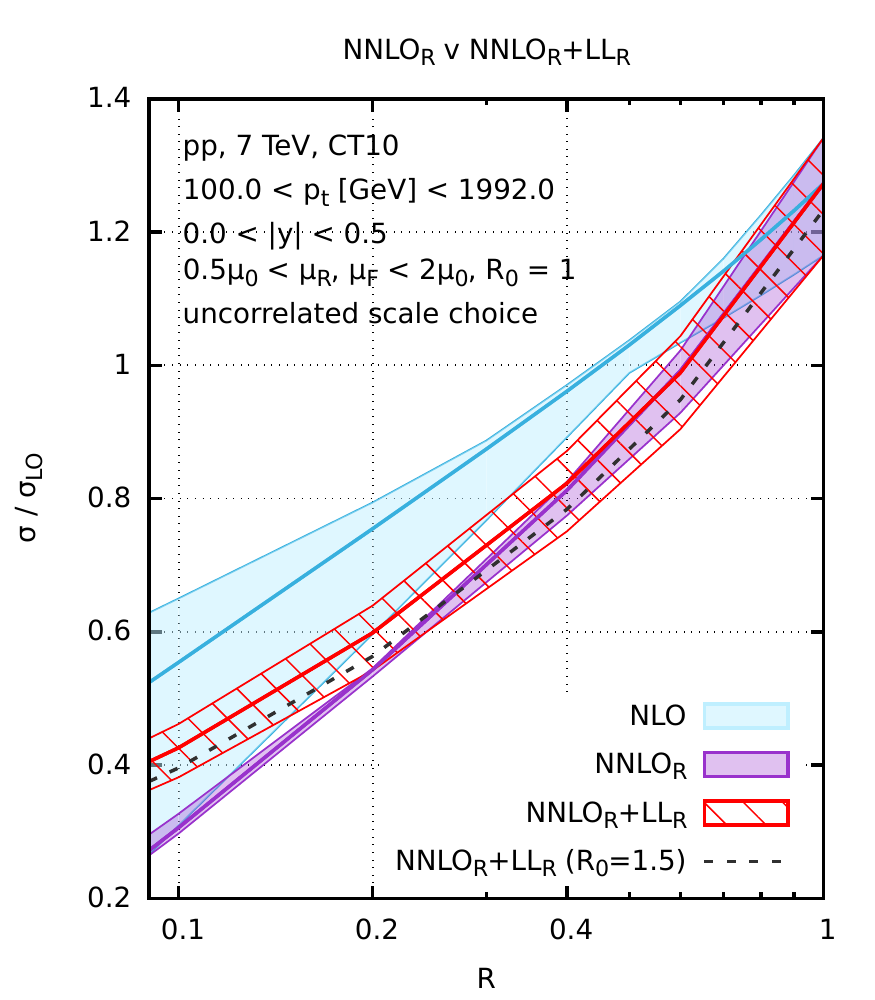}
  \caption{Left: comparison of the NLO, \NNLOR and \NNLOR-mult.\
    results for the inclusive jet cross section for $p_t > 100 \GeV$,
    as a function of $R$, normalised to the LO result.
    Right, corresponding comparison of NLO, \NNLOR and \NNLORLLR
    together with the central curve for \NNLORLLR when $R_0$ is
    increased to $1.5$.
    In both plots, for the \NNLOR-mult.\ and \NNLORLLR results 
    the scale-dependence has been evaluated separately in the normalisation and
    fragmentation contributions and added in quadrature to obtain
    the final uncertainty band.
  }
  \label{fig:correl-v-uncorrel-scale-choice-nnlo}
\end{figure}

Let us start by examining the \NNLOR result, shown versus
$R$ as the purple band in Fig.~\ref{fig:correl-v-uncorrel-scale-choice-nnlo}
(left), together with the \NNLOR-mult.\ results using
Eq.~(\ref{eq:multiplicative-NNLO}) and the NLO band.
One sees that the $R$ dependence of the \NNLOR result is 
steeper than in the NLO result, especially for $R\gtrsim 0.2$.
This pattern is qualitatively in line with one's expectations from
Fig.~\ref{fig:smallR-validation} (right) and will hold also for the
full NNLO calculation, which differs from \NNLOR only by an
$R$-independent (but $p_t$ and scale-dependent) additive constant.
The point of intersection between the NLO and \NNLOR results, at
$R=1$, is instead purely a consequence of our choice of $R_m=1$ in
Eq.~(\ref{eq:sigma-nnlo}). 
Thus at $R=1$, both the central value and scale dependence are by
construction identical to those from the NLO calculation.

The left-hand plot of
Fig.~\ref{fig:correl-v-uncorrel-scale-choice-nnlo} also shows the
\NNLOR-mult.\ result. 
Relative to what we saw when comparing NLO and NLO-mult., the most
striking difference here is the much better agreement between \NNLOR
and \NNLOR-mult., with the two generally coinciding to within a few
percent.
For $R \gtrsim 0.4$, this good agreement between different approaches
carries through also to the comparison between \NNLOR and \NNLORLLR.
However, for yet smaller values of $R$, the \NNLORLLR result starts to
be substantially above the \NNLOR and \NNLOR-mult.\ ones.
This is because the \NNLOR and \NNLOR-mult.\ results have unresummed
logarithms that, for very small-$R$, cause the cross section to go
negative, whereas the resummation ensures that the cross section
remains positive 
(modulo the potential issue with unresummed \NLLR terms that remain
after matching).

Comparing the \NNLORLLR result to the \NLOLLR of
Fig.~\ref{fig:correl-v-uncorrel-scale-choice} (right), one finds that
the central value of the \NNLORLLR prediction starts to lie outside
the \NLOLLR uncertainty band for $R \lesssim 0.5$.
This highlights the importance of the NNLO corrections, and in
particular of terms with subleading $\ln 1/R^2$ enhancements. 
Finally, the dependence on the choice of $R_0$ is slightly reduced at
\NNLORLLR compared to \NLOLLR and it remains within the
scale-variation uncertainty band.

\begin{figure}
  \centering
  \includegraphics[width=0.49\textwidth,page=1]{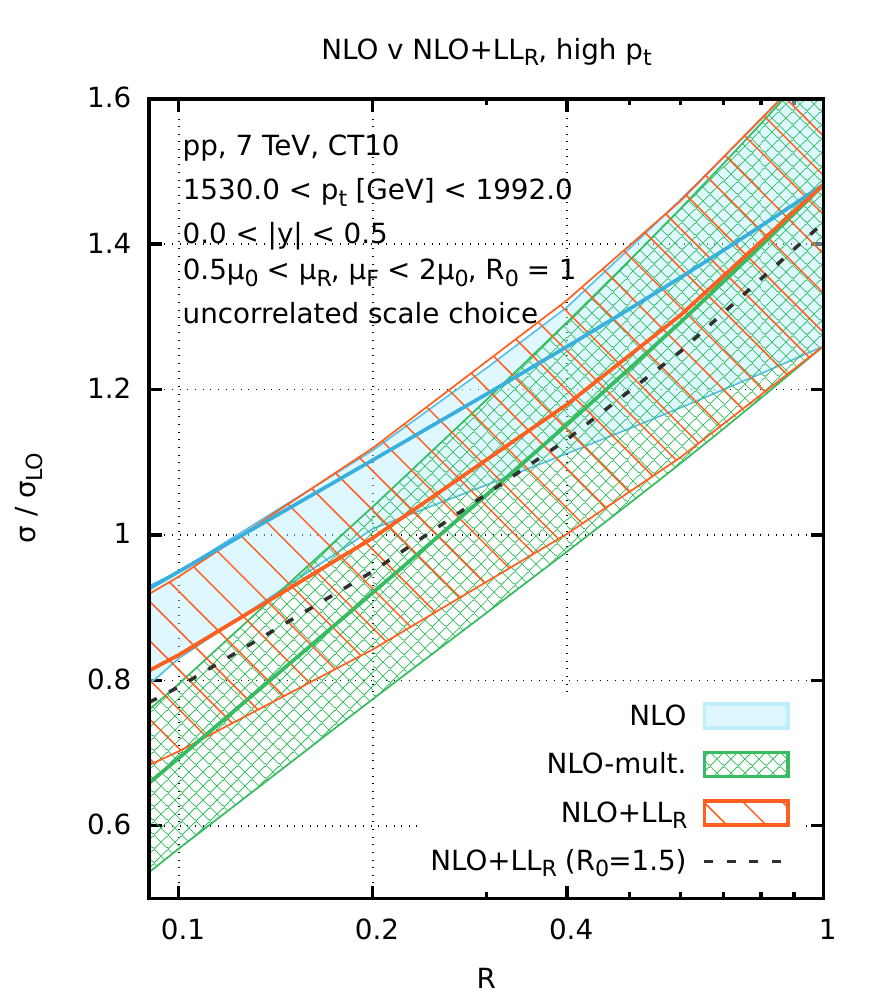}
  \hfill
  \includegraphics[width=0.49\textwidth,page=1]{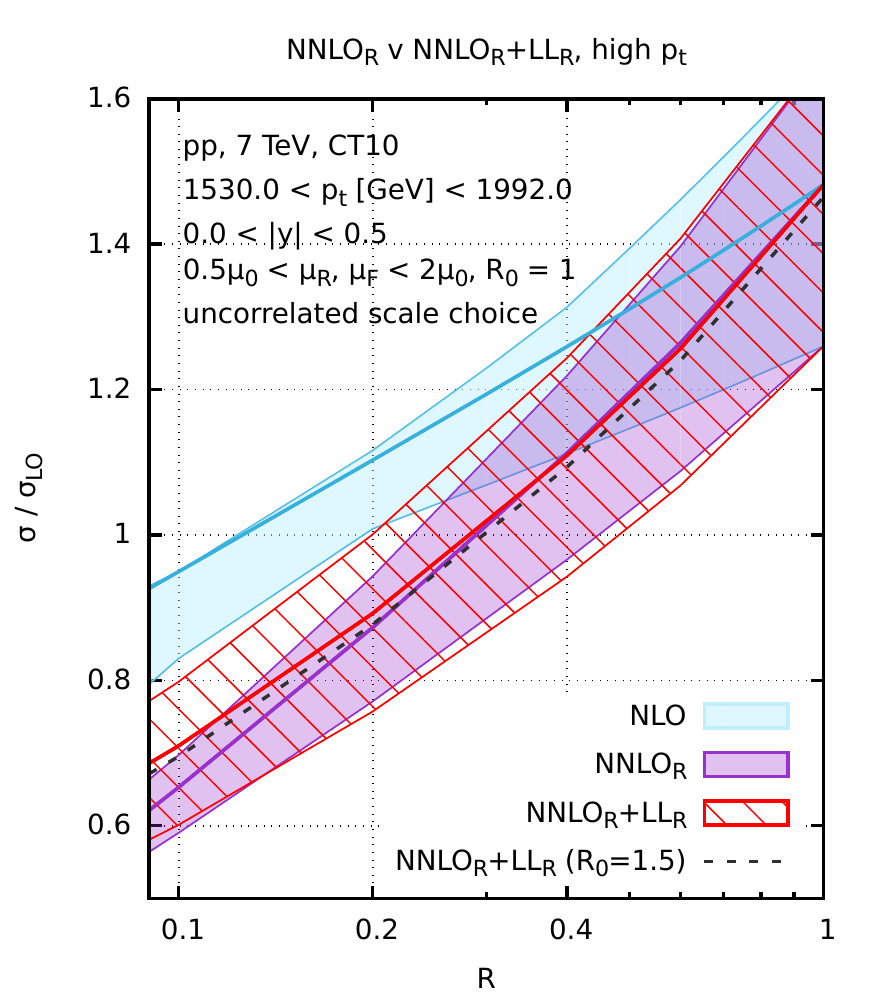}
  \caption{Same as figure \ref{fig:correl-v-uncorrel-scale-choice},
    but focusing only on the high $p_t$ bin.
    Both plots use an uncorrelated scale variation in the
    normalisation and fragmentation factors.
  }
  \label{fig:correl-v-uncorrel-scale-choice-highpt}
\end{figure}

\begin{figure}
  \centering
  \includegraphics[width=0.5\textwidth,page=1]{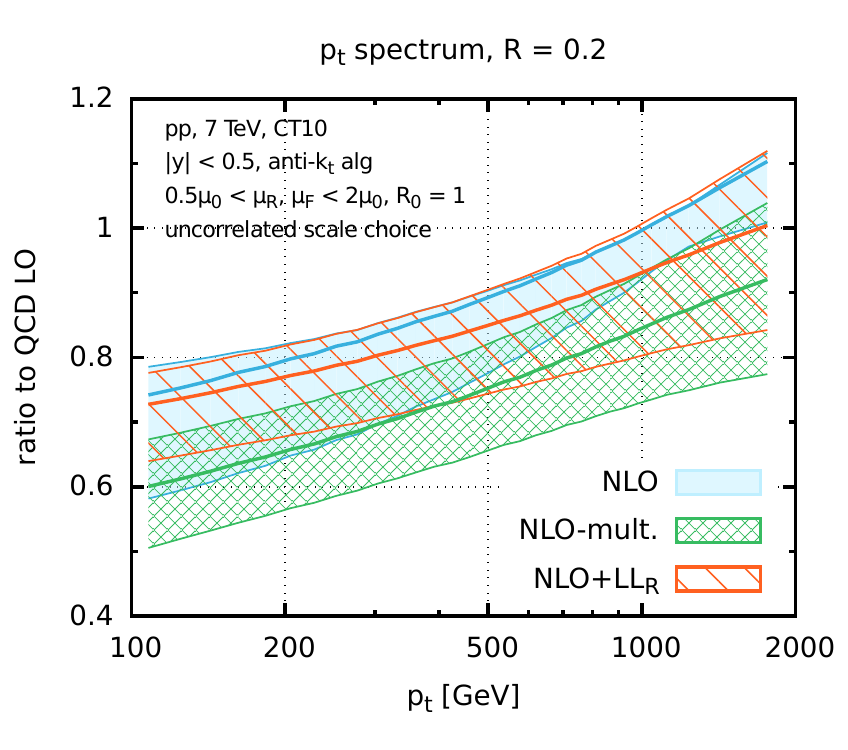}%
  \includegraphics[width=0.5\textwidth,page=1]{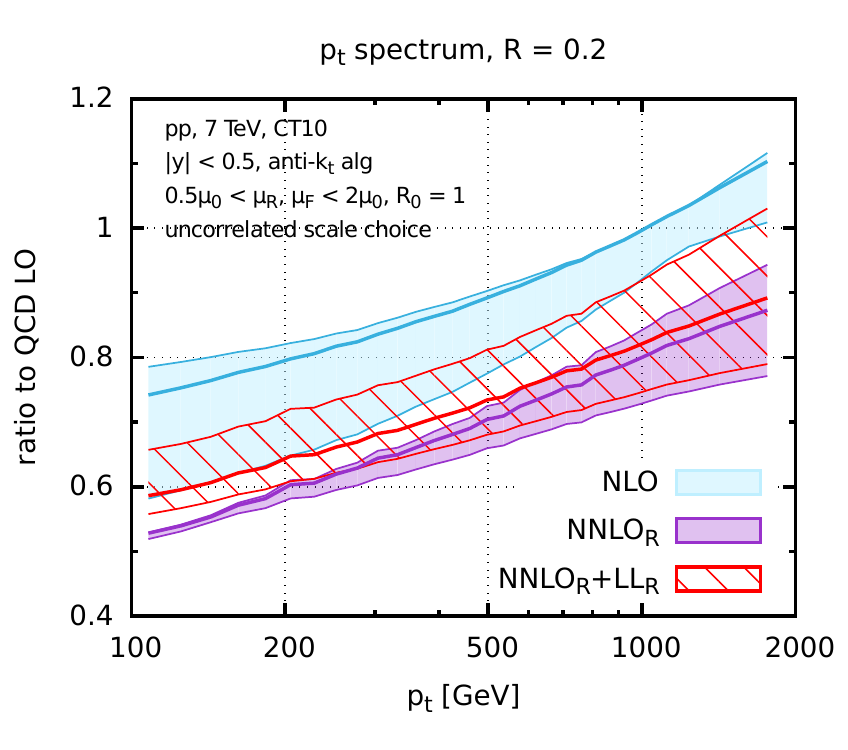}
  \caption{Left: the NLO inclusive jet spectrum as a function of $p_t$, normalised to LO, 
    together with the ``NLO-mult.'' result and the NLO+\LLR matched 
    results for $R=0.2$.
    The cross section is shown for the rapidity bin $|y|<0.5$.
    The bands give the scale uncertainty obtained using an
    uncorrelated scale choice.
    Right: analogous plots with the \NNLOR{} and \NNLOR{}+\LLR
    predictions.}
  \label{fig:correl-v-uncorrel-spectrum}
\end{figure}

\begin{figure}
  \centering
  \includegraphics[width=0.5\textwidth,page=1]{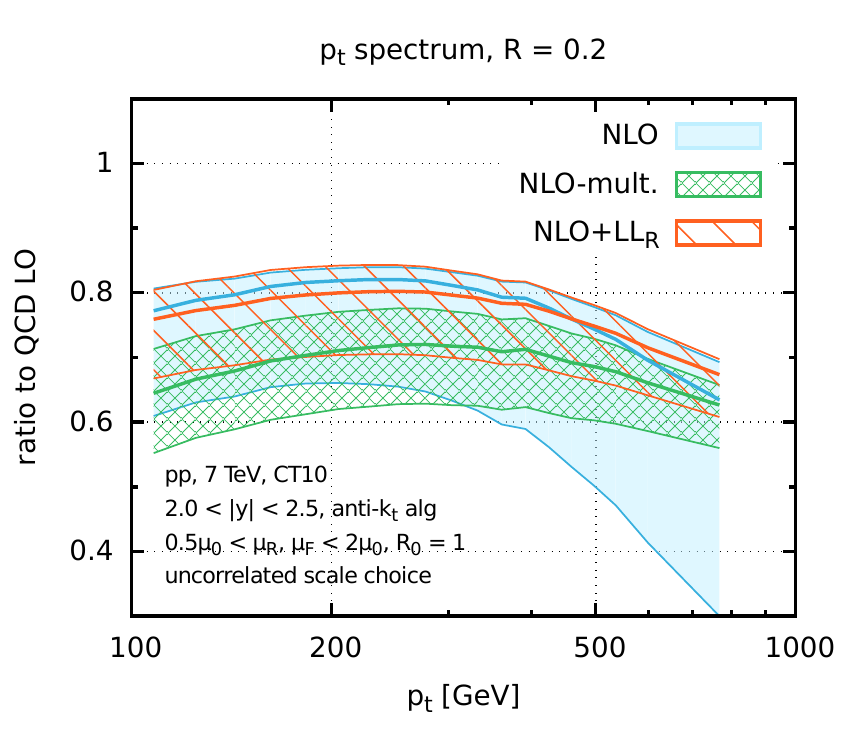}%
  \includegraphics[width=0.5\textwidth,page=1]{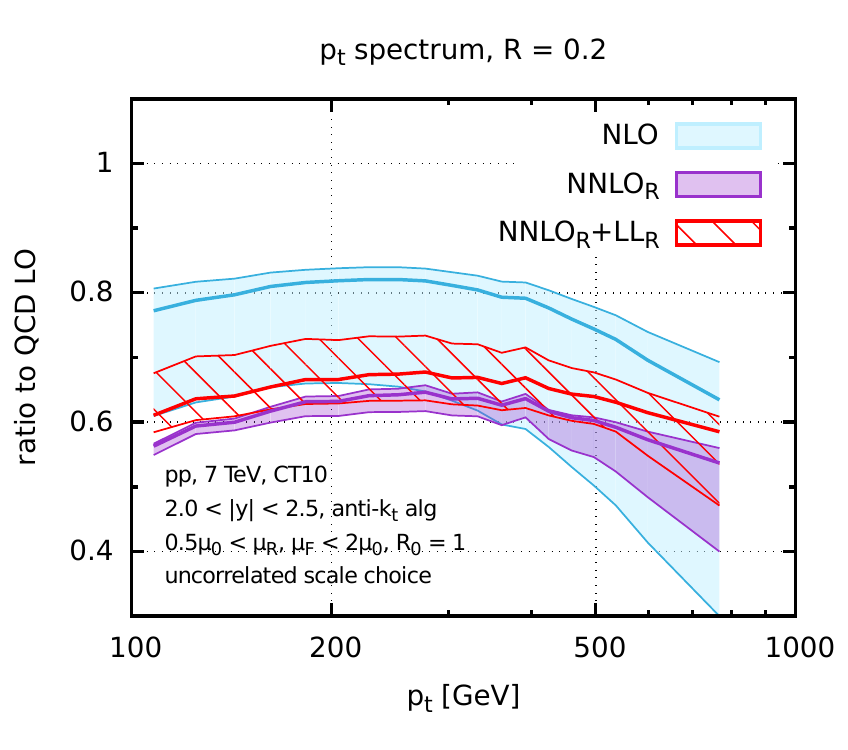}
  \caption{Same as figure \ref{fig:correl-v-uncorrel-spectrum} but
    showing the cross section in the rapidity bin $2<|y|<2.5$.}
  \label{fig:correl-v-uncorrel-spectrum-highrap}
\end{figure}

To help complete the picture, we also show results as a function of
$R$ in a high-$p_t$ bin, $1530 < p_t < 1992 \GeV$ in
Fig.~\ref{fig:correl-v-uncorrel-scale-choice-highpt}.
Most of the qualitative observations that we discussed above remain
true also for high $p_t$.
The main difference relative to the $p_t > 100 \GeV$ results is that
scale uncertainty bands generally grow larger.
This is perhaps due to threshold effects and might call for the
inclusion of threshold resummation, see e.g.\
Ref.~\cite{deFlorian:2013qia} and references therein.
Figs.~\ref{fig:correl-v-uncorrel-spectrum} and
\ref{fig:correl-v-uncorrel-spectrum-highrap} show the jet spectrum as
a function of $p_t$, normalised to the LO result, for $R=0.2$ and two
rapidity bins.
Again, the conclusions are similar.

All of the predictions shown here have been obtained with the choice
$R_m = 1$ in Eq.~(\ref{eq:sigma-nnlo}), equivalent to the assumption
that $\sigma_2(R_m=1) = 0$ in Eq.~(\ref{eq:full-NNLO-from-NNLOR}).
For a discussion of how the predictions change if $\sigma_2(R_m=1)$ is
non-zero, the reader is referred to section~\ref{sec:double-virtual-impact}.

To conclude this section, our main observation is that \LLR and NNLO
terms both have a significant impact on the $R$ dependence of the
inclusive jet spectrum, with the inclusion of both appearing to be
necessary in order to obtain reliable predictions for $R\lesssim
0.4$. 
In particular, if NNLO and NLO coincide for $R=1$, then for $R=0.4$
the NNLO results will be about $20\%$ below the NLO ones.
Going down to $R=0.2$, one sees that even with NNLO corrections
resummation of small-$R$ logarithms is important, having a further
$10\%$ effect.

\subsection{Impact of finite two-loop corrections}

\label{sec:double-virtual-impact}

In our \NNLOR-based predictions, we have all elements of the full NNLO
correction except for those associated with 2-loop and squared 1-loop
diagrams (and corresponding counterterms).

Here, we examine how our results depend on the size of
those missing contributions.
We introduce a factor $K$ that corresponds to the NNLO/NLO ratio for a
jet radius of $R_m$:
\begin{equation}
  \label{eq:app:rescaled-xsect-Rm}
  \sigma^{\NNLO_{R,K}}(R_m) = K \times \sigma^\NLO(R_m)\,.
\end{equation}
For other values of the jet radius, we have
\begin{equation}
  \label{eq:app:rescaled-xsect-R}
  \sigma^{\NNLO_{R,K}}(R) = \sigma_0\left[
    1 + \frac{\sigma_1(R)}{\sigma_0} + \Delta_2(R,R_m) + 
     (K-1) \times \left(1 + \frac{\sigma_1(R_m)}{\sigma_0}\right)
  \right]\,.
\end{equation}
As before, we will take $R_m = 1.0$.
One could attempt to estimate $K$ from the partial NNLO calculation of
Ref.~\cite{Currie:2013dwa}, however given that this calculation is not
yet complete, we prefer instead to leave $K$ as a free parameter and
simply examine the impact of varying it.

\begin{figure}
  \centering
  \includegraphics[width=0.49\textwidth,page=1]{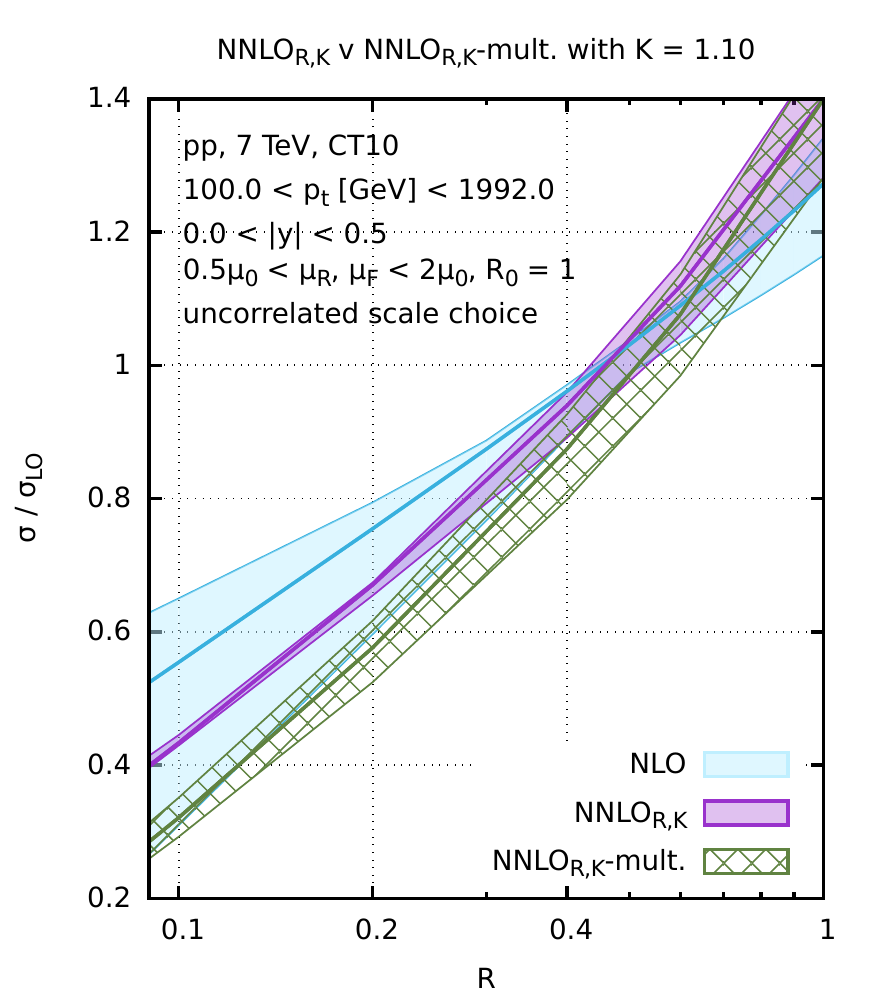}
  \hfill
  \includegraphics[width=0.49\textwidth,page=1]{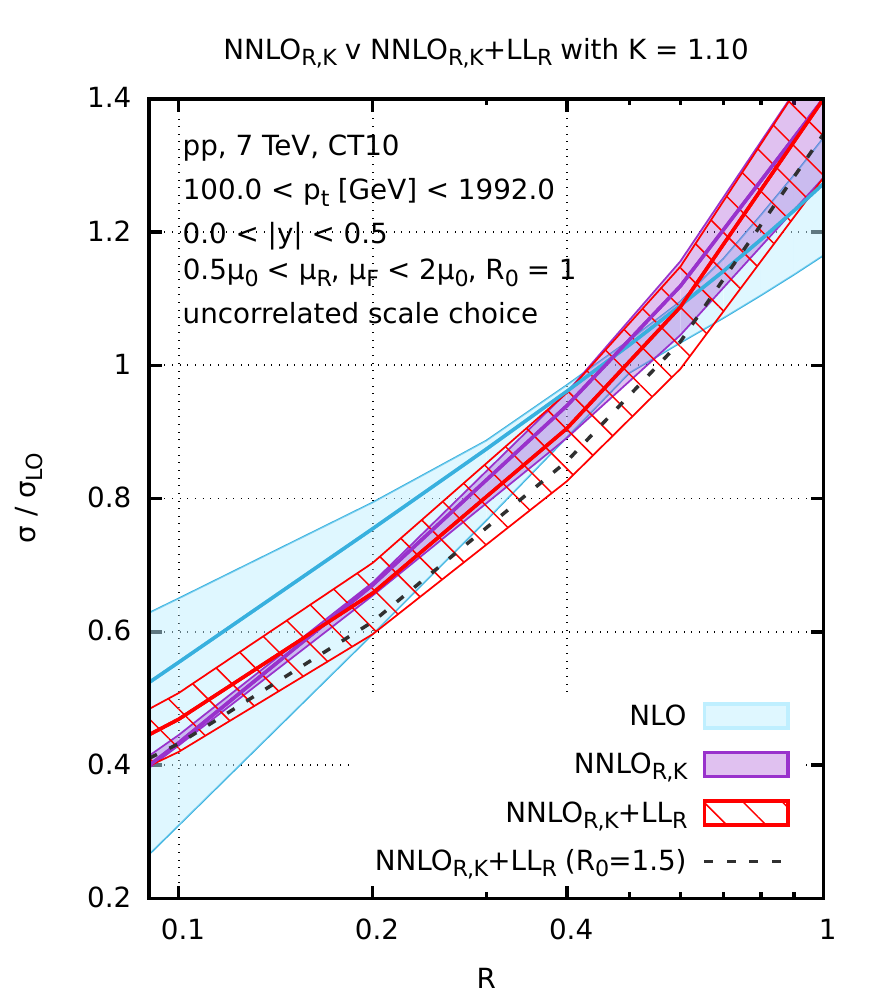}
  \caption{The same as
    figure~\ref{fig:correl-v-uncorrel-scale-choice-nnlo},
    but applying a K-factor of 1.10 to the
    \NNLOR{} prediction, as an estimate of the potential impact of the
    full NNLO calculation. 
  }
  \label{fig:comparison-kfactor}
\end{figure}

In Fig.~\ref{fig:comparison-kfactor}, we show the impact of taking
$K=1.10$, to be compared to
Fig.~\ref{fig:correl-v-uncorrel-scale-choice-nnlo}, which corresponds
to $K=1$.
As $K$ is increased, one sees that \NNLORK and \NNLORKLLR start to agree
over a wider range of $R$.
This behaviour can be understood by observing there are two effects
that cause \NNLORK and \NNLORKLLR to differ: on one hand the small-$R$
resummation prevents the cross section from going negative at very
small radii, raising the prediction in that region relative to \NNLOR
and reducing the overall $R$ dependence.
On the other hand, the normalisation (first) factor in
Eq.~(\ref{eq:multiplicative-matching-nnloB}), which is larger than
$1$, multiplies the full NNLO $R$ dependence that is present in the
fragmentation (second) factor, thus leading to a steeper $R$
dependence than in pure \NNLORK. 
With $K=1$, the first effect appears to dominate.
However as $K$ is increased, the second effect is enhanced and then
the two effects cancel over a relatively broad range of $R$
values.

To put it another way, in the \NNLORK result the $K$
factor acts additively, shifting the cross section by the same
amount independently of $R$. In the \NNLORKLLR result, the $K$
factor acts multiplicatively, multiplying the cross section by a
constant factor independently of $R$. 
By construction, the two always agree for $R=R_0=1$.
With $K=1$, \NNLORK is below
\NNLORKLLR at small $R$, but the additive shift for $K>1$ brings
about a larger increase of \NNLORK than the multiplicative factor
for \NNLORKLLR, because $\sigma/\sigma(R_0)$ is smaller than one.

Another point to note is that while in
Fig.~\ref{fig:correl-v-uncorrel-scale-choice-nnlo} the \NNLOR-mult.\
and \NNLOR results agreed over the full range of $R$, that is no
longer the case with $K=1.1$: this is because \NNLOR-mult.\ acquires a
multiplicative correction, as compared to the additive correction for
\NNLORK. 
Therefore one strong conclusion from our study is that independently
of the size of the NNLO $K$-factor, plain fixed order calculations at
NNLO are likely to be insufficient for $R\lesssim 0.4$.

\subsection{Comparison to \powheg}
\label{sec:powheg-comparison}

One widely used tool to study the inclusive jet spectrum is \powheg's
dijet implementation~\cite{Alioli:2010xa}.
Insofar as parton showers should provide \LLR accuracy and \powheg
guarantees NLO accuracy, \powheg together with a shower should provide
NLO+\LLR accuracy.
It is therefore interesting to compare our results to those from the
\powhegbox's dijet process (v3132), which are obtained here using a
generation cut $\texttt{bornktmin}$ of $50\GeV$ and a suppression
factor $\texttt{bornsuppfact}$ of $500\GeV$.\footnote{We also carried
  out a run with $\texttt{bornktmin}$ of $25\GeV$ and
  $\texttt{bornsuppfact}$ of $300\GeV$ and found results that are
  consistent with those shown here to within the statistical errors,
  which at low $p_t$ are of the order of $1\%$.}
We have used it with \pythia 8 (v8.186 with tune
4C~\cite{Corke:2010yf}) for the parton shower, \pythia~6 (v6.428 the
Perugia~2011 tune~\cite{Skands:2010ak}) and with \herwig~6 (v6.521
with the AUET2-CTEQ6L1 tune~\cite{ATLAS:2011gmi}).
We examine the results at parton level, with multiple-parton
interaction (MPI) effects turned off.
Since the \pythia~6 and \pythia~8 results are very similar we will
show only the latter.
In the case of \pythia~8, we include an uncertainty band from the
variation of scales in the generation of the \powheg events.

\begin{figure}
  \centering
  \includegraphics[width=0.49\textwidth,page=1]{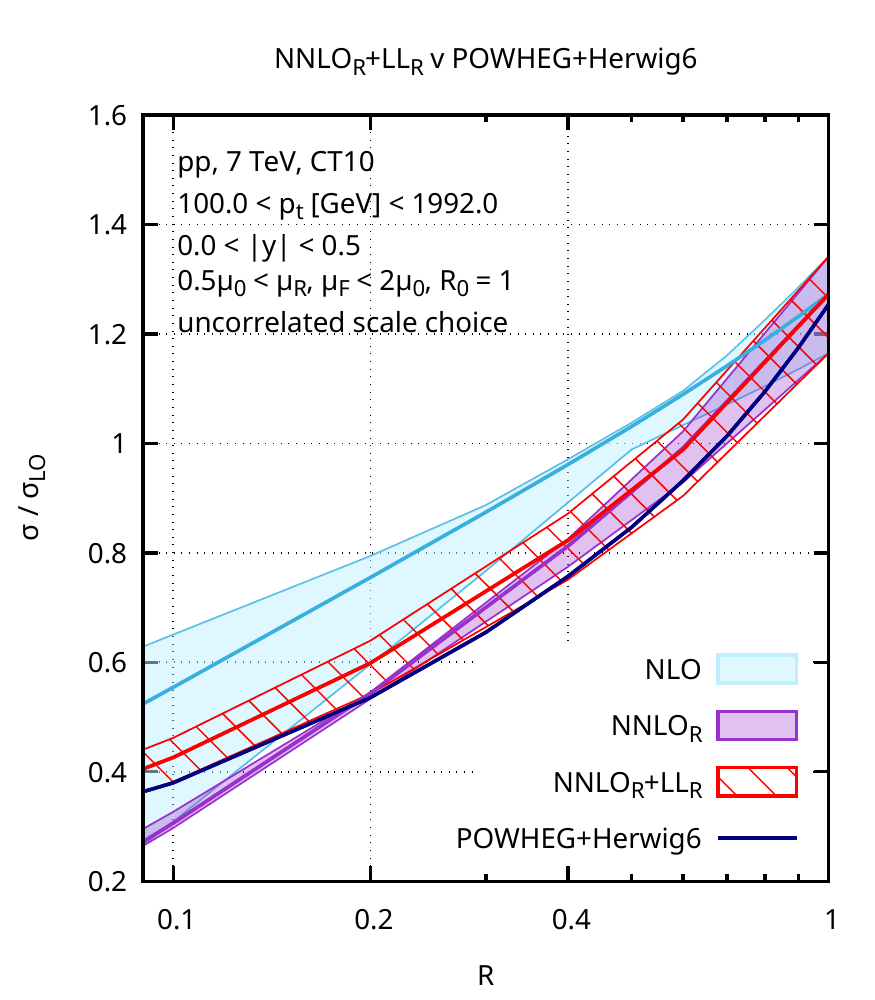}%
  \hfill%
  \includegraphics[width=0.49\textwidth,page=1]{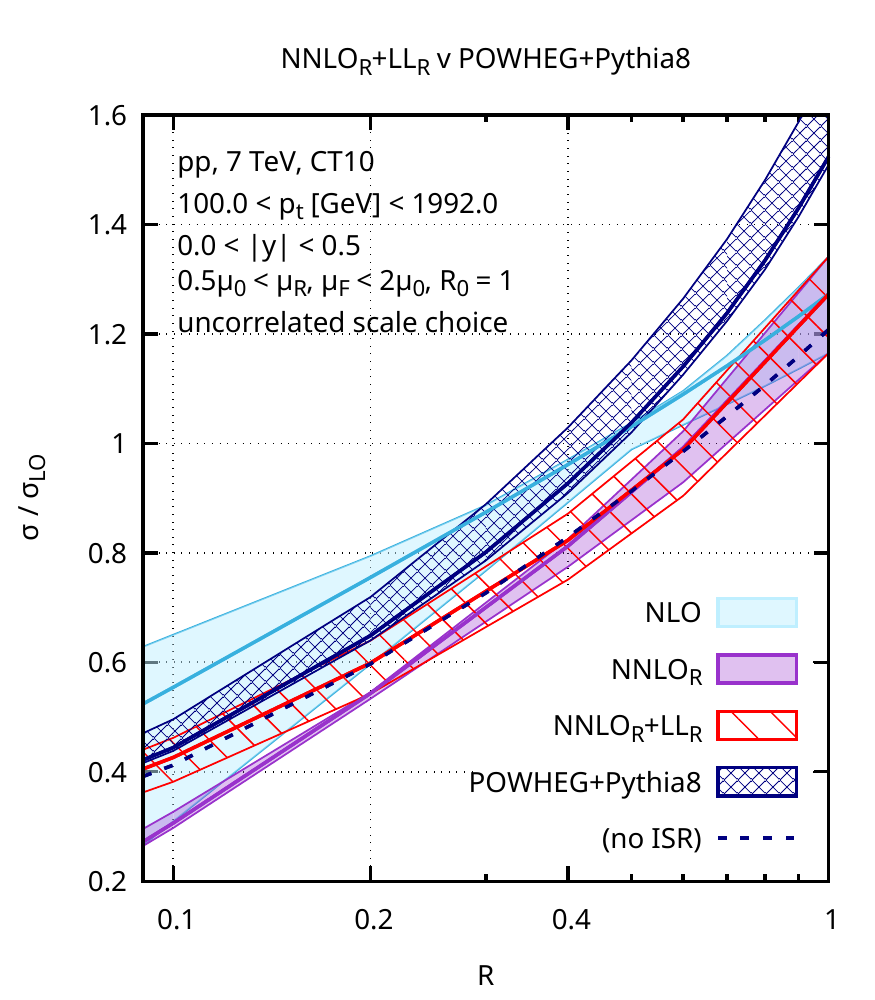}%
  \\
  \begin{minipage}[c]{0.45\linewidth}
    \caption{Top: comparison between the \NNLOR-based results and
      \powheg{}+\mbox{\herwig6} (left) and \powheg{}+\mbox{\pythia8}
      (right), shown as a function of $R$, integrated over $p_t$ for
      $p_t > 100 \GeV$.
      Bottom right: comparison of \powheg{}+\mbox{\pythia8} with
      \NNLOR-based results, where the latter have an additional NNLO
      $K$-factor of $1.15$. 
      \label{fig:comparison-powheg}
  }
\end{minipage}\hfill
\begin{minipage}[c]{0.49\linewidth}
  \includegraphics[width=\textwidth,page=1]{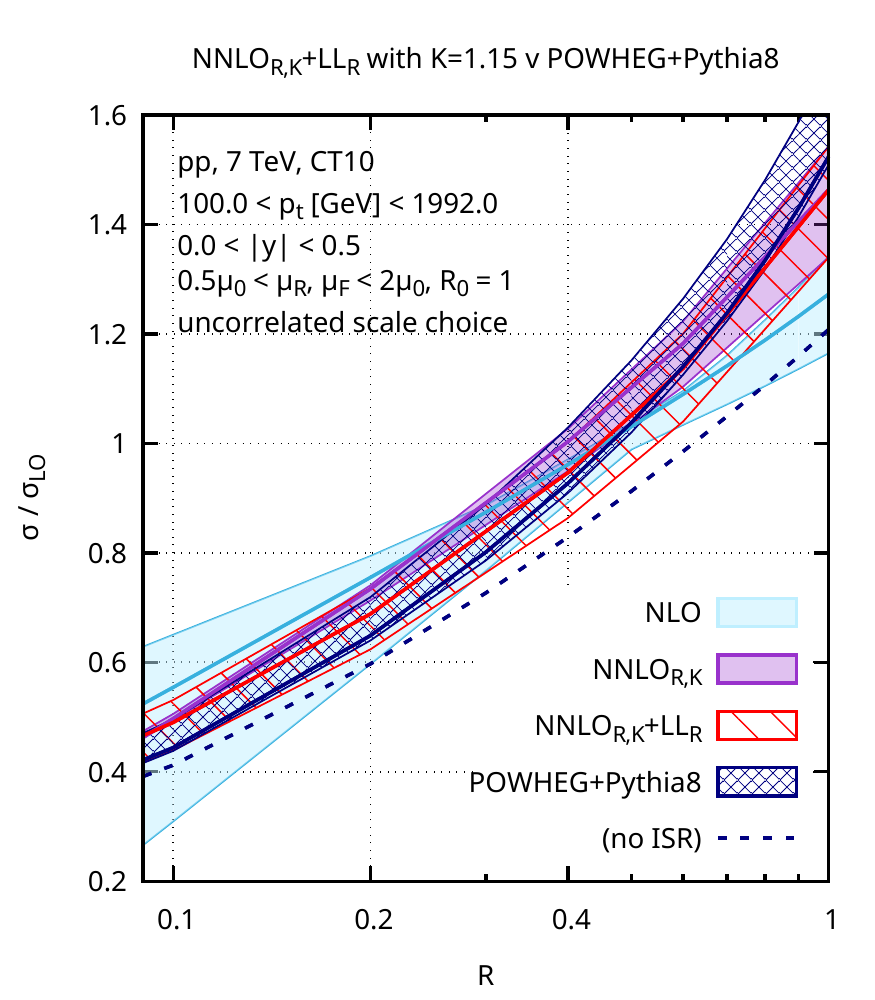}
\end{minipage}
\end{figure}

In Fig.~\ref{fig:comparison-powheg}, we show the $p_t$-integrated
cross section as a function of $R$.
The dark blue band (or line) shows the predictions obtained from \powheg.
In the top left-hand plot one sees a comparison with
\powheg{}+\mbox{\herwig6}, which agrees with the \NNLORLLR result to
within the latter's uncertainty band, albeit with a slightly steeper
$R$ dependence at large $R$ values. 
In the top right-hand plot, one sees a comparison with
\powheg{}+\mbox{\pythia8}.
There is reasonable agreement for small radii, however the
\powheg{}+\pythia8 prediction has much steeper $R$ dependence and is
substantially above the \NNLORLLR result for $R=1$.
Differences between \herwig and \pythia results with \powheg have been
observed before~\cite{Aad:2011fc}, though those are at hadron level,
including underlying-event effects, which can introduce further
sources of difference between generators.

%
%

%
%

One difference between the \NNLORLLR results and those from \powheg{}
with a shower-generator is an additional resummation of running scales
and Sudakov effects for initial-state radiation (ISR).
To illustrate the impact of ISR, the dark-blue dashed curve shows how
the \powheg{}+\pythia8 prediction is modified if one switches off
initial-state radiation (ISR) in the shower.
Though not necessarily a legitimate thing to do (and the part of the
ISR included in the \powheg-generated emission has not been switched
off), it is intriguing that this shows remarkably good agreement with
the \NNLORLLR results over the full $R$ range.
This might motivate a more detailed future study of the interplay
between ISR and the jet spectrum.
Note that, as shown in \cite{Alioli:2010xa}, nearly all the $R$
dependence of the \powheg{}+parton-shower result comes from the parton
shower component.
It is not so straightforward to examine \herwig with ISR turned off so
we have not included this in our study.

Given the differences between 
\powheg+\pythia8 and our \NNLORLLR results, it is also of interest to
examine what happens for $K\neq 1$.
We can tune $K$ so as to produce reasonable
agreement between \NNLORKLLR and \powheg{}+\pythia8 for $R=1$ and this
yields $K\simeq 1.15$, which we have used in the bottom-right plot.
Then it turns out that both predictions agree within uncertainty bands
not just at $R=1$, but over the full $R$ range.
In this context it will be particularly interesting to see what
effective value of $K$ comes out in the full NNLO calculation.
Note that the patterns of agreement observed between different
predictions depend also on $p_t$ and rapidity. 
For a more complete
picture we refer the reader to our online tool~\cite{OnlineTool}.

\section{Hadronisation}
\label{sec:hadronisation}

Before considering comparisons to data, it is important to examine also the
impact of non-perturbative effects.
There are two main effects: hadronisation, namely the effect of
the transition from parton-level to hadron-level; and the underlying
event (UE), generally associated with multiple interactions between partons
in the colliding protons.
Hadronisation is enhanced for small radii so we
discuss it in some detail.

One way of understanding the effect of hadronisation and the
underlying event is to observe that they bring about a shift in $p_t$.
This can to some extent be calculated analytically and applied to the
spectrum~\cite{Dasgupta:2007wa}.
An alternative, more widespread approach is to use a Monte Carlo
parton shower program to evaluate the ratio of hadron to parton level
jet spectra and multiply the perturbative prediction by that ratio.
One of the advantages of the analytical hadronisation approaches is
that they can matched with the perturbative calculation, e.g. as
originally proposed in Ref.~\cite{Dokshitzer:1995zt}.
In contrast, a drawback of the Monte Carlo hadronisation estimates is
that the definition of parton-level in a MC simulation is quite
different from the definition of parton level that enters a
perturbative calculation: in particular showers always include a
transverse momentum cutoff at parton level, while perturbative
calculations integrate transverse momenta down to zero.

To help guide our choice of method, we shall first compare the $p_t$
shift as determined in Ref.~\cite{Dasgupta:2007wa} with what is found
in modern Monte Carlo tunes. 
We first recall that the average shift should scale as $1/R$ (see also
Refs.~\cite{Korchemsky:1994is,Seymour:1997kj}) for hadronisation and
as $R^2$ for the 
underlying event (see also Ref.~\cite{Cacciari:2008gn}).
For small-$R$ jets, hadronisation should therefore become a large
effect, while the underlying event should vanish.
By relating the hadronisation in jets to event-shape measurements in DIS and
$e^+e^-$ collisions in a dispersive-type
model~\cite{Dokshitzer:1995zt,Dokshitzer:1995qm},
Ref.~\cite{Dasgupta:2007wa} argued that the average $p_t$ shift should
be roughly
\begin{equation}
  \label{eq:avg-pt-shift-hadronisation}
  \mean{\Delta p_t} \simeq - \frac{C}{C_F}\left( \frac{1}{R} +
    \order{1} \right) \times 0.5\GeV\,,
\end{equation}
where $C$ is the colour factor of the parton initiating the jet, $C_F
= \frac43$ for a quark and $C_A = 3$ for a gluon.
Those expectations were borne out by Monte Carlo simulations at the
time, with a remarkably small $\order{1}$ term.
Eq.~(\ref{eq:avg-pt-shift-hadronisation}) translates to a $-6\GeV$
shift for $R=0.2$ gluon-initiated jets. 
On a steeply falling spectrum, such a shift can modify the spectrum
significantly.

\begin{figure}
  \centering
  \includegraphics[page=1,width=0.49\textwidth]{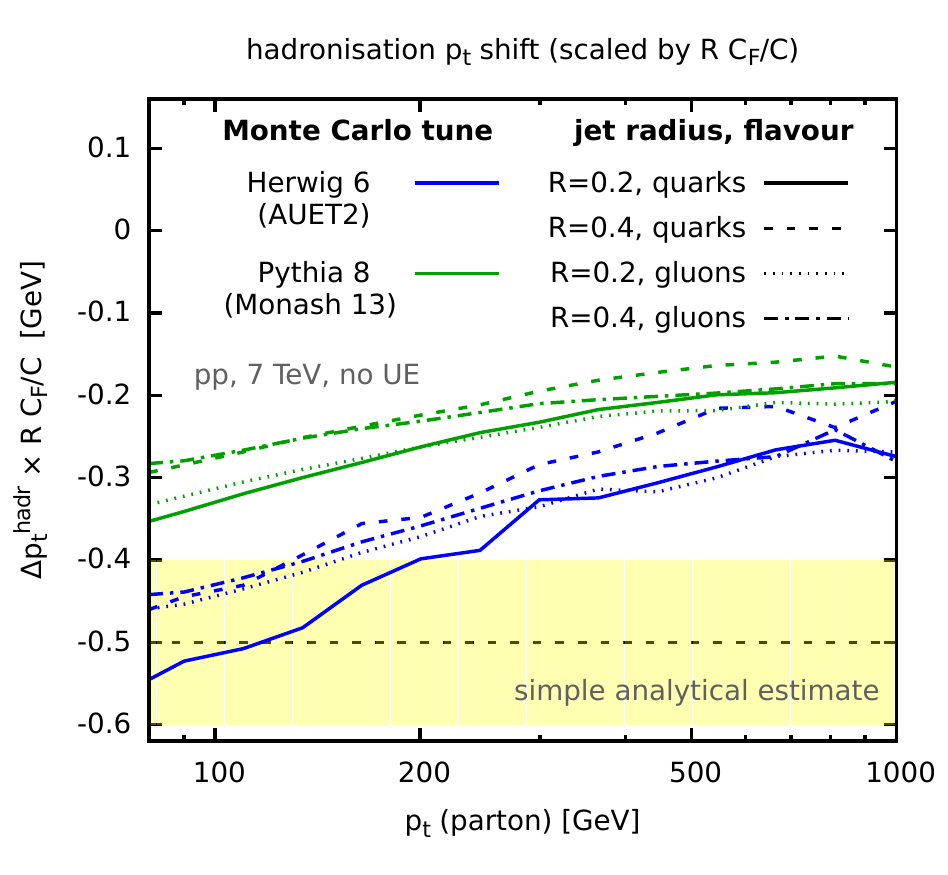}
  \hfill
  \includegraphics[page=2,width=0.49\textwidth]{Hjet-MC/dijet-plot-v4}
  \caption{The average shift in jet $p_t$ induced by hadronisation in a range
    of Monte Carlo tunes, for $R=0.4$ and $R=0.2$ jets, both quark and
    gluon induced.
    The shift is shown as a function of jet $p_t$ and is rescaled by a
    factor $R C_F/C$ ($C = C_F$ or $C_A$) in order to test the scaling
    expected from Eq.~(\ref{eq:avg-pt-shift-hadronisation}).
    The left-hand plot shows results from the AUET2~\cite{ATLAS:2011gmi}
    tune of \herwig~6.521~\cite{Corcella:2000bw,Corcella:2002jc} and 
    the Monash~13 tune~\cite{Skands:2014pea} of \pythia
    8.186~\cite{Sjostrand:2007gs}, while the right-hand plot shows 
    results from the Z2~\cite{Field:2010bc} and 
    Perugia~2011~\cite{Skands:2010ak,Cooper:2011gk} tunes of
    \pythia~6.428~\cite{Sjostrand:2006za}. 
    The shifts have been obtained by clustering each Monte Carlo event
    at both parton and hadron level, matching the two hardest jets
    in the two levels and determining the difference in their $p_t$'s.
    The simple analytical estimate of $0.5\GeV \pm 20\%$ is shown as
    a yellow band.
  }
  \label{fig:rescaled-hadr-shift}
\end{figure}

Fig.~\ref{fig:rescaled-hadr-shift} shows the shift in $p_t$ in going
from parton-level jets to hadron level jets, as a function of the jet
$p_t$.
Four modern Monte Carlo generator tunes are shown
\cite{ATLAS:2011gmi,Corcella:2000bw,Corcella:2002jc,Skands:2014pea,
  Sjostrand:2007gs,Skands:2010ak,Cooper:2011gk,Sjostrand:2006za}, two in
each plot.
For each generator tune (corresponding to a given colour), there are
four curves, corresponding to two values of $R$, $0.2$ and $0.4$ and
both quark and gluon jets.
The shifts have been rescaled by a factor $R C_F /C$.
This means that if radius and colour-factor dependence in
Eq.~(\ref{eq:avg-pt-shift-hadronisation}) are exact, then all
lines of a given colour will be superposed. 
This is not exactly the case, however lines of any given colour do
tend to be quite close, giving reasonable confirmation of the expected
trend of $C/R$ scaling.

A further expectation of Eq.~(\ref{eq:avg-pt-shift-hadronisation})
is that the lines should cluster around $0.5\GeV$ and be $p_t$
independent. 
This, however, is not the case.
Firstly, there is almost a factor of two difference between different
generators and tunes, with \pythia~6 Perugia 2011 and \pythia~8 Monash
2013 both having somewhat smaller than expected hadronisation
corrections.
Secondly there is a strong dependence of the shift on the initial jet
$p_t$, with a variation of roughly a factor of two between
$p_t = 100\GeV$ and $p_t = 1\TeV$.
Such a $p_t$ dependence is not predicted within simple approaches to
hadronisation such as
Refs.~\cite{Korchemsky:1994is,Dokshitzer:1995zt,Dokshitzer:1995qm,Dasgupta:2007wa}.
It was not observed in Ref.~\cite{Dasgupta:2007wa} because the Monte
Carlo study there
restricted its attention to a limited range of jet $p_t$, $55-70\GeV$.
The event shape studies that provided support for the analytical
hadronisation were also limited in the range of scales they probed,
specifically, centre-of-mass energies in the range $40-200\GeV$ (and
comparable photon virtualities in DIS).
Note, however, that scale dependence of the hadronisation has been
observed at least once before, in a Monte Carlo study shown in Fig.~8
of Ref.~\cite{Salam:2001bd}: effects found there to be associated
with hadron masses generated precisely the trend seen here in
Fig.~\ref{fig:rescaled-hadr-shift}.
The $p_t$ dependence of those effects can be understood analytically,
however we leave their detailed study in a hadron-collider context to
future work.\footnote{Hadron-mass effects have been discussed also in
  the context of Ref.~\cite{Mateu:2012nk}.}
Experimental insight into the $p_t$ dependence of hadronisation might
be possible by examining jet-shape
measurements~\cite{Aad:2011kq,Chatrchyan:2012mec} over a range of
$p_t$, however such a study is also beyond the scope of this work.

In addition to the issues of $p_t$ dependence, one further concern
regarding the analytical approach is that it has limited predictive
power for the fluctuations of the hadronisation corrections from
jet to jet.
Given that the jet spectrum falls steeply, these fluctuations can
have a significant impact on the final normalisation of the jet
spectrum. 
One might address this with an extension of our analytical approach to
include shape functions, e.g.\ as discussed in
Ref.~\cite{Korchemsky:1999kt}. 

In light of the above discussion, for evaluating hadronisation effects
here, we will resort to the standard approach of rescaling spectra by
the ratio of hadron to parton levels derived from Monte Carlo
simulations.%

\begin{figure}
  \centering
  \includegraphics[width=0.48\textwidth]{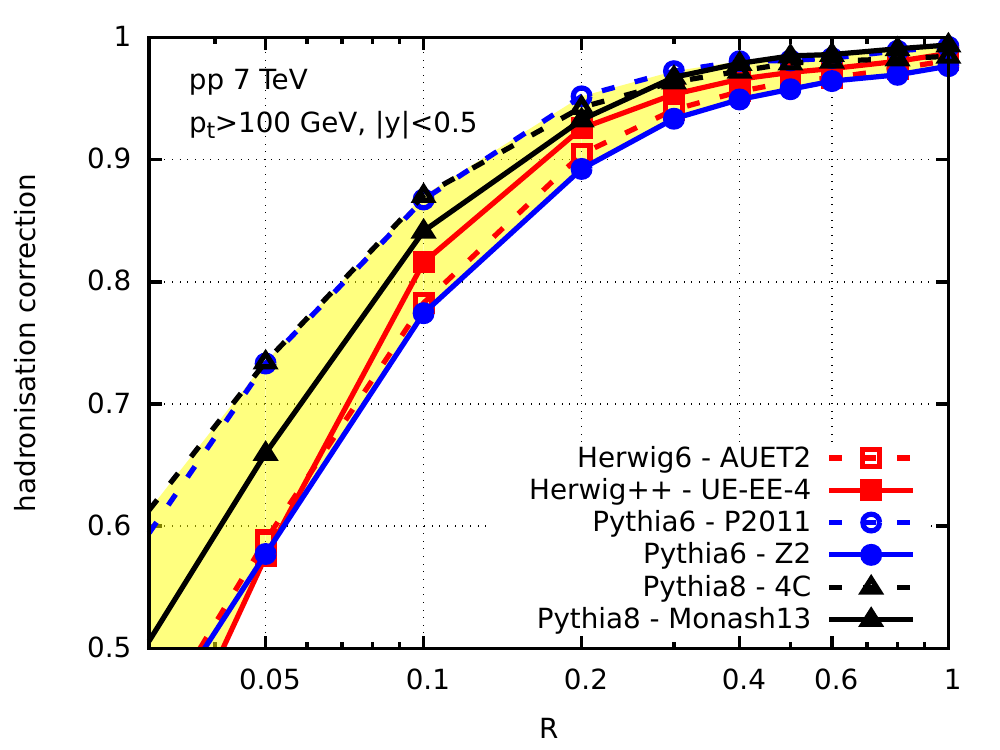}%
  \hfill
  \includegraphics[width=0.48\textwidth]{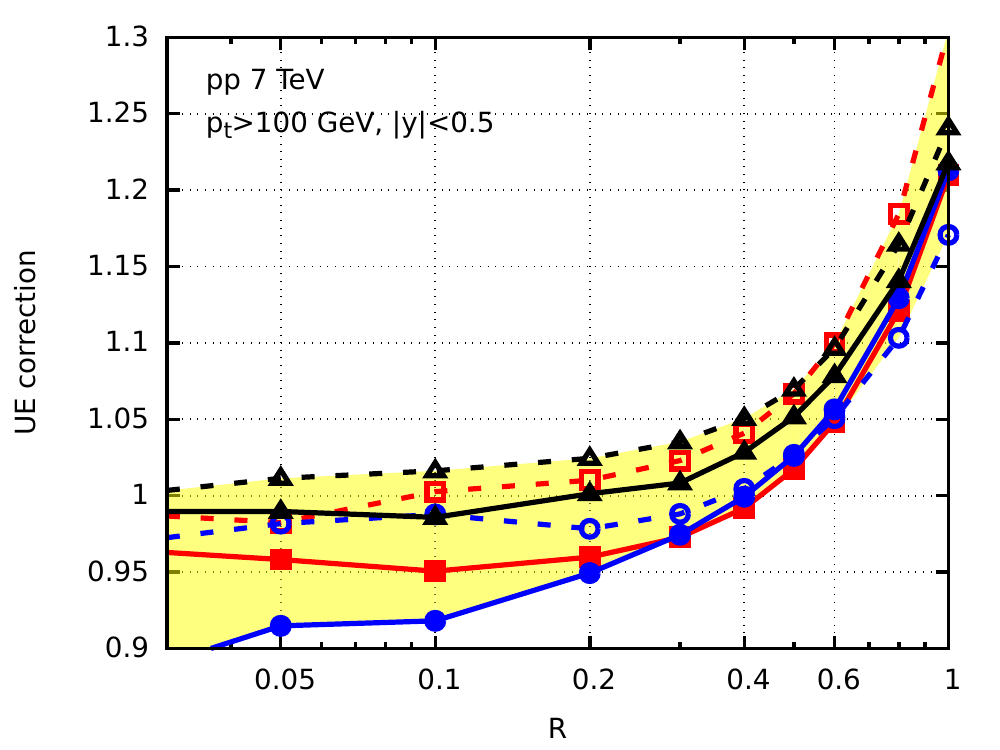}%
  \\
  \includegraphics[width=0.48\textwidth]{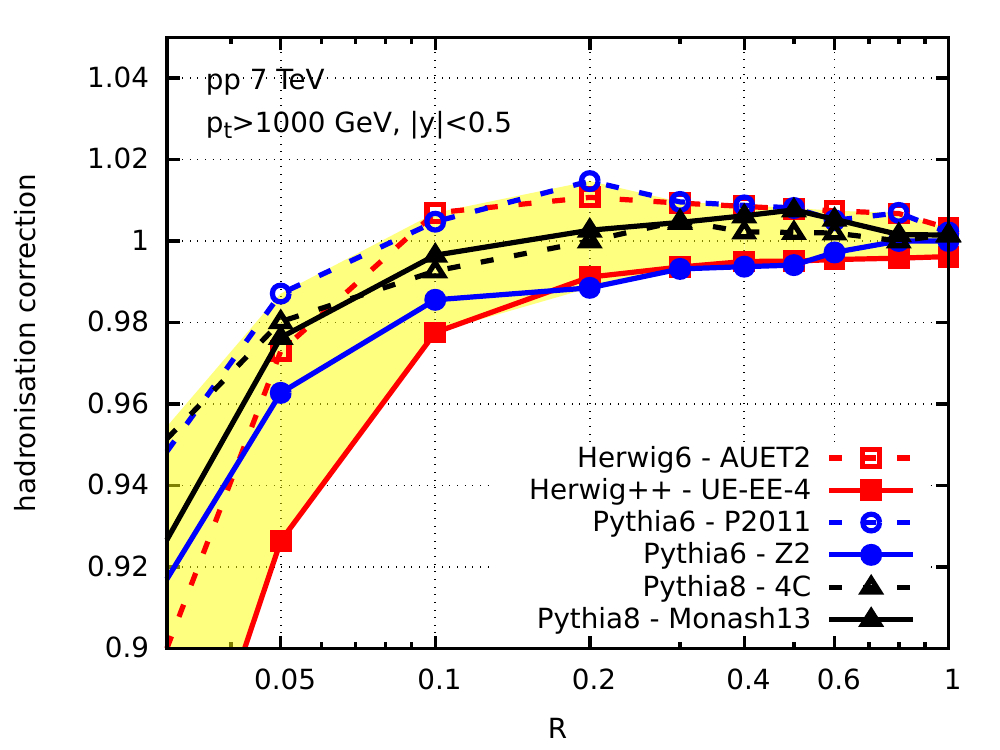}%
  \hfill
  \includegraphics[width=0.48\textwidth]{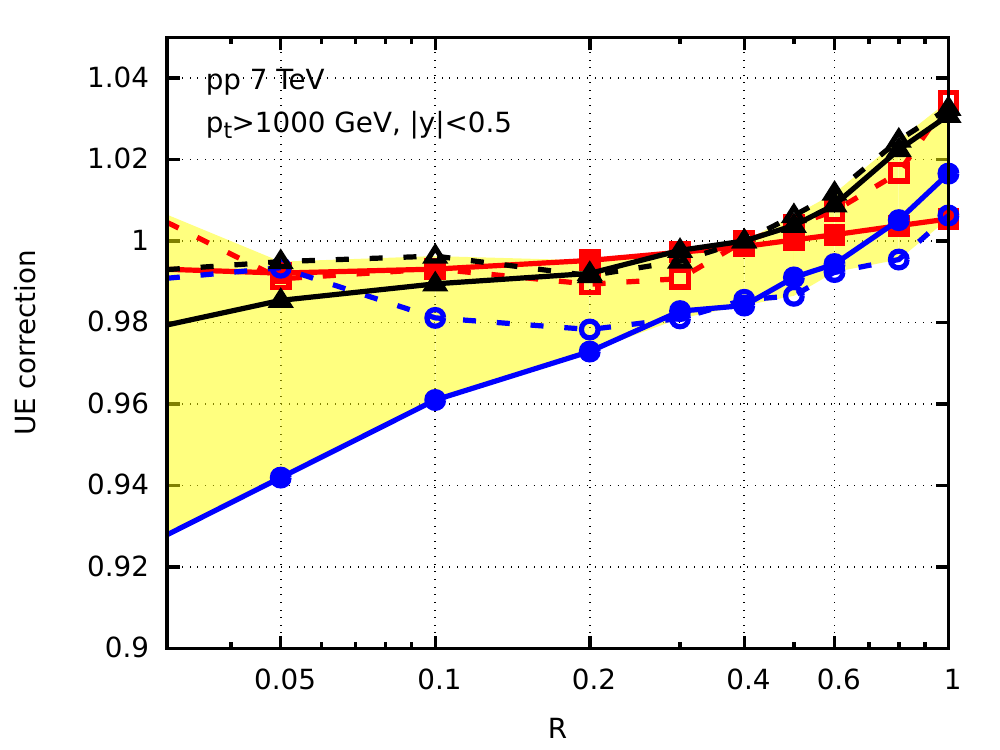}%
  \caption{ Hadronisation (left) and underlying event (right)
    multiplicative corrections to the jet spectrum, as a function of
    $R$ for pp collisions at $7\TeV$.  
    The top row shows results for $p_t>100\GeV$ and $|y|<0.5$, while
    the bottom row is for $p_t>1\TeV$.
    Six combinations of generator and tune are shown, and the yellow
    band corresponds to the envelope of the tunes.
}
  \label{fig:hadr+UE-v-R}
\end{figure}

Fig.~\ref{fig:hadr+UE-v-R} shows, as a function of $R$, the ratio of
hadron-level without UE to parton-level (left) and the ratio of hadron
level with UE to hadron level without UE (right), for a range of Monte
Carlo tunes.
The results are shown for $p_t > 100 \GeV$ in the upper row and $p_t >
1\TeV$ in the lower row.
A wide range of $R$ values is shown, extending well below
experimentally accessible values.
Beyond the tunes shown in Fig.~\ref{fig:rescaled-hadr-shift}, here we
also include the UE-EE-4~tune~\cite{Gieseke:2012ft} of 
\herwigpp 2.71~\cite{Bahr:2008pv,Bellm:2013hwb} and
tune 4C~\cite{Corke:2010yf} of \pythia~8.186~\cite{Sjostrand:2007gs}.
To investigate the issue of possible mismatch between our analytic
parton-level calculations and parton-level as defined in Monte Carlo
simulations, we have considered a modification of Monte Carlo parton
level where the transverse moment cutoff was taken to zero
(an effective cutoff still remains, because of the use finite parton
masses and $\Lambda_\text{QCD}$ in the shower, however this method can
arguably still give a rough estimate of the size of the effect one is
dealing with).
One finds that taking the cutoff to zero changes the parton-level
spectrum by a few percent effect.
As this is somewhat smaller than the differences that we will shortly
observe between tunes, it seems that for the time being it may not be
too unreasonable to neglect it.

While there is a substantial spread in results between the different
tunes in Fig.~\ref{fig:hadr+UE-v-R}, the observed behaviours are
mostly as expected, with hadronisation 
reducing the jet spectrum, especially at the smallest $R$ values,
while the UE increases it, especially at large $R$ values.
The magnitude of these effects is strongly $p_t$ dependent, with
(roughly) a factor of ten reduction at not-too-small $R$ values when
going from $p_t>100\GeV$ to $p_t> 1\TeV$.
Such a scaling is consistent with a rough $1/(R p_t)$ behaviour for
hadronisation and $R^2/p_t$ behaviour for the UE (ignoring the slow
changes in quark/gluon fraction and steepness of the spectrum as $p_t$
increases).

One surprising feature concerns the behaviour of the UE corrections at
very small radii: firstly, in a number of the tunes the corrections
tend to be smaller than $1$, suggesting that the multi-parton
interactions (MPI) that are responsible for the UE \emph{remove} energy from
the core of the jet. 
For $R$ values in the range $0.4{-}1$, the effect of MPI is instead to
add energy to the jet, as expected.
Secondly, this loss of energy from the jet is not particularly
suppressed at high $p_t$.
The most striking example is the Z2 tune where there can be
corrections of up to $7\%$ at $R=0.03$ (and even more at yet smaller
$R$ values).
The effect is somewhat reduced in the Z2-LEP tune, which has modified
fragmentation parameters.
One wonders if the mechanism for MPI generation might be inducing
some form of factorisation breaking.
A simple context in which to study this might be the high-$p_t$
inclusive hadron spectrum, where factorisation would imply that MPI
should have no effect.
While $k_t$-factorisation is believed to be broken for the inclusive
hadron spectrum~\cite{Collins:2007nk}, we are not aware of definite
statements concerning breaking of collinear factorisation. 

\section{Comparisons to data}
\label{sec:data-comparisons}

Having formulated and studied the perturbative and non-perturbative
contributions to the inclusive jet spectrum, we now consider
comparisons with data.
The purpose of this section is to highlight the relative sizes of
different physical effects as compared to the precision of the data.

We will compare our predictions to the two datasets that have the
smallest $R$ values: that from ALICE at centre-of-mass energy
$\sqrt{s}=2.76\TeV$ with $R=0.2$ and $0.4$~\cite{Abelev:2013fn} and
that from ATLAS at $\sqrt{s}=7\TeV$ with $R=0.4$ and
$0.6$~\cite{Aad:2014vwa}.\footnote{The CMS collaboration has also
  published inclusive jet spectrum
  results~\cite{Chatrchyan:2012bja,Khachatryan:2015luy}, however the
  smallest $R$ considered there is slightly larger, $R=0.5$.  }

All our results are obtained with CT10 NLO PDFs. 
This is the case also for our LO and \NNLOR results.
For the latter,
since \NNLOR does not correspond to full NNLO, it is justifiable to
use NLO PDFs.\footnote{In interpreting the plots, one may wish to keep
  in mind the potential impact of $K\neq 1$, which is illustrated
  explicitly in Section~\ref{sec:data-theory-K}. The plots use a
  $p_t$-independent NNLO $K$ factor, however the true $K$ factor would
  depend on $p_t$.}
One should also be aware that most modern PDF sets include inclusive
jet-data in their fit.
Accordingly they may have a bias associated with the theory choice
that was used in their determination.
With an updated theoretical framework, such as that used here, the
PDFs would conceivably change and a complete study would benefit from
refitting the PDFs.
That is beyond the scope of this work and anyway more appropriately
done once full NNLO results become available.
For completeness, we have nevertheless briefly examined the impact of
changing PDFs in the context of a pure \LLR calculation, examining
also CT10nnlo~\cite{Gao:2013xoa}, CT14nlo,
CT14nnlo~\cite{Dulat:2015mca}, MSTW2008nlo~\cite{Martin:2009iq},
MMHT2014nlo, MMHT2014nnlo~\cite{Harland-Lang:2014zoa},
NNPDF30\_nlo\_as\_0118 and and
NNPDF30\_nnlo\_as\_0118~\cite{Ball:2014uwa}.
For $p_t$'s below $500\GeV$, most of these PDFs give results slightly
above those from CT10, but by no more than $6\%$, which is modest
relative to other uncertainties and differences that we will see
below.
%

All fixed-order results are obtained with version~4.1.3 of the \nlojet
program~\cite{nlojet}.
Our central renormalisation and factorisation scale choice is
$\mu_0 = p_{t,\max}^{R=1}$, the transverse momentum of the hardest
jet in the event as clustered with $R=1$.
The envelope of independent variations of $\mu_R$ and $\mu_F$ by a factor of two
(while maintaining $\frac12\le \mu_R/\mu_F\le 2$) provides the
perturbative uncertainty estimate.
In the case of NLO-mult. and (N)NLO$_{(R)}$+{\LLR} results,
the scale variation is
performed independently for the normalisation and fragmentation
factors and the uncertainty from the two factors is then added in
quadrature.
As explained in section~\ref{sec:matching}, this is intended to avoid
spuriously small scale uncertainties associated with cancellations
between different physical contributions.

Non-perturbative corrections are taken as the average of the
parton-to-hadron Monte Carlo correction factors (including
hadronisation and UE) as obtained with the six different tunes
discussed in section~\ref{sec:hadronisation}.
The envelope of that set of six corrections provides our estimate
of the uncertainty on the non-perturbative corrections, which is
added in quadrature to the perturbative uncertainty.

In the case of the ATLAS data we will explore transverse momenta well
above the electroweak (EW) scale, where EW corrections become substantial.
The ATLAS collaboration accounted for these using the calculation of
tree-level ($\order{\as \alpha_\text{EW}}$) and loop ($\order{\as^2
    \alpha_\text{EW}}$) EW effects from Ref.~\cite{Dittmaier:2012kx}.
Here, since we concentrate on QCD effects, when showing the data we
divide it by the EW corrections quoted by ATLAS.\footnote{
  %
  Those
  corrections don't account for real $W$ and $Z$ emission.
  The first estimate of real EW emission effects was given by
  Baur~\cite{Baur:2006sn}, but at the time only $14 \TeV$ collisions
  were envisaged.
  The real contributions for 7 TeV collisions have been evaluated in
  Ref.~\cite{Meric:2013dua}.
  At high $p_t$'s they grow to become up to $3{-}4\%$, however in this
  region statistical and systematic uncertainties on the data are
  substantially larger and so we believe it is reasonable to
  neglect them.}

\subsection{Comparison to ALICE data}
\label{sec:alice-comparisons}

As a first application of small-$R$ resummation in comparisons to
data, we look at the inclusive jet cross section in proton-proton
collisions at $\sqrt{s}=2.76\GeV$ reported by the ALICE collaboration
\cite{Abelev:2013fn}.
The measurements are in the $|y|<0.5$ rapidity range, with jets
obtained using the anti-$k_t$ algorithm with a boost-invariant
$p_t$ recombination scheme, for radii $R=0.2$ and $0.4$.

\begin{figure}
  \centering
  \includegraphics[width=0.5\textwidth]{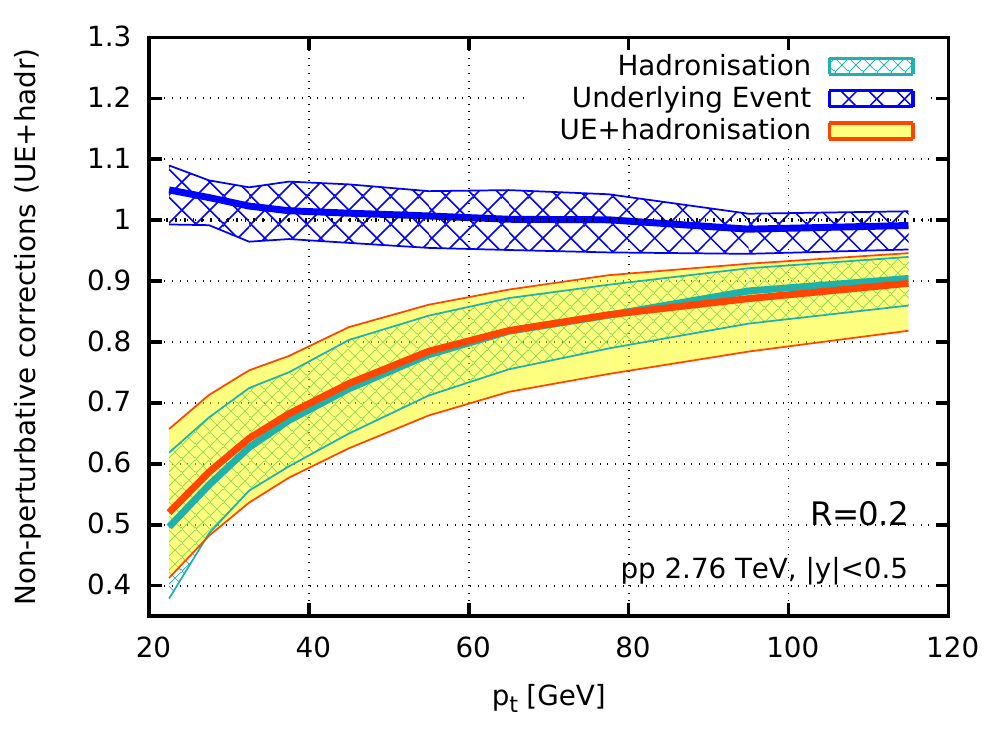}%
  \includegraphics[width=0.5\textwidth]{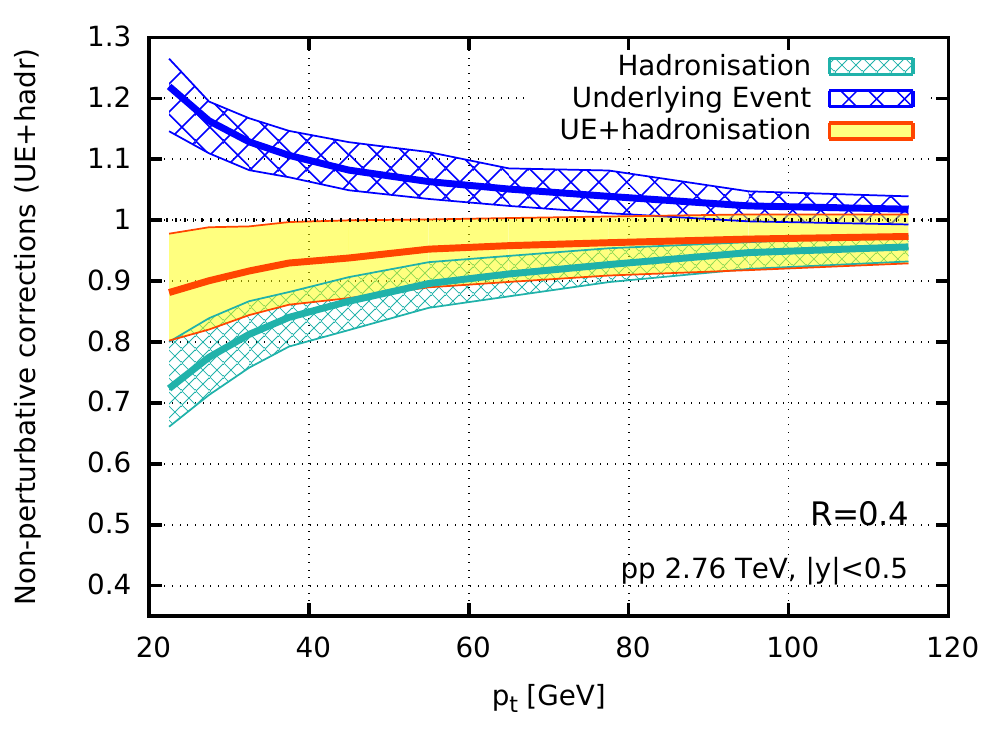}
  \caption{Non-perturbative corrections to the inclusive jet spectrum
    for the $p_t$ range, rapidity and centre-of-mass energy
    corresponding to the ALICE data~\cite{Abelev:2013fn} for $R=0.2$
    (left) and $R=0.4$ (right).
    The results are shown separately for hadronisation, UE and the
    product of the two, and in each case include the average and
    envelope of the corrections from the six tunes discussed in
    section~\ref{sec:hadronisation}. 
}
  \label{fig:non-pert-alice}
\end{figure}

\begin{figure}
  \centering
  \includegraphics[width=0.5\textwidth]{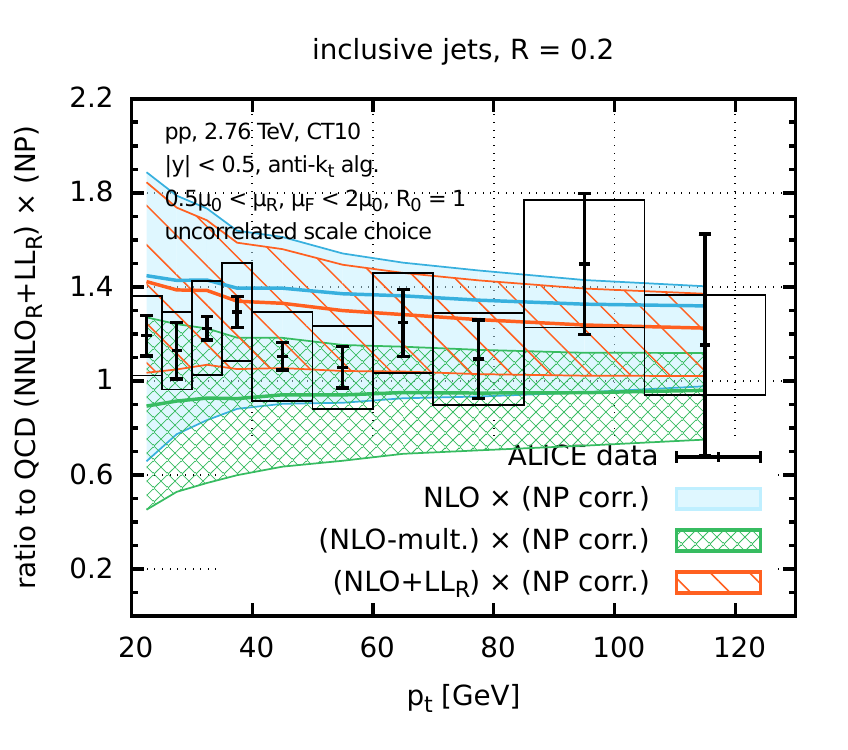}%
  \includegraphics[width=0.5\textwidth]{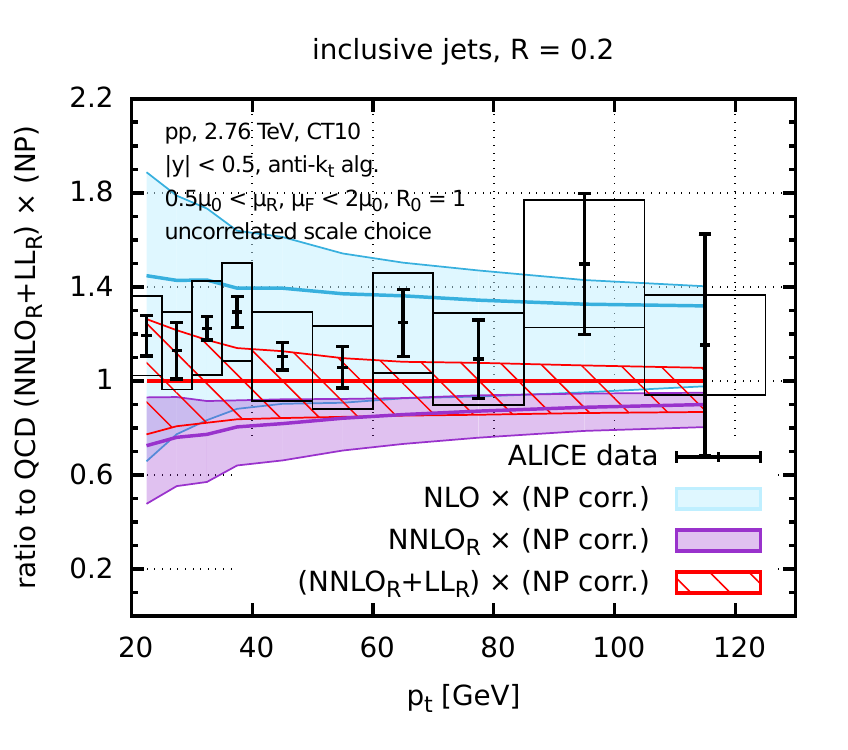}\\
  \includegraphics[width=0.5\textwidth]{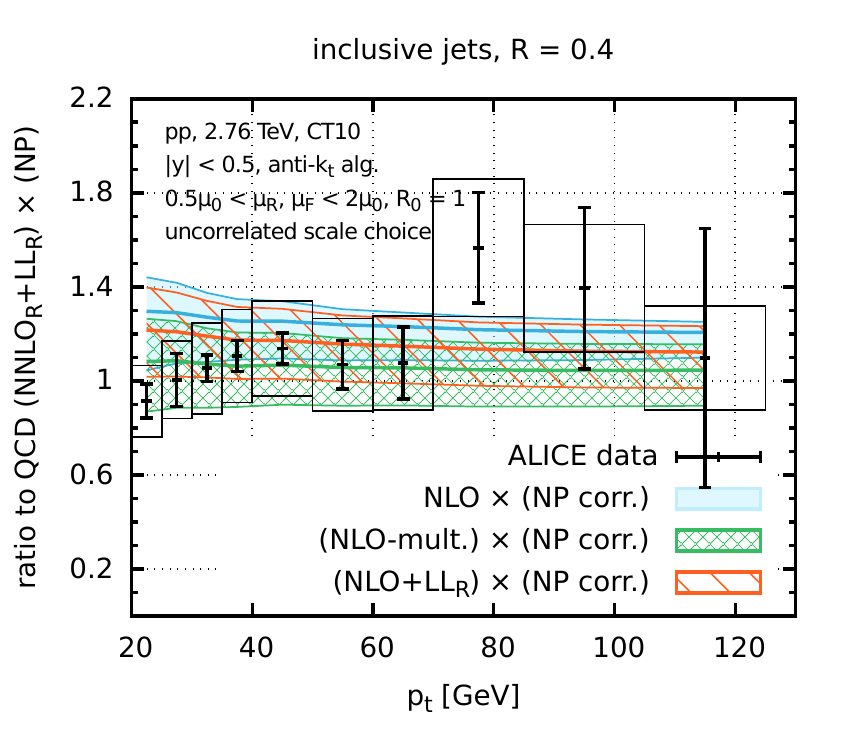}%
  \includegraphics[width=0.5\textwidth]{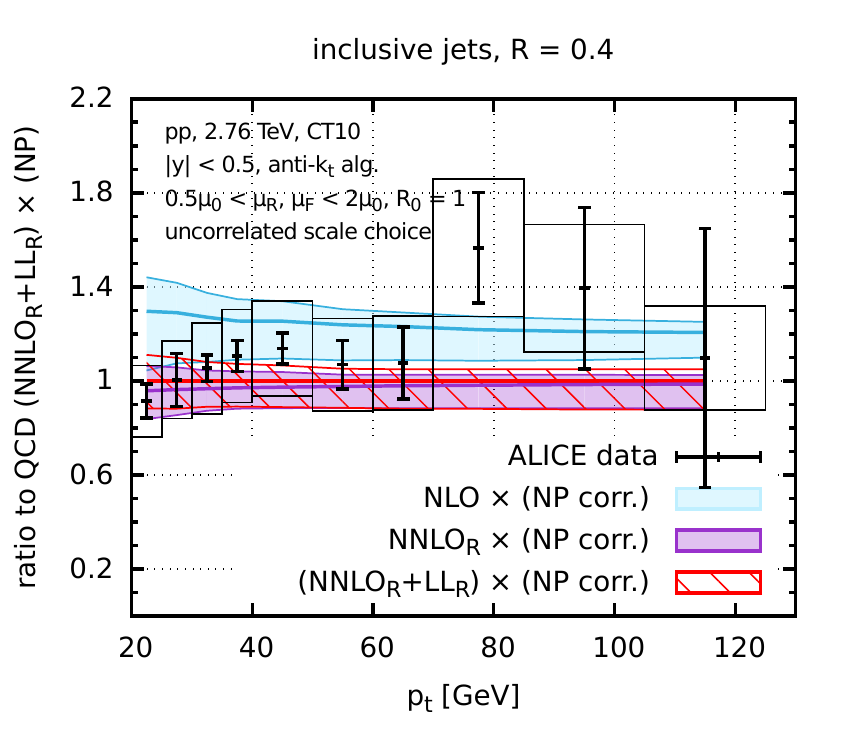}\\
  \caption{Comparison between a range of theoretical predictions for
    the inclusive jet spectrum and data from
    ALICE~ at $\sqrt{s}=2.76\TeV$~\cite{Abelev:2013fn}.
    The upper row is for $R=0.2$ and the lower one for $R=0.4$.
    The left-hand column shows NLO-based comparisons, while the
    right-hand one shows \NNLOR-based comparisons.
    Rectangular boxes indicate the size of systematic uncertainties on
    the data points, while the errors bars correspond to the
    statistical uncertainties.
    Results are normalised to the central \NNLORLLR prediction
    (including non-perturbative corrections).
  }
  \label{fig:alice-comparison}
\end{figure}

The non-perturbative corrections for hadronisation and underlying event
are shown in Fig.~\ref{fig:non-pert-alice}.
For $R=0.2$, non-perturbative corrections are
largely dominated by hadronisation, with underlying event being a
small effect, as expected for sufficiently small $R$.
The net non-perturbative correction is about $-50\%$ at the lowest
$p_t$ of $20\GeV$, while it decreases to about $-10\%$ at $100\GeV$.
For $R=0.4$ there is a partial cancellation between hadronisation and
UE, with a net impact of about $-10\%$ percent at low $p_t$ and a
$5{-}10\%$ uncertainty.

The comparison of our full results to the ALICE data is given in
Fig.~\ref{fig:alice-comparison}, as a ratio to the \NNLORLLR theory
prediction (including non-perturbative corrections).
The top row shows the jet spectrum for $R=0.2$, while the lower row
corresponds to $R=0.4$.
The left-hand plots show NLO-based theory results.
They all appear to be consistent with the data within their large
uncertainties.
The
right-hand plots show \NNLOR-based theory (with plain NLO retained to
facilitate cross-comparisons).
In general the \NNLORLLR results appear to provide the best match for
the data, though they are slightly low.
In particular, for $R=0.2$ where the differences between \NNLORLLR and
\NNLOR are substantial, almost $30\%$ at low $p_t$, there seems to be a
preference for \NNLORLLR.
In contrast, at $R=0.4$ there is little difference between the two
predictions though both are significantly more compatible with the
data than is the plain NLO.
In considering these statements, it is however important to keep
several caveats in mind: 
the systematic uncertainties on the data and on the non-perturbative
corrections (especially for $R=0.2$) are not negligible and a
one-$\sigma$ shift could somewhat affect the conclusions.
Furthermore, the currently unknown finite NNLO contribution (the
difference between \NNLOR and full NNLO) may also have a relevant
impact.

To further evaluate the compatibility of our results and the data we
examine the ratio of the inclusive jet spectra at the two $R$ values,
${\cal R}(p_t;R_1,R_2)=\sigma(p_t;R_1)/\sigma(p_t;R_2)$ with $R_1=0.2$ and
$R_2=0.4$.
This ratio is of interest because it allows us to directly study the 
$R$ dependence of the results and also because certain components of
the uncertainties cancel in the ratio, in both the data and the
theoretical prediction.
In the experimental results, for example, the luminosity
uncertainty cancels, as should part of the jet energy scale and
resolution uncertainties.
In the theoretical prediction, PDF uncertainties cancel.
The ALICE collaboration's results~\cite{Abelev:2013fn} explicitly
include a determination of the ratio.

Earlier studies that focused on the $R$ ratios~\cite{Soyez:2011np}
directly used the perturbative expansion for the cross-section ratio,
rather than the ratio of perturbative predictions for the cross
sections.
That approach could be extended also to matched ratios, and one
example of a NNLO+\LLR matching formula for the ratio would be
\begin{multline}
  \label{eq:ratio-matching-nnlo}
  {\cal R}^{\NNLO+\LLR,\text{expand}} = 
  \frac{\sigma^\LLR(R_1)}{\sigma^\LLR(R_2)}
  \times
  \Bigg(1 + \Delta_{1+2}(R_1,R_2)
  - \Delta_{1+2}(R_1,R_2)\frac{\sigma_1(R_2)}{\sigma_0}\\\hspace*{24mm}
  - \frac{\sigma_1^\LLR(R_1)+\sigma_2^\LLR(R_1)
    -\sigma_1^\LLR(R_2)-\sigma_2^\LLR(R_2)}{\sigma_0}\\
  + \left(\frac{\sigma_1^\LLR(R_1)}{\sigma_0}-\Delta_1(R_1,R_2)\right)\frac{\sigma_1^\LLR(R_1)-\sigma_1^\LLR(R_2)}{\sigma_0}\Bigg)\,.
\end{multline}
However, we prefer here to simply take the ratios of the
relevant theory prediction (NLO, \NNLOR, \NNLORLLR, etc.) at the two
$R$ values, e.g.
\begin{equation}
  \label{eq:ratio-matching-nnlo-us}
  {\cal R}^{\NNLOLLR} =
  \frac{\sigma^{\NNLORLLR}(R_1)}{\sigma^{\NNLORLLR}(R_2)} \,.
\end{equation}
This simple ratio has the same formal accuracy as
(\ref{eq:ratio-matching-nnlo}) and fits better our primary goal, which
is to predict inclusive jet cross sections and only examine their
ratios for different $R$ values as a supplementary test.
%
In the case of the results matched to \LLR resummation and of the
(N)NLO-mult.\ results, the normalisation factor (with the cross section
at radius $R_0$) cancels in the ratio, leaving only the fragmentation
factor. 
For the \NNLOR-mult.\ and \NNLORLLR results in particular, this means
that any dependence on the unknown full NNLO $K$-factor (or,
equivalently, the choice of $R_m$ in Eq.~(\ref{eq:sigma-nnlo})) is
eliminated, and the prediction for the ratio is identical to that
which would be obtained with the full NNLO result.
Accordingly, we will drop the subscript $_R$ label in these cases,
i.e.\ writing ${\cal R}^{\NNLOLLR}$ in
Eq.~(\ref{eq:ratio-matching-nnlo-us}) rather than ${\cal
  R}^{\NNLORLLR}$.

To estimate the perturbative theoretical uncertainties on the ratio,
we take the envelope of the ratios as determined for our seven
renormalisation and factorisation scale choices.
In the case of (N)NLO-mult.\ and (N)NLO+\LLR results, since the
normalisation factor cancels, we only consider the component of the
perturbative uncertainties associated with the fragmentation factor.
We have verified that the effect of $R_0$ variation is contained
within the scale-variation envelope.
%
For the non-perturbative uncertainties, we take the envelope of
the ratios of the corrections factors from different Monte Carlo
tunes.
The perturbative and non-perturbative uncertainties on the ratio are
added in quadrature.

\begin{figure}
  \centering
  \includegraphics[width=0.5\textwidth,page=1]{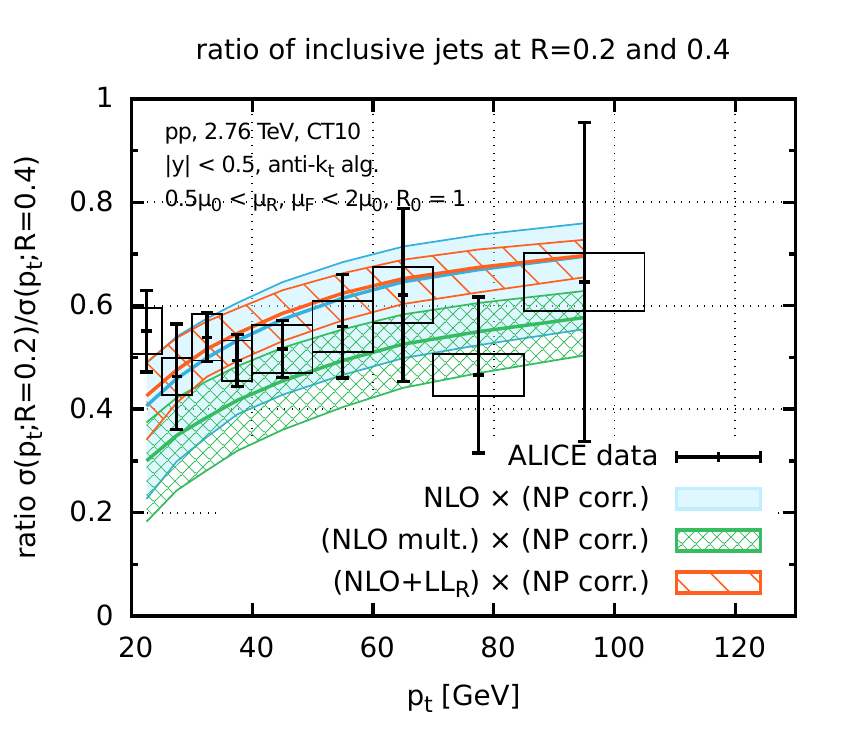}%
  \includegraphics[width=0.5\textwidth,page=1]{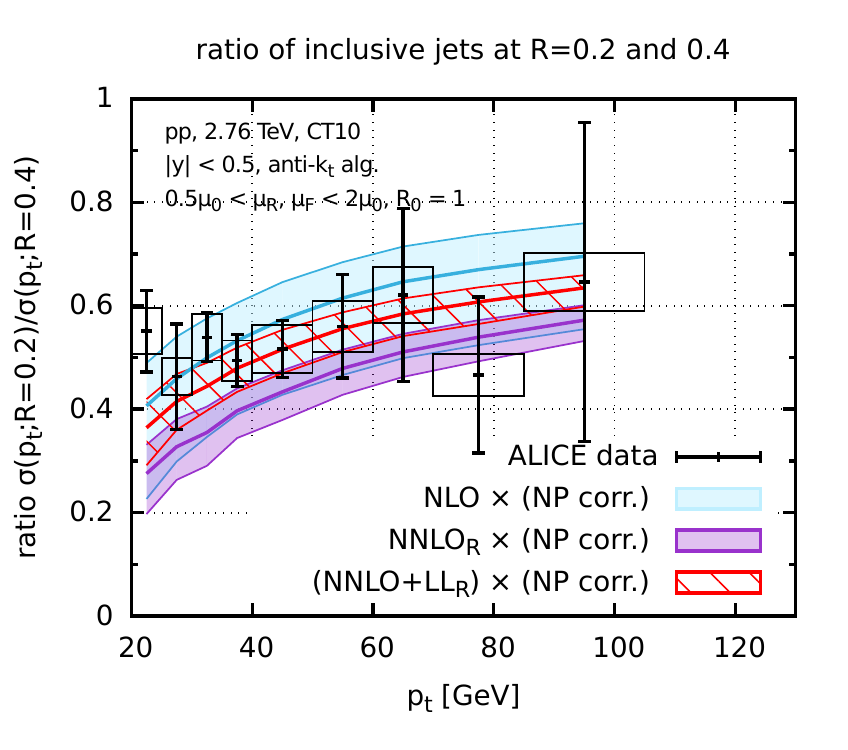}
  \caption{Comparison between a range of theoretical predictions for
    the inclusive jet cross-section ratio and data from
    ALICE~ at $\sqrt{s}=2.76\TeV$~\cite{Abelev:2013fn}.
    The left-hand column shows NLO-based comparisons, while the
    right-hand one shows \NNLOR-based comparisons.
    Rectangular boxes indicate the size of systematic uncertainties on
    the data points, while the errors bars correspond to the
    statistical uncertainties.
  }
  \label{fig:alice-comparison-ratio}
\end{figure}

The comparison of the theory predictions with the measurements of the
ALICE collaboration is presented in
Fig.~\ref{fig:alice-comparison-ratio}, at \NLO accuracy on the left and
at NNLO$_{(R)}$-based accuracy on the right.
At first sight, it appears that the data have a considerably flatter
$p_t$ dependence than any of the theory predictions.
The latter all grow noticeably with increasing $p_t$, a
consequence mainly of the $p_t$ dependence of the non-perturbative
correction factor, cf.\ Fig.~\ref{fig:non-pert-alice}.
Nevertheless, on closer inspection one sees that if one ignores the
left-most data point then the remaining data points are compatible
with the predicted $p_t$ dependence.
The overall agreement is then best with the \NNLO\LLR-based
prediction.
However, the sizes of the experimental uncertainties are such that it
is difficult to draw firm conclusions.

We have also examined the impact of using
Eq.~(\ref{eq:ratio-matching-nnlo}) instead of
(\ref{eq:ratio-matching-nnlo-us}) and find that the difference is
small, no more than $5\%$.
We have also examined the pure NNLO expansion of the ratio of cross
sections, as used in Ref.~\cite{Soyez:2011np} and find that this too
is quite similar to Eq.~(\ref{eq:ratio-matching-nnlo-us}), much more
so than the direct ratio of NNLO results,
$\sigma^{\NNLOR}(R_1)/\sigma^{\NNLOR}(R_2)$.
Thus our finding that we obtain reasonable agreement between
Eq.~(\ref{eq:ratio-matching-nnlo-us}) and the data is consistent with
the observations of Ref.~\cite{Soyez:2011np}, which were based on
expanded NNLO ratios.%
\footnote{ 
  Note,
  that Ref.~\cite{Soyez:2011np} used an analytical rather than
  Monte-Carlo based approach to estimating hadronisation corrections.
}

\subsection{Comparison to ATLAS data}
\label{sec:atlas-comparisons}

Let us now turn to a comparison with the inclusive jet cross-sections
reported by the ATLAS collaboration~\cite{Aad:2014vwa},
obtained from $4.5 \text{ fb}^{-1}$ of 
proton-proton collisions at $\sqrt{s}=7\TeV$.
Jets are identified with the anti-$k_t$ algorithm, this time with a
usual $E$-scheme, taking radii $R=0.4$ and $0.6$.
The measurements are doubly-differential, given as a function of jet
$p_t$ and rapidity, and performed for $p_t > 100 \GeV$ and $|y| < 3$.
Note that given the difference in centre-of-mass energy, the lower
$p_t$ for the ATLAS data, $100\GeV$, involves the same partonic $x$
range as $p_t=40\GeV$ for the ALICE data.

\begin{figure}
  \centering
  \includegraphics[width=0.5\textwidth]{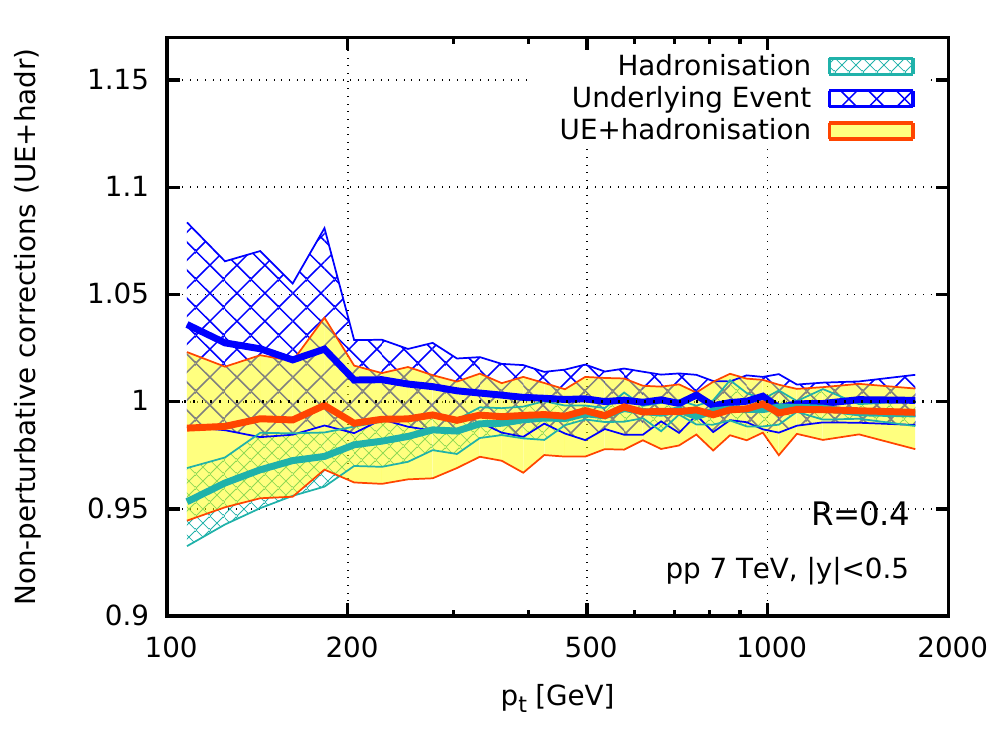}%
  \includegraphics[width=0.5\textwidth]{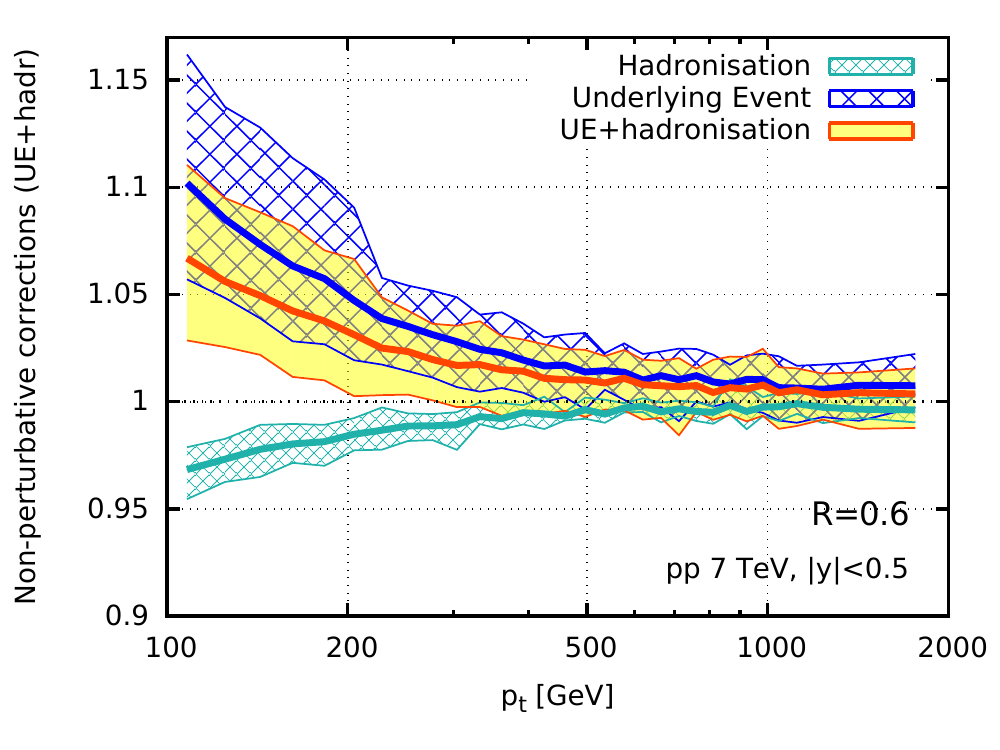}
  \caption{Non-perturbative corrections to the inclusive jet spectrum
    for the $p_t$ range, rapidity and centre-of-mass energy
    corresponding to the ATLAS data~\cite{Aad:2014vwa} for $R=0.4$
    (left) and $R=0.6$ (right).}
  \label{fig:non-pert-atlas}
\end{figure}
\begin{figure}
  \centering
  \includegraphics[width=0.5\textwidth]{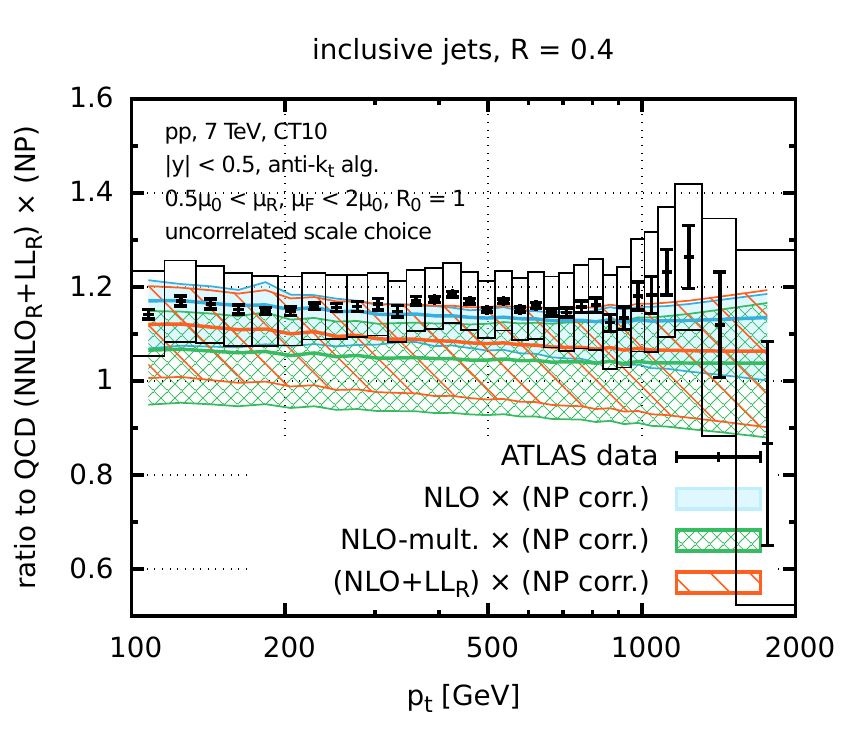}%
  \includegraphics[width=0.5\textwidth]{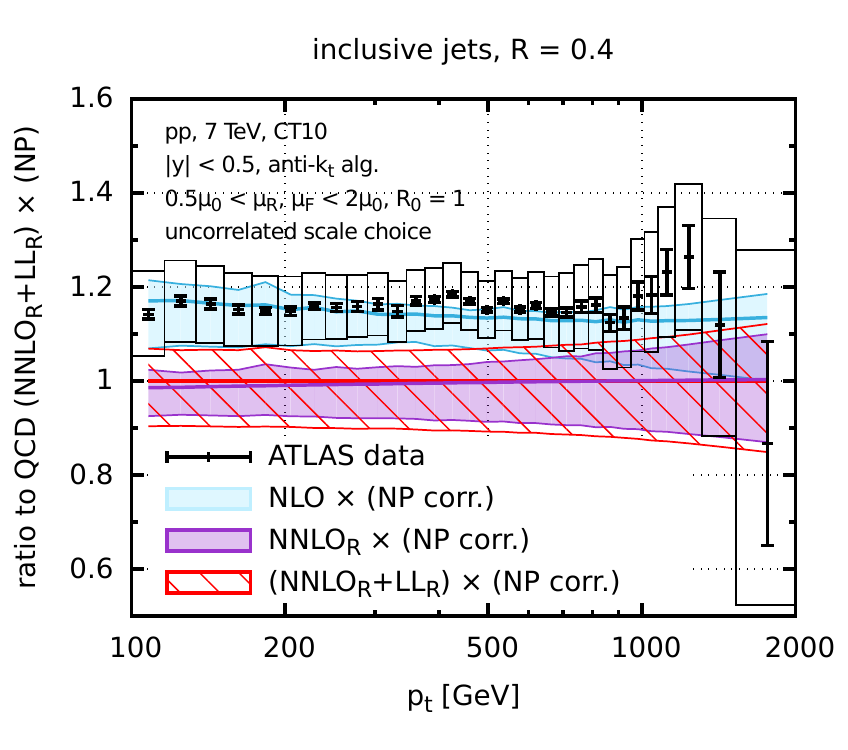}\\
  \includegraphics[width=0.5\textwidth]{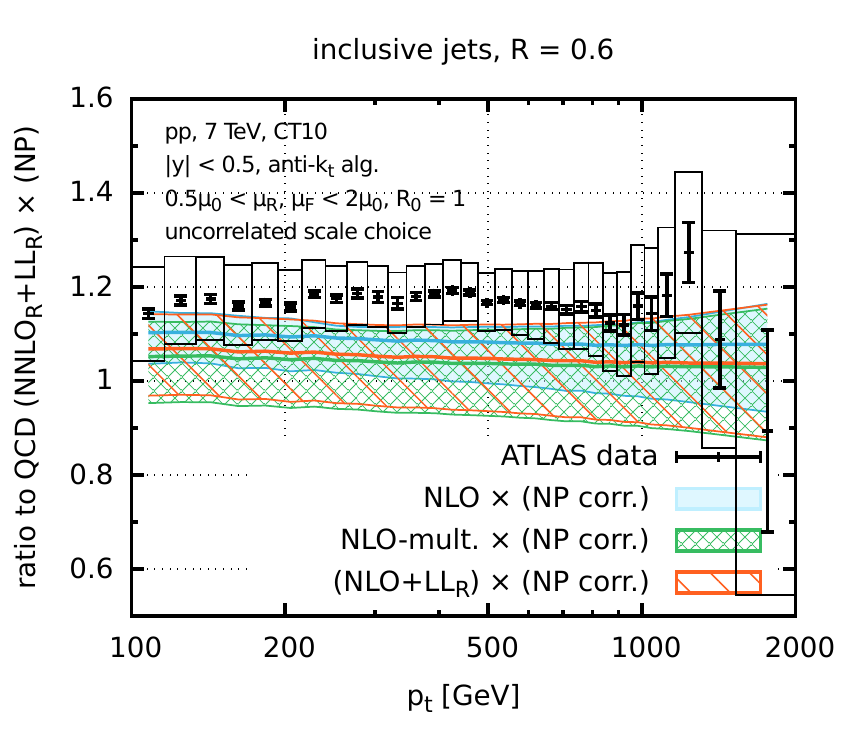}%
  \includegraphics[width=0.5\textwidth]{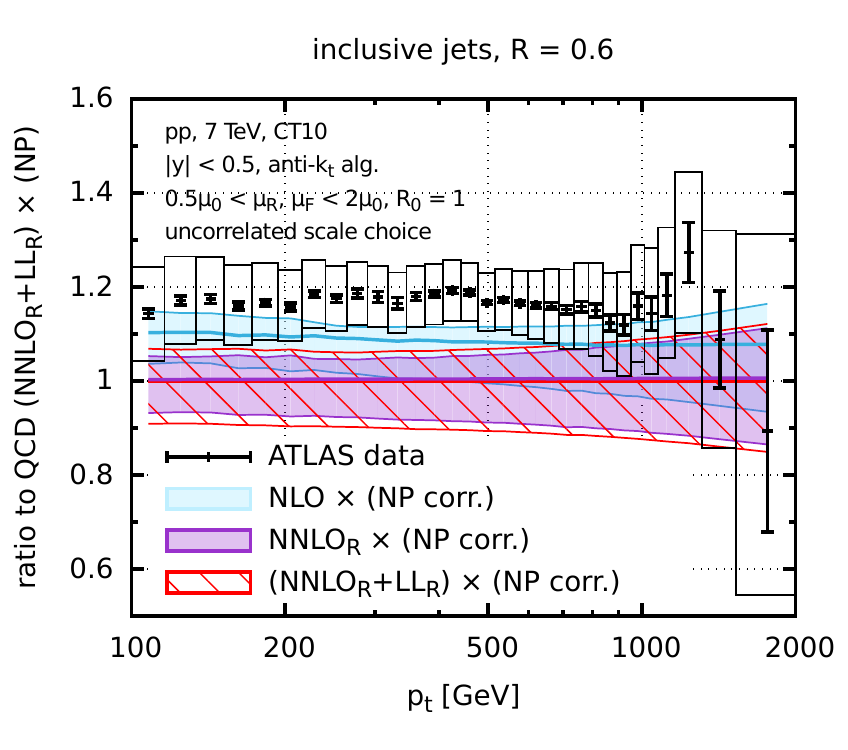}\\
  \caption{Comparison between a range of theoretical predictions for
    the inclusive jet spectrum and data from
    ATLAS at $\sqrt{s}=7\TeV$~\cite{Aad:2014vwa} 
    in the rapidity bin $|y|<0.5$.
    The upper row is for $R=0.4$ and the lower one for $R=0.6$.
    The left-hand column shows NLO-based comparisons, while the
    right-hand one shows \NNLOR-based comparisons.
    Rectangular boxes indicate the size of systematic uncertainties on
    the data points, while the errors bars correspond to the
    statistical uncertainties.
    Results are normalised to the central \NNLORLLR prediction
    (including non-perturbative corrections).
  }
  \label{fig:atlas-comparison}
\end{figure}
\begin{figure}
  \centering
  \includegraphics[width=0.5\textwidth]{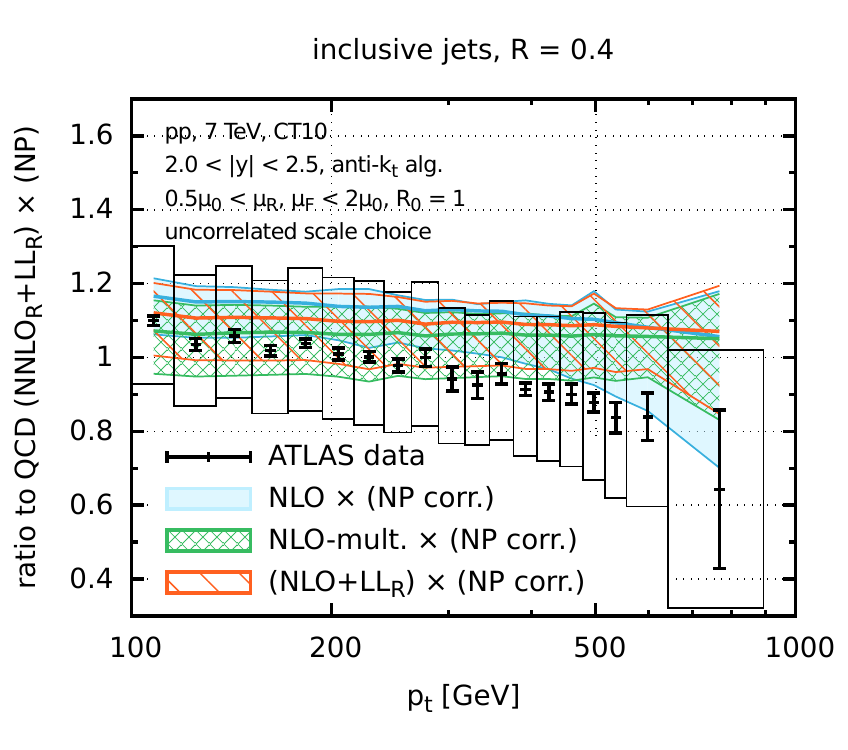}%
  \includegraphics[width=0.5\textwidth]{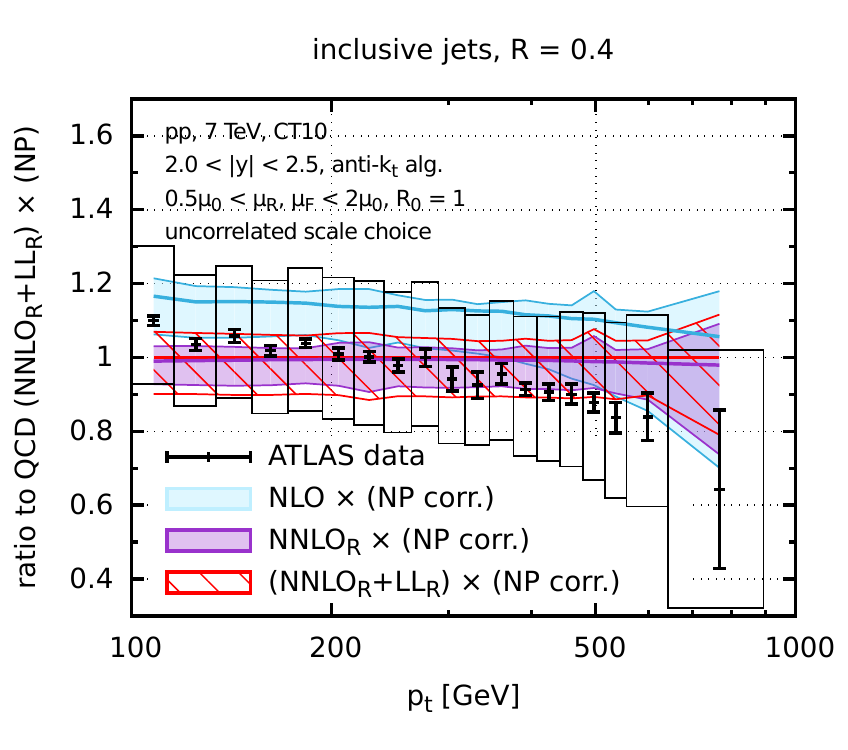}\\
  \includegraphics[width=0.5\textwidth]{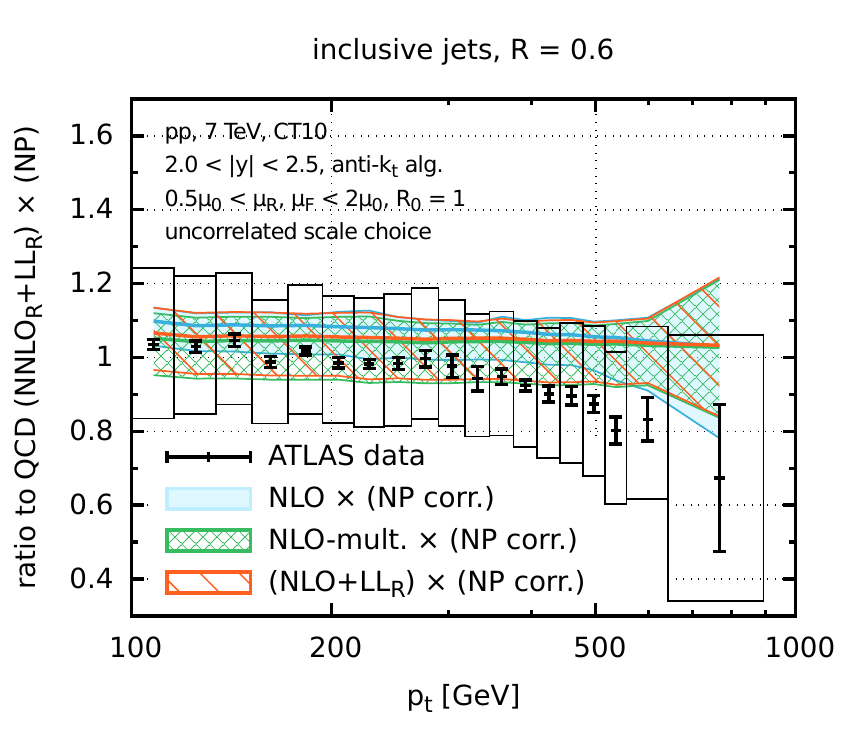}%
  \includegraphics[width=0.5\textwidth]{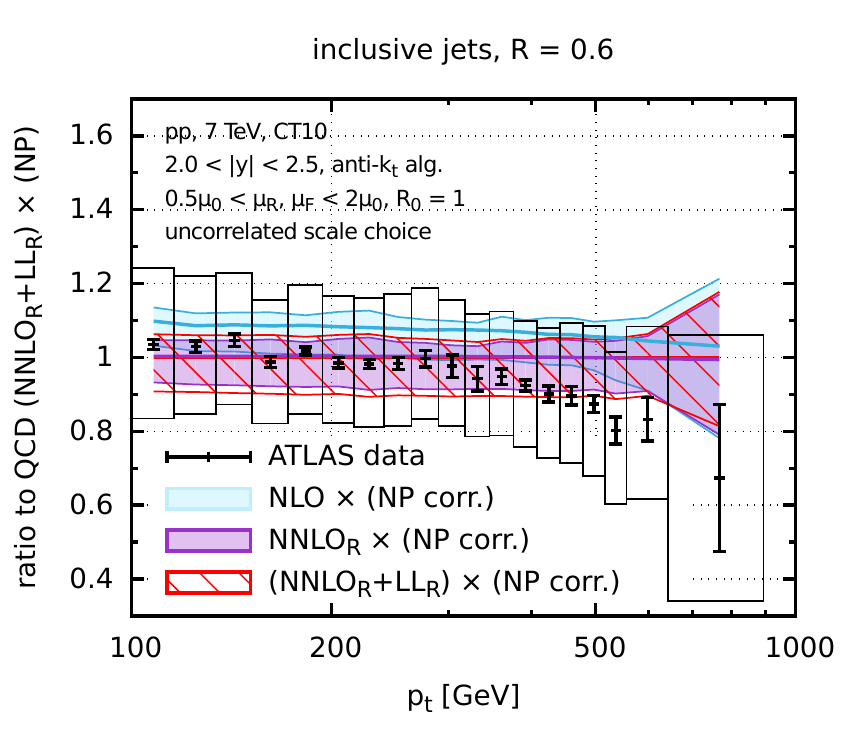}\\
  \caption{Analogue of Fig.~\ref{fig:atlas-comparison}, but for the
    rapidity bin $2 < |y| < 2.5$.}
  \label{fig:atlas-comparison-highrap}
\end{figure}

The hadronisation and underlying event corrections applied are shown
in Fig.~\ref{fig:non-pert-atlas}.
As in the case of the ALICE data, for $R=0.4$ these two classes of correction mostly
cancel.
When increasing the jet radius to $R=0.6$, the hadronisation
corrections shrink, while the UE corrections increase and now
dominate, leaving a net effect of up to $6{-}7\%$ at the lowest $p_t$'s.

Figs.~\ref{fig:atlas-comparison} and
\ref{fig:atlas-comparison-highrap} show comparisons between data and
theory for two rapidity bins, $|y|<0.5$ and $2.0<|y|<2.5$.
At central rapidities the situation here contrasts somewhat with that
for the ALICE data
and in particular the inclusion of \NNLOR corrections worsens the
agreement with data:
over most of the $p_t$ range, the data points are about $15{-}20\%$
higher than than either \NNLOR or \NNLORLLR (which are close to each
other, as expected for $R \gtrsim 0.4$).
Nevertheless, one encouraging feature of the \NNLOR-based predictions
is that there is now a consistent picture when comparing $R=0.4$ and
$R=0.6$, insofar as the ratio of data to \NNLOR-theory is essentially
independent of $R$.
This is not the case when comparing data and NLO predictions (cf.\
Fig.~\ref{fig:correl-v-uncorrel-scale-choice-nnlo}, which shows the
steeper $R$ dependence of \NNLOR-based results as compared to NLO).
We return to the question of $R$ dependence in more detail below.

In the forward rapidity bin, over most of the $p_t$ range, the data
instead favours the \NNLOR-based predictions over NLO, while at high
$p_t$ the data falls below all of the predictions.
However the systematic uncertainties on the data are slightly larger
than the difference with any of the theory predictions, making it
difficult to draw any solid conclusions.

A significant positive 2-loop correction (cf.\ the discussion in
sections~\ref{sec:double-virtual-impact}, \ref{sec:powheg-comparison}
and~\ref{sec:data-theory-K}) would bring overall better agreement at
central rapidities, but would worsen the agreement at forward
rapidities.
However, the finite 2-loop effects can be $p_t$ and rapidity
dependent, making it difficult to draw any conclusions at this stage.
Furthermore, one should keep in mind that adjustments in PDFs could
affect different kinematic regions differently.

We close this section with an explicit comparison of the ratio of the
jet spectra for the two different $R$ values.
For the theoretical prediction, we proceed as discussed in the
previous subsection, when we made a comparison with the ALICE data for
such a ratio.
We will not include EW effects, since in the ratio they appear to be
at a level well below $1\%$.

Concerning the experimental results, the central value of the ratio
can be obtained directly from the ATLAS data at the two $R$ values.
However the ATLAS collaboration has not provided information on the
uncertainties in the ratio.
It has provided information~\cite{ATLASHepData} to facilitate the determination of
correlations between $p_t$ and rapidity bins, specifically $10000$
Monte Carlo replicas of their data to aid in estimating statistical
correlations, as well as a breakdown of systematic uncertainties into
$\order{70}$ sources that are individually $100\%$ correlated across
bins and totally uncorrelated with each other.
The information is presented in a format such that, technically, it can
also be used to estimate the uncertainties in the ratio of cross section
for two $R$ values.
However, we have been advised by the ATLAS collaboration that the
degree of correlation between systematic uncertainties at different
$R$ values is not well known.
Accordingly, we label the uncertainties obtained in this way as
``approx. uncert.'' to emphasise that we do not have full knowledge of
the experimental uncertainties in the ratio and that they are potentially
larger than our estimate.

\begin{figure}
  \centering
  \includegraphics[width=0.5\textwidth,page=1]{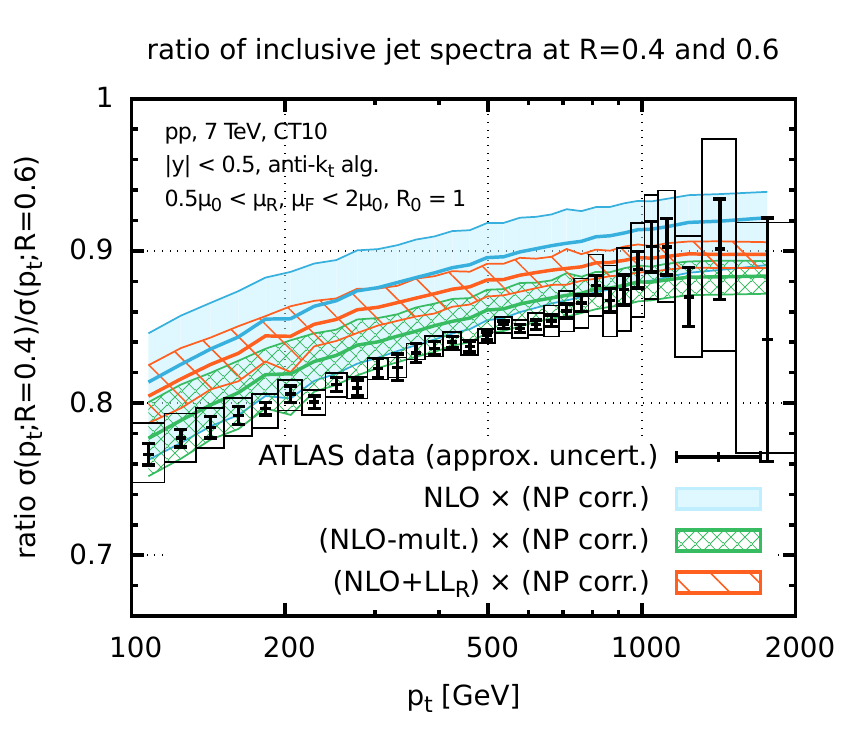}%
  \includegraphics[width=0.5\textwidth,page=1]{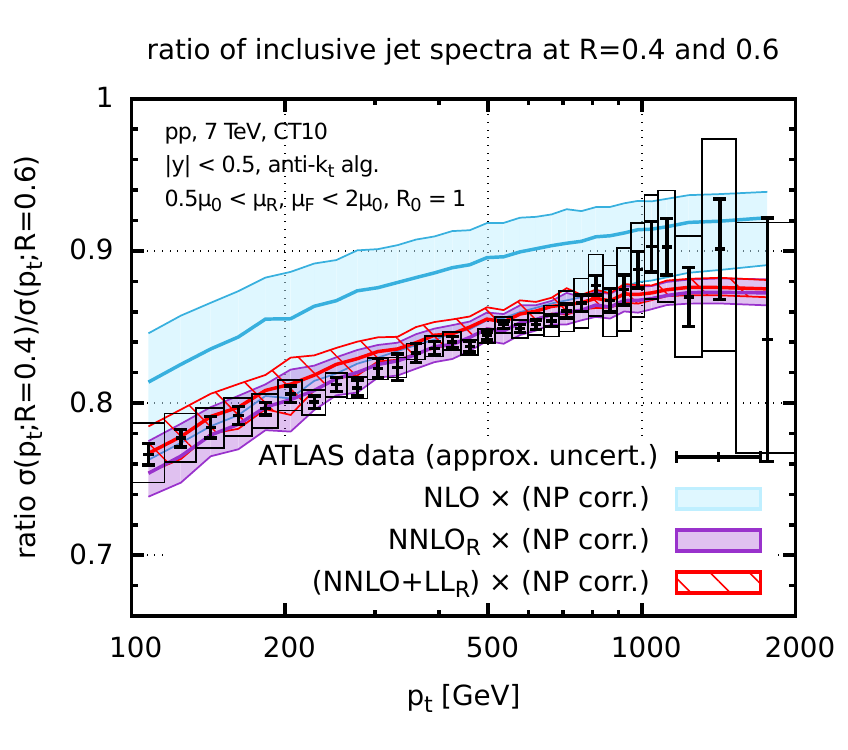}
  \caption{Comparison between a range of theoretical predictions for
    the inclusive jet cross-section ratio and data from
    ATLAS~ at $\sqrt{s}=7\TeV$~\cite{Aad:2014vwa}.
    The left-hand column shows NLO-based comparisons, while the
    right-hand one shows NNLO$_{(R)}$-based comparisons.
    Rectangular boxes indicate our estimated systematic uncertainties
    on the data points, while the errors bars correspond to the
    statistical uncertainties. 
    Note that these estimates are known to be incomplete, insofar as the
    information provided by the ATLAS collaboration on its results is
    not intended to be used for the determination of uncertainties on
    cross section ratios at different radii.  }
  \label{fig:atlas-comparison-ratio}
\end{figure}

Keeping in mind this caveat, we show in
Fig.~\ref{fig:atlas-comparison-ratio} a comparison between various
theoretical predictions for the cross section ratio at $R=0.4$
relative to $R=0.6$, together with the experimental data.
One sees overall very good agreement with both the \NNLOR and
\NNLOLLR-based results, and substantially worse accord with NLO-based
predictions (albeit consistent with pure NLO and \NLO-mult.\ within
their larger uncertainties).

\subsection{Brief comparisons with an NNLO $K$-factor}
\label{sec:data-theory-K}

\begin{figure}
  \centering
  \includegraphics[width=0.5\textwidth]{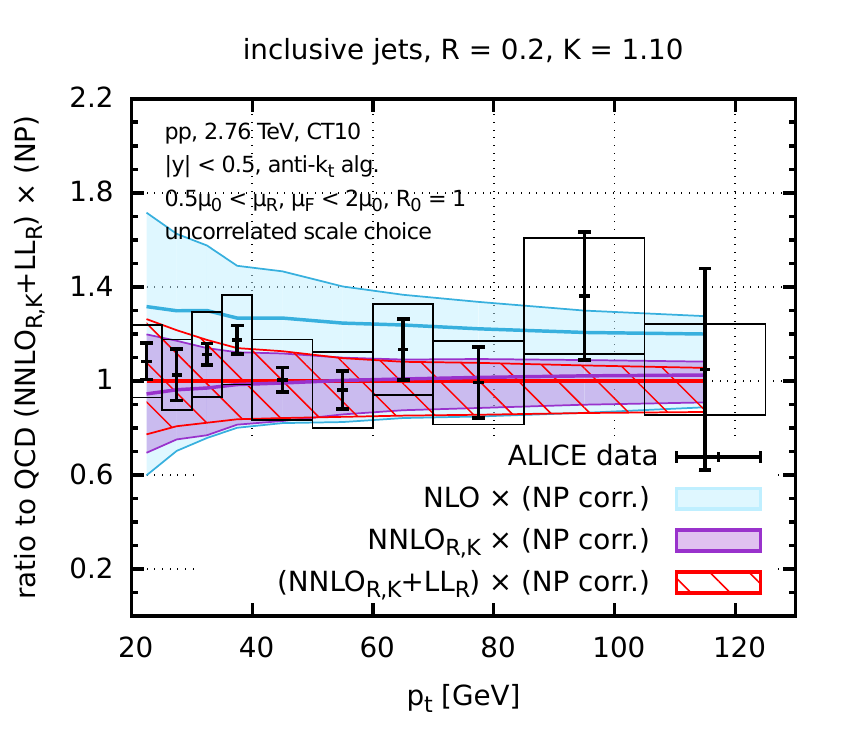}%
  \hfill
  \includegraphics[width=0.5\textwidth]{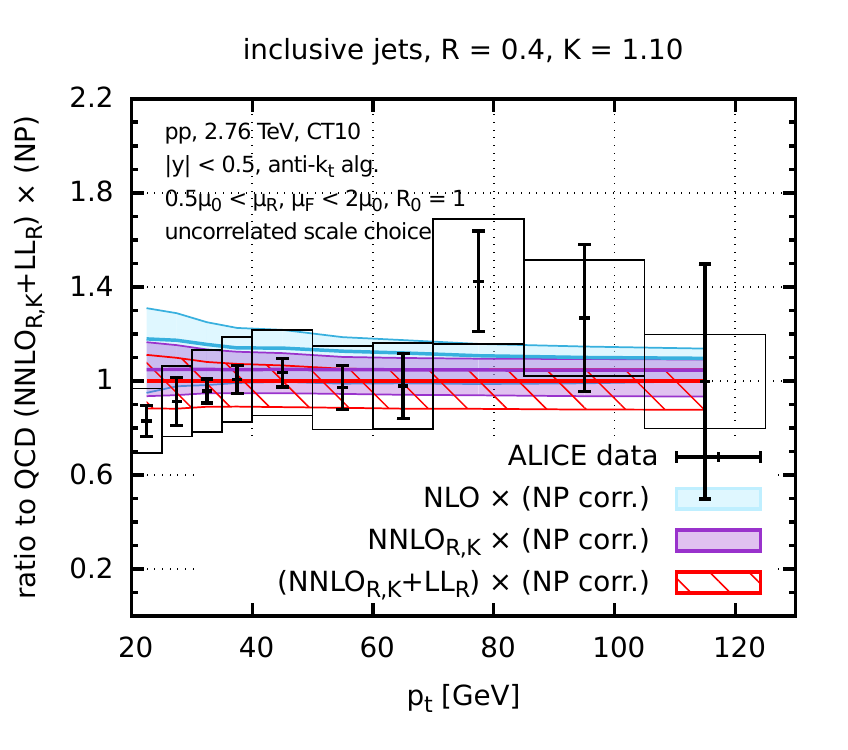}
  \caption{Comparison between theoretical predictions with 
    a NNLO $R_m=1$ correction factor $K=1.10$ and data from 
    ALICE at $\sqrt{s}=2.76\TeV$~\cite{Abelev:2013fn} at 
    $R=0.2$ and $R=0.4$.
    Rectangular boxes indicate the size of systematic uncertainties on
    the data points, while the errors bars correspond to the
    statistical uncertainties.
    Results are normalised to the central \NNLORKLLR prediction
    (including non-perturbative corrections).
  }
  \label{fig:alice-comparison-Kfact}
\end{figure}

\begin{figure}
  \centering
  \includegraphics[width=0.5\textwidth]{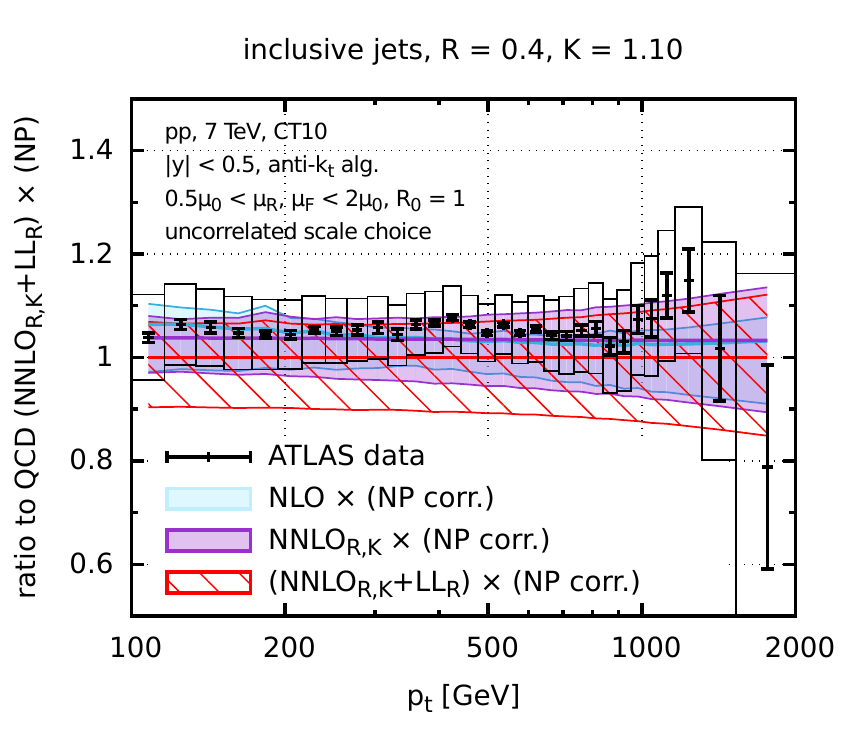}%
  \hfill
  \includegraphics[width=0.5\textwidth]{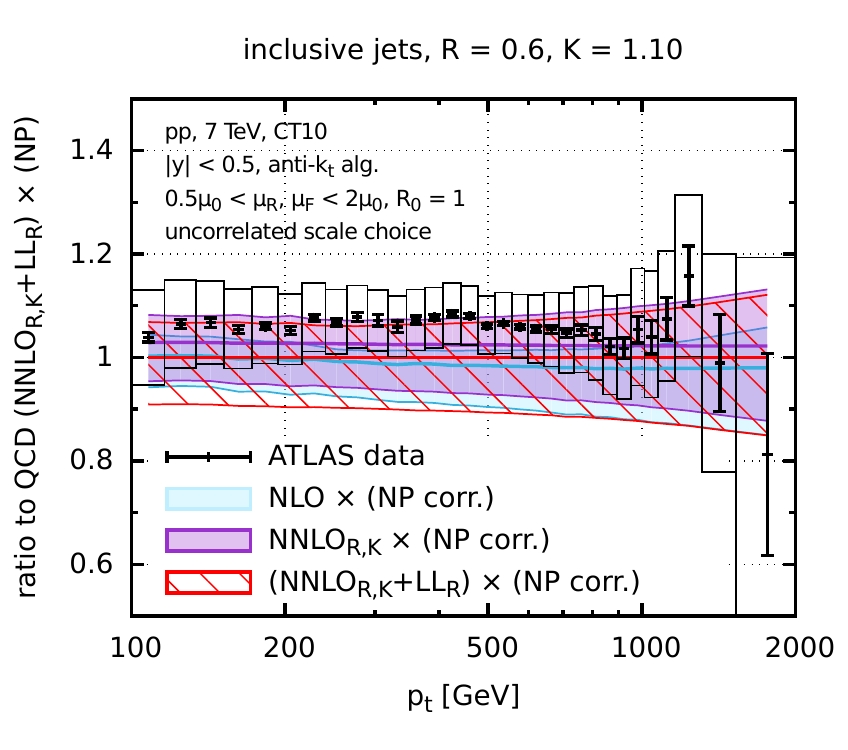}
  \caption{Comparison between theoretical predictions with a NNLO
    $R_m=1$ correction factor 
    $K = 1.10$ and 
    data from ATLAS at $\sqrt{s}=7\TeV$~\cite{Aad:2014vwa} 
    in the rapidity bin $|y|<0.5$, for $R=0.4$ and $R=0.6$.
    Rectangular boxes indicate the size of systematic uncertainties on
    the data points, while the errors bars correspond to the
    statistical uncertainties.
    Results are normalised to the central \NNLORKLLR prediction
    (including non-perturbative corrections).  }
  \label{fig:atlas-comparison-Kfact}
\end{figure}

\begin{figure}
  \centering
  \includegraphics[width=0.5\textwidth]{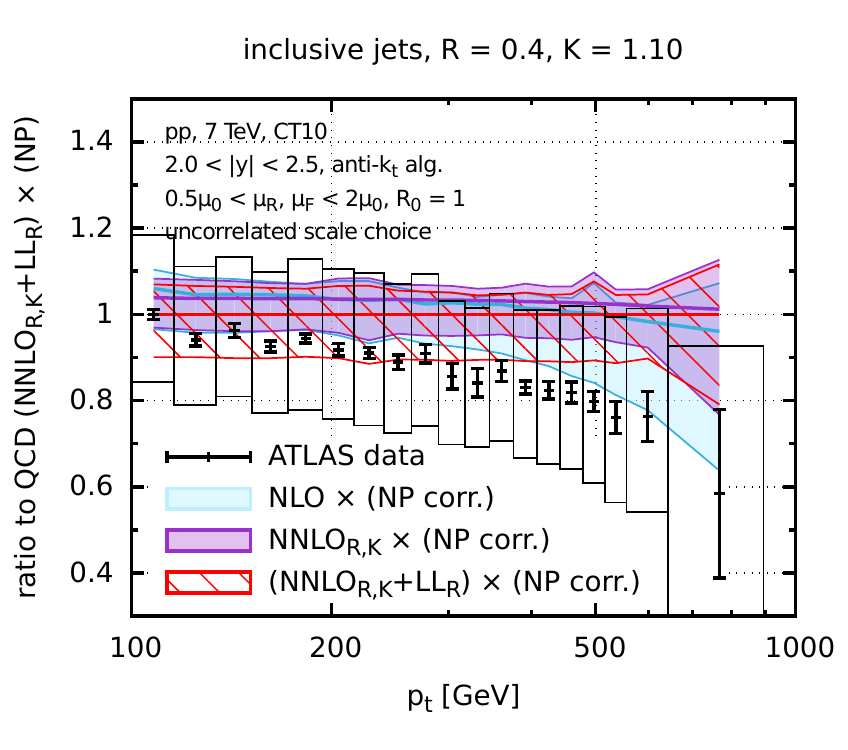}%
  \includegraphics[width=0.5\textwidth]{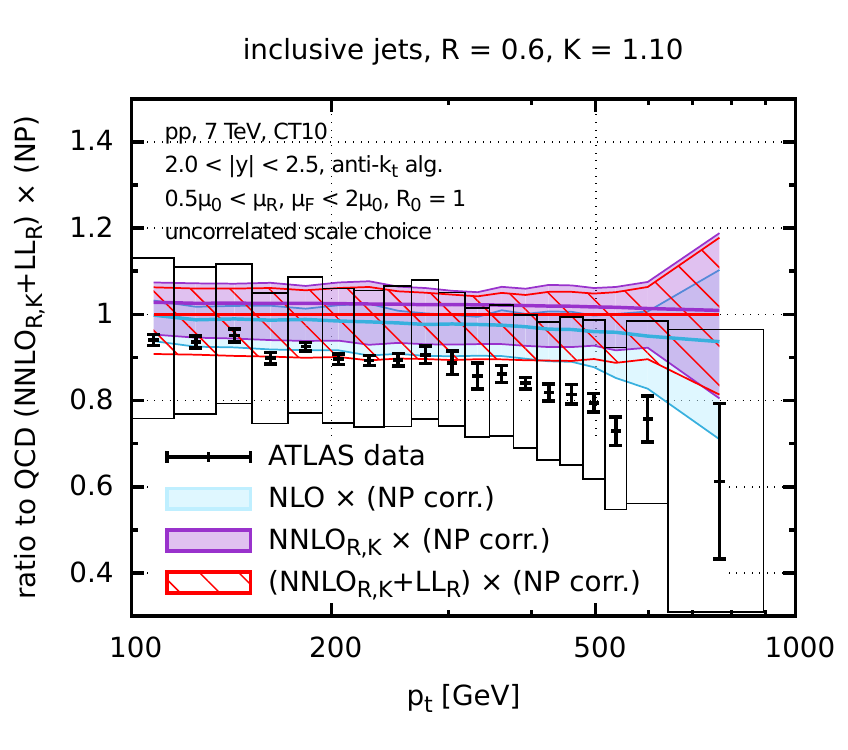}
  \caption{Analogue of Fig.~\ref{fig:atlas-comparison-Kfact}, but for the
    rapidity bin $2 < |y| < 2.5$.}
  \label{fig:atlas-comparison-highrap-Kfact}
\end{figure}

For completeness, here we show the comparisons between theoretical
predictions and data change when we introduce a two-loop $K$-factor for
$R= R_m$, as described in section~\ref{sec:double-virtual-impact}.
Figures \ref{fig:alice-comparison-Kfact},
\ref{fig:atlas-comparison-Kfact} and
\ref{fig:atlas-comparison-highrap-Kfact} are to be compared to their
counterparts in sections~\ref{sec:alice-comparisons}
and~\ref{sec:atlas-comparisons} i.e. Figs.~\ref{fig:alice-comparison},
\ref{fig:atlas-comparison} and \ref{fig:atlas-comparison-highrap}.
In most cases, the changes that one observes are largely as expected,
with a corresponding trivial rescaling of the observed
data--theory ratio.
One exception is in the case of the $R=0.2$ comparison to ALICE data,
Fig.~\ref{fig:alice-comparison-Kfact} (left), where with $K=1.10$ one
observes that the \NNLORK results are now in very close accord with
the \NNLORKLLR results.
This is to be contrasted with the situation in
Fig.~\ref{fig:alice-comparison}. 
The difference is due to the fact that the $K$ factor acts additively
on the \NNLORK result, but multiplicatively on the \NNLORKLLR result,
as discussed already in section~\ref{sec:double-virtual-impact}.

Note that for the ATLAS comparison, while a $K$-factor of $K=1.10$
improves agreement with the data at central rapidities, it appears to
worsen it somewhat at high rapidities, as can be seen in 
Fig.~\ref{fig:atlas-comparison-highrap-Kfact}.
One should, however, keep in mind that the true $K$-factor will depend
both on rapidity and $p_t$, and also that modifications associated
with changes in PDFs can affect forward and central rapidities
differently.


\section{Conclusion}
\label{sec:conclusion}

In this paper we have used the limit of small-radius jets to explore a
variety of features of the most basic of jet observables, the
inclusive jet spectrum.

A first observation, in section~\ref{sec:resum}, was that the
small-$R$ approximation starts to reproduce fixed-order $R$ dependence
quite well already for $R$ just below $1$, giving us confidence in the
usefulness of that approximation for phenomenologically relevant $R$
values.

In seeking to combine small-$R$ resummation with NLO predictions, in
section~\ref{sec:matching}, it was natural to write the cross section as a
product of two terms: an overall normalisation for elementary partonic
scattering, together with a factor accounting for fragmentation of
those partons into small-$R$ jets. 
Such a separation can be performed also at fixed order.
There appear to be spurious cancellations between higher-order
contributions for the two factors and this led us to propose that one
should estimate their scale uncertainties independently and then add
them in quadrature.
This procedure has similarities with methods used for jet vetoes in
Higgs physics~\cite{Stewart:2011cf,Banfi:2012yh}.

We also saw that there are large $R$-dependent terms at NNLO that are
beyond the control of our \LLR resummation
(sections~\ref{sec:smallR-validation} and \ref{sec:matching-NNLO}).
To account for them in the absence of the full NNLO calculation, we
introduced a stand-in for NNLO that we called \NNLOR. 
This is defined to be identical to NLO for $R=1$ but includes full
NNLO $R$ dependence, which can be obtained from a NLO 3-jet
calculation.
Once complete NNLO predictions become available, it will be trivial to
replace the \NNLOR terms with \NNLO ones. 

For an accurate description of the inclusive jet spectrum one must
also account for non-perturbative effects.
In section~\ref{sec:data-comparisons} we revisited the analytical
hadronisation predictions of Ref.~\cite{Dasgupta:2007wa}. 
We found that the predicted scaling with $R$ and the parton flavour
was consistent with what is observed in Monte Carlo simulations.
However such simulations additionally show a non-trivial $p_t$
dependence that is absent from simple analytical estimates.
Accordingly we decided to rely just on Monte Carlo simulations to
evaluate non-perturbative corrections.

We compared our results to data from the ALICE and ATLAS
collaborations in section~\ref{sec:data-comparisons}.
For the smallest available $R$ value of $0.2$, both the \NNLOR and the \LLR
corrections beyond \NNLOR play important roles and at the lower end of
ALICE's $p_t$ range, the effect of \NNLOR corrections was almost
$50\%$, while further \LLR corrections mattered at the $20\%$ level.
For $R =0.4$, \NNLOR corrections still mattered, typically at the
$10{-}30\%$ level, depending on the $p_t$.
However \LLR resummation then brought little additional change. 
Overall, for the ALICE data and the forward ATLAS data, \NNLORLLR
brought somewhat better agreement than NLO, while
for central rapidities, the ATLAS data were substantially
above the \NNLORLLR predictions.
It will be important to revisit the pattern of agreement once the full
NNLO corrections are known, taking into account also aspects such as
correlated experimental systematic uncertainties and PDF uncertainties.

Where the \NNLOR and \NNLORLLR predictions clearly make the most
difference is for reproducing the $R$-dependence of the cross
sections.
For the inclusive spectrum plots, once one goes to \NNLORLLR, the
picture that emerges is consistent across different values of $R$. 
That was not the case at NLO.
This is visible also in the ratios of cross sections at different $R$
values. 
In particular, for the reasonably precise ATLAS data, \NNLOR and
\NNLOLLR are in much better agreement with the data than the NLO-based
predictions.
For the ALICE data, the uncertainties are such that it is harder to
make a definitive statement. 
Nevertheless \NNLOLLR performs well and notably better than plain
\NNLOR. 


Overall, the substantial size of subleading $R$-enhanced terms in the NNLO
corrections also motivates studies of small-$R$ resummation beyond
\LLR accuracy and of small-$R$ higher order effects in other jet
observables.
 
A final comment concerns long-term prospects. 
We have seen here that the availability of data at multiple $R$ values
provides a powerful handle to cross-check theoretical predictions.
As the field moves towards ever higher precision, with improved
theoretical predictions and reduced experimental systematic
uncertainties, cross checks at multiple $R$ values will, we believe,
become increasingly important.
In this respect, we strongly encourage measurements at three different
radii. 
Small radii, $R\simeq 0.2{-}0.3$, are particularly sensitive to
hadronisation effects; large radii, $R\simeq 0.6{-}0.8$ to underlying
event effects;  the use of an intermediate radius $R\simeq0.4$
minimises both and provides a good central choice.
Only with the use of three radii do we have a realistic chance of
disentangling the three main sources of theoretical uncertainties, namely
perturbative effects, hadronisation, the underlying event.


\section*{Acknowledgements}

MD and GPS are grateful to the MITP for hospitality and support
during part of this work.
We would all like to thank Matteo Cacciari for numerous useful
discussions and comments throughout this work.
We are grateful to Simone Alioli and Emanuele Re for assistance and
cross-checks with \powheg and to them and the remaining authors of
Ref.~\cite{Alioli:2010xa} for discussions about the results.
We thank Peter Skands for discussions about the non-perturbative
effects.
We also thank Jan Kretzschmar, Mark Sutton and Pavel Starovoitov for
discussions about ATLAS's non-perturbative correction factors and the
first of them for discussions about the determination of uncertainties
on ratios of cross sections at different radius values.
Finally we are grateful to an anonymous referee for constructive
comments on the structure of the paper.

MD's work is supported in part by the Lancaster-Manchester-Sheffield
Consortium for Fundamental Physics under STFC grant ST/L000520/1. 
FD is supported by the ILP LABEX (ANR-10-LABX-63) supported by French
state funds managed by the ANR within the Investissements d'Avenir
programme under reference ANR-11-IDEX-0004-02.
GPS and GS's work is supported by ERC advanced grant Higgs@LHC.
GS's work is also supported by the Paris-Saclay IDEX.


\appendix


\section{Differences in $\as$ and $t$ expansions}
\label{sec:convergence-t-v-as}

In our original work on small-$R$ effects~\cite{Dasgupta:2014yra}, we
compared the \LLR resummed results to an expansion in powers of $t$,
with $t$ as defined in Eq.~(\ref{eq:t}).
However $t$ is a non-trivial function of $\as$ and expansions in $\as$
and $t$ can have different convergence properties.

\begin{figure}
  \centering
  \includegraphics[width=0.48\textwidth,page=2]{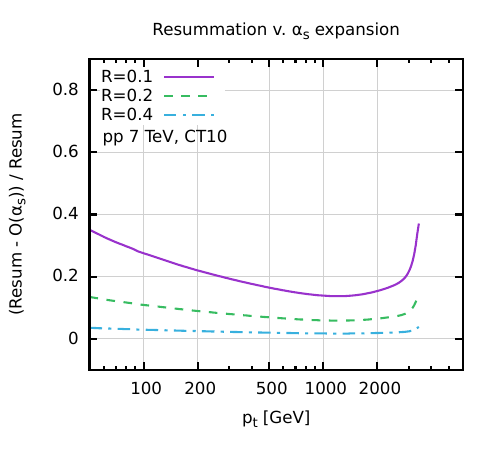}\hfill
  \includegraphics[width=0.48\textwidth,page=4]{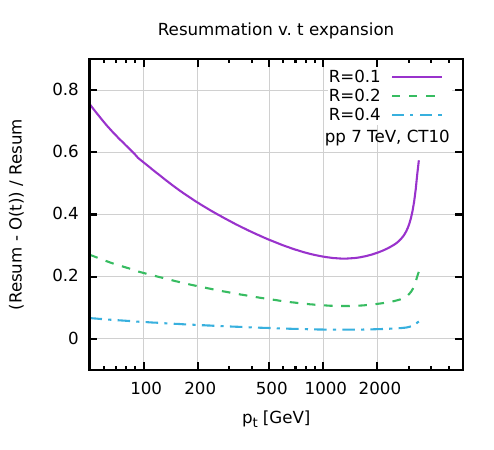}
  \caption{Comparison of the convergence of the $\as$ expansion  of
    the small-$R$ resummation (left) relative to that of the $t$
    expansion (right).
  }
  \label{fig:expansion-alphas}
\end{figure}

This is illustrated in Fig.~\ref{fig:expansion-alphas}.
The left-hand plot shows the difference between the resummation and
its expansion to NLO in powers of $\as$, normalised to the full
resummed result.
The right-hand plot shows the difference between the resummation and
its expansion to NLO in powers of $t$, with the same normalisation.
Three different $R$ values are shown.
One sees that the $t$ expansion converges more slowly than the $\as$
expansion.

At first sight, this observation is somewhat surprising, insofar as
$t$ would appear to be the natural variable for considering small $R$
effects.
Part of the explanation is as follows: the $t$ expansion has
alternating sign coefficients, at least for the first couple of
orders. 
This means that the NLO $\order{t}$ correction (relative to LO) is
larger than the overall effect of resummation.
The $\order{\as}$ correction has the same coefficient as the
$\order{t}$ correction (modulo an overall normalisation factor of
$\frac1{2\pi} \ln R_0^2/R^2$).
However, $t$ involves the average of the coupling over a range of
scales between $p_t$ and $R p_t$, which is larger than $\as(p_t)$.
Consequently, the NLO $\order{\as}$ term is not as large as
the NLO $\order{t}$ term, and it overshoots the resummation by less.
Though not illustrated in Fig.~\ref{fig:expansion-alphas}, a similar
phenomenon occurs also when comparing to expansions at NNLO.

One should keep in mind that for observables other than
the inclusive jet spectrum, it may no longer be true that a $t$
expansion converges more slowly than an $\as$ expansion.
Rather, when discussing fixed-order convergence properties compared to
full small-$R$ resummation, one should simply be aware that the
convergence properties of the $t$ and $\as$ expansions will be
sometimes be noticeably different.

Note also that the above discussion holds specifically for the
expansion of the \LLR result.
As we have seen in section~\ref{sec:smallR-validation}, N\LLR effects
are large and at NNLO are of opposite sign to the \LLR contribution. 
This further complicates the discussion of the convergence properties
of the inclusive jet spectrum.

\section{Scale choice beyond leading order}
\label{sec:central-scale-choice}

When making fixed-order predictions for the inclusive jet cross
section, there are two widely used prescriptions for the choice of a
central renormalisation and factorisation scale. 
One prescription is to use a single scale for the whole event, set by
the $p_t$ of the hardest jet in the event, $\mu_0 = p_{t,\max}$.
This was adopted, for example, in Ref.~\cite{Aad:2014vwa}.
Another prescription is to take instead a different scale for each
jet, specifically that jet's $p_t$, $\mu_0 = p_{t,\text{jet}}$. 
This was adopted for example in
Ref.~\cite{Abelev:2013fn}.\footnote{Note that yet other scale choices
  have been used in the literature, notably in predictions for dijet
  masses~\cite{Ellis:1992en,Aad:2011fc}.}

At LO, the two prescriptions give identical results, since there are
only two jets in the event and they have the same $p_t$.
However, starting from NLO the prescriptions can differ substantially.
Interestingly, a study of the small-radius limit can provide
considerable insight into which choice is more appropriate.

\begin{figure}
  \centering
  \includegraphics[width=0.5\textwidth]{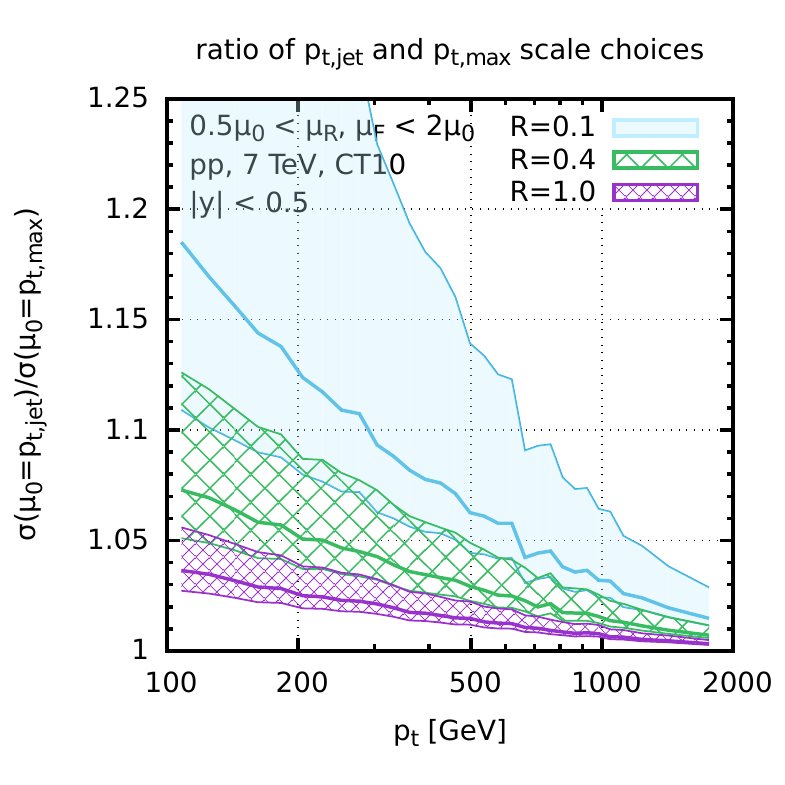}\hfill
  \includegraphics[width=0.5\textwidth,page=1]{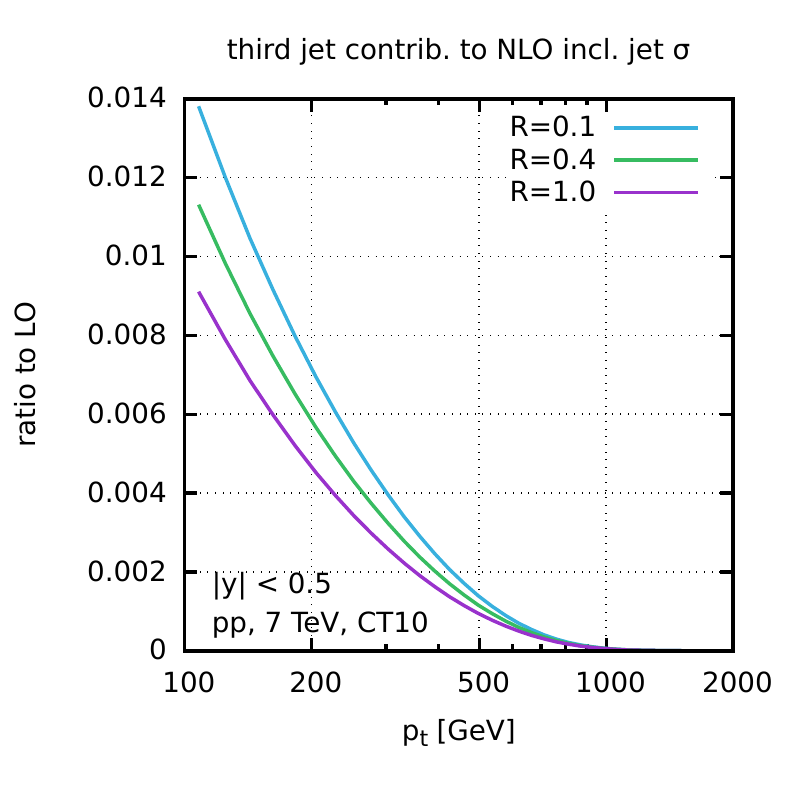}
  \caption{
    Left: ratio of NLO predictions for the inclusive spectrum when
    using the per-jet scale choice $\mu_0 = p_{t,\text{jet}}$ versus the
    per-event choice $\mu_0 = p_{t,\max}$.
    The results are shown as a function of jet $p_t$ for three jet radius choices, $R=0.1,0.4$
    and $1.0$ and have been obtained with \nlojet.
    The bands correspond to the effect of scale variation, where the
    scales are varied upwards and downwards by a factor of two
    simultaneously for the numerator and denominator.
    Right: fraction of the inclusive jet spectrum (for
    $|y|<0.5$) that comes from jets beyond the two hardest.
    The 3-jet rate and the overall normalisation are both evaluated at LO.
  } 
  \label{fig:scale-choice}
\end{figure}

Figure~\ref{fig:scale-choice} (left) shows the ratio of the NLO result as
obtained with $\mu_0 = p_{t,\text{jet}}$ to that with
$\mu_0 = p_{t,\text{max}}$, as a function of the jet $p_t$, for three
different jet radii.
The main observation is that the $\mu_0 = p_{t,\text{jet}}$ prescription
increases the cross section, especially at small radii: it brings an
increase of almost $20\%$ for $R=0.1$ at low $p_t$, versus
$\lesssim 4\%$ for $R=1.0$ (in both cases for a central scale choice). 
As we saw in section~\ref{sec:matching-NNLO}, for reasonably small $R$,
the NNLO corrections suppress the cross section.
Therefore the choice $\mu_0 = p_{t,\text{jet}}$ takes us in the wrong
direction.

In order to understand this better, it is useful to make a number of
observations:
\begin{enumerate}
\item For the virtual part of the NLO calculation, the two scale
  prescriptions give identical results, so the deviation of the ratio
  from $1$ in Fig.~\ref{fig:scale-choice} (left) can come only from the real
  part.
\item The real part itself involves two different pieces: that from
  binning either of the two leading jets, and that from binning the
  3rd jet.
  The right-hand plot of Fig.~\ref{fig:scale-choice} shows that the
  leading-order 3rd-jet contribution is at the level of $1{-}2\%$ of
  the leading-order dijet result and so it is reasonable to neglect it
  in our discussion.\footnote{The 3rd jet is produced with a
    probability $\order{\as}$, however because its $p_t$ is lower than
    that of the two leading jets, its contribution to the (steeply
    falling) jet spectrum is substantially suppressed.}
\item When a real emission is within an angle $R$ of its nearest other
  parton, there are only two jets in the event and the two
  scale-choice prescriptions are identical.
\item Differences between the prescriptions arise when the softest
  parton falls outside one of the two leading jets.
  Then one of those jets has a reduced $p_t$ and the choice $\mu_0 =
  p_{t,\text{jet}}$ gives a smaller scale than $\mu_0 =
  p_{t,\text{max}}$.
  This occurs with a probability that is enhanced by $\ln 1/R$.
\item At $p_t\sim 100\GeV$, where the effects are largest,
  renormalisation scale ($\mu_R$) variations play a much larger role
  than factorisation scale ($\mu_F$) variations.
  Therefore a smaller scale translates to a larger value of $\as$ and
  thus a larger cross section for the real contribution (which is
  always positive).
  Consequently, the prescription $\mu_0 = p_{t,\text{jet}}$ leads to a cross
  section that is larger than the prescription $\mu_0 =
  p_{t,\text{max}}$ and the difference is enhanced by a factor $\ln
  1/R$ for small $R$.
\end{enumerate}
This qualitatively explains the behaviour seen in Fig.~\ref{fig:scale-choice}
(left). 
The $\mu_0 = p_{t,\text{jet}}$ scale choice introduces a
correction that goes in the wrong direction because it leads to a
smaller scale (and larger $\as$) for the real part, but without a
corresponding modification of the virtual part.
Thus it breaks the symmetry between real and virtual corrections.

The above reasoning leads us to prefer the $\mu_0 = p_{t,\text{max}}^{R=1}$
prescription. 
To make it a unique event-wide choice, independent of $R$, we define
always define $\mu_0=p_{t,\text{max}}$ using jets with a radius equal to one,
regardless of the $R$ value used in the measurement.

We note that $\mu_0=p_{t,\text{max}}$ has a potential linear sensitivity
to initial-state radiation, i.e.\ initial state radiation of
transverse momentum $p_{t,i}$ shifts $\mu_0$ by an amount $p_{t,i}$.
A yet more stable choice might be $\mu_0=\frac12(p_{t,1}+p_{t,2})$, the
average transverse momentum of the two hardest jets (again defined
with a radius of one).
For this choice, the shift of $\mu_0$ would be limited
$\order{p_{t,i}^2/(p_{t,1}+p_{t,2})}$.
We leave its study to future work.

Yet another option is the use of MINLO type
procedures~\cite{Hamilton:2012np}.
For dijet systems, this should be rather similar to
$\mu_0=\frac12(p_{t,1}+p_{t,2})$.

\clearpage

\end{document}